\documentclass[prb,twocolumn,longbibliography,superscriptaddress]{revtex4-2}
\usepackage{graphicx}
\usepackage{textgreek}

\usepackage{amsmath}
\usepackage{amssymb}
\usepackage{hyperref}
\usepackage[capitalize]{cleveref} 
\usepackage{braket} 
\usepackage{mathtools} 
\usepackage{cancel}

\AtBeginDocument{
	\renewcommand{\Re}{\operatorname{Re}}
	
}

\DeclareMathOperator{\tr}{tr}
\DeclareMathOperator{\Tr}{Tr}
\DeclareMathOperator{\sgn}{sgn}
\DeclareMathOperator{\diag}{diag}
\DeclarePairedDelimiter\abs{\lvert}{\rvert}

\begin{document}
\title{Superfluid stiffness within Eliashberg theory: the role of vertex corrections}
\author{Zachary M. Raines}
\author{Shang-Shun Zhang}
\author{Andrey V. Chubukov}
\affiliation{School of Physics and Astronomy and William I. Fine Theoretical Physics Institute, University of Minnesota, Minneapolis, MN 55455, USA}
\begin{abstract}
	In this work we consider the superfluid stiffness of a generically non-Galilean invariant interacting system and investigate under what conditions the stiffness may nonetheless approach the Galilean-invariant value $n/m$. Within Eliashberg theory we find that the renormalized stiffness is approximately given by $n/m$ in the case when the
	$l=0$
	and $l=1$ components of the effective Fermi-surface projected interaction are approximately equal over a range of frequencies.
	This holds, in particular, when the interaction is peaked at zero momentum transfer.
	We examine this result through three complementary lenses: the $\delta (\omega)$ term in the conductivity, the phase dependence of the Luttinger-Ward free energy, and the coupling of the amplitude and phase sectors in the Hubbard-Stratonovich collective mode action.
	From these considerations we show that the value of the stiffness is determined by the strength of renormalization of the current vertex and that the latter can be
	interpreted as the shift of the self-consistent solution due to flow of the condensate, or alternatively as coupling of the phase mode to $l=1$ fluctuations of the order parameter.
	We highlight that even though the superfluid stiffness in some non-Galilean systems approaches the Galilean value, this is not enforced by symmetry, and in general the stiffness may be strongly suppressed from its BCS value.
	As a corollary we obtain the generic form of the phase action within Eliashberg theory
	and charge and spin Ward identities for a superconductor with frequency dependent gap function.
\end{abstract}

\maketitle

\section{Introduction}

The superfluid stiffness $D_s (T)$ is one of the key characteristics of a superconductor -- it determines the strength of the
$\delta$-functional contribution to the optical conductivity and the energy cost of phase fluctuations.
In two dimensions (2D), $D_s$ has the dimension of energy and we explicitly define it via $\sigma (\omega \to 0) = e^2 \pi D_s (T) \delta (\omega) + \ldots$ or equivalently, via
$E_\text{cond} = (1/8) D_s (\nabla \phi (r))^2$, where $E_\text{cond}$ is the condensation energy per unit volume, and
$\phi (r)$  is the phase of a superconducting order parameter $\Delta (r) =\Delta e^{i \phi (r)}$~\cite{Scalapino1993a,Paramekanti1998}.

In a clean BCS superconductor, $D_s (T=0) = E_F/\pi$, where $E_F$ is the Fermi energy~\cite{Benfatto2001,Chubukov2016}.
For a parabolic dispersion
this reduces to $D_s (T=0) = n/m$, where $n$ is the total electron density and $m$ the bare electron mass.
In a dirty BCS superconductor, $D_s$ is reduced and can be substantially smaller than in the clean case~\cite{Paramekanti1998,Ghosal2001,Seibold2012a}.
At a small $E_F$ (the low density limit), $D_s$ can become smaller than the bound state energy of two fermions in a vacuum, $E_0$.
In this situation the system displays, even at weak coupling, BEC behavior where bound pairs of fermions are formed at $T_p \sim E_0$, while actual superconductivity with a macroscopic phase coherence sets in at smaller $T_c \approx (\pi/8) D_s (T_c) \sim E_F$~\footnotetext[99]{A more accurate expression is  $T_c \sim E_F/\log{\log{E_0/E_F}}$}\cite{Note99,Randeria1989,Chubukov2016}.

The subject of this paper is the analysis of the superfluid stiffness at $T=0$ in strongly coupled clean superconductors, with special attention to systems
in the vicinity of a quantum critical point (QCP), where
superconductivity emerges out of a non-Fermi liquid.
We will not discuss here disorder effects~\cite{[For an investigation of the phase stiffness in disordered quantum critical superconductivity see e.g.\ ] [] Valentinis2023} nor the behavior at small $E_F$.
We assume that $E_F$ is larger than the fermion-boson interaction strength and analyze the behavior of the stiffness within Eliashberg theory.
To shorten notations, below we label $D_s (T=0)$ simply as $D_s$.

Our primary goal is to understand the interplay between the contributions to $D_s$ from the quasiparticle residue $1/Z$ and from the renormalization of the current vertex.  Without vertex renormalization, $D_s$ is renormalized down from the BCS value
to  $D_s \sim E_F/Z$
and is substantially reduced at strong coupling, when
the quasiparticle residue is small.
It was argued, however~\cite{Leggett1965}, that in a Galilean invariant system, $1/Z$ is exactly cancelled out by vertex renormalization due to a special Ward identity, which states that the renormalization factor for the current
vertex is exactly $Z$.
As a result, $D_s = D^{Gal}_s$ is unaffected by interactions and remains the same as
for a BCS superconductor ($D^{Gal}_s = n/m$ at $T =0$).

Our goal is to understand the interplay between $1/Z$ and vertex renormalizations in systems near a QCP.
A frequently used low-energy model for such systems is one of fermions near the Fermi surface Yukawa-coupled to
soft dynamical bosonic collective fluctuations in the corresponding spin or charge channel.
The bosonic dynamics plays a crucial role for the pairing and non-Fermi-liquid behavior in the normal state~\cite{Metlitski2015a,Abanov2020}.
This dynamical model is, however, non-Galilean-invariant,
even if a fermionic dispersion can be approximated as parabolic, because  the
dynamical term in the bosonic propagator is not invariant under a Galilean boost in which
the momentum $q$ of a boson remains unchanged while the frequency $\omega$ shifts
to $\omega + v q$, where $v$ is the velocity of the boost.
Meanwhile, a QCP towards spin/charge order and superconductivity near it
can develop already in a Galilean-invariant
system.
For the latter, one then has to add additional 4-fermion interactions,
e.g., an effective interaction mediated by two dynamical bosons (the Aslamazov-Larkin-type terms).
It  a priori unclear how these additional interactions, which are generally less singular near a QCP than the direct Yukawa coupling with a soft boson, account for the cancellation of the $1/Z$ factor in the stiffness.

We argue that near-cancellation happens already without the additional terms.
The key here is the observation that
for a Galilean-invariant system, a spin/charge order
emerges with $q=0$~\cite{[{Here we do not consider the non-analyticities in the spin/charge response of a Galilean-invariant system at small $q$, as this likely leads to first-order superconducting transition. See e.g. }] [{}] Chubukov2004}, hence
soft bosonic excitations carry small momenta $q$.
We argue that in this situation,
the leading term in the renormalization of the current vertex is  the same as in the renormalization of
one of the components of
the
the spin vertex ($\sigma^i_{\alpha \beta} c^\dagger_{k,\alpha} c_{k,\beta}$ at the bare level).
The fully renormalized spin vertex is related to the self-energy by the Ward identity, associated with the
global spin conservation,
and cancels out $1/Z$.
This holds for both Galilean-invariant and Galilean-non-invariant systems.
The subleading terms, which distinguish between
the renormalizations of the current and spin vertices, are small in $q$ and
remain non-singular at a QCP.
For a Galilean-invariant system, these subleading terms
cancel out by
additional, less singular interaction terms in the
fermion-boson Hamiltonian.

In this communication we discuss the interplay between $Z$ and the renormalization of the current vertex near a $q=0$ QCP in some detail.
We obtain a generic expression for $D_s$ for interacting fermions
near a QCP and show under which condition it reduces to the BCS result $D_s = n/m$.
This condition (the equivalence between two functions of Matsubara frequency) is satisfied for a Galilean-invariant system, but also approximately holds for a non-Galilean-invariant system.
We call these systems \emph{effectively Galilean}.
We obtain how $D_s$ changes once the condition is violated and illustrate this for the case of fermions interacting with a boson with propagator $\chi (q, \Omega_m) \propto 1/(\omega^2_D + \omega^2_m + (cq)^2)$.
We use the boson velocity $c$ as a control parameter and show within Eliashberg theory how the renormalization of $D_s$ evolves between the limits of large $c$, when the scattering is predominantly in forward direction, and small $c$, when scattering by any $q$ is equally probable.
We show that in the first case
$D_s \approx n/m$, while in the second case
$D_s$ is reduced to $\sim (n/m Z)$.
This last result holds for the interaction mediated by an Einstein phonon.

These results appear naturally when the superfluid stiffness is identified with the prefactor for the $\delta (\omega)$ term in the optical conductivity.
We also show how the fully dressed $D_s$ emerges in the Luttinger-Ward  (LW)
description of a superconductor with coordinate-dependent phase $\phi = \mathbf{q} \cdot \mathbf{r}$, or, equivalently, of a superconductor with a non-zero total momentum $q$ of a pair.
The key issue we discuss here is how the corrections
to the current vertex
emerge in this approach.
We show that they originate
from the change of the fermionic self-energy due to the phase twist
$\exp({i \mathbf{q} \cdot \mathbf{r}})$
and that the existence of such corrections
is a general feature of linear response in the LW
formalism.
We also show how the fully renormalized superfluid stiffness can be derived within the Hubbard-Stratonovich (HS) formalism in the context of the phase action.

We restrict our analysis primarily to the cases when superconductivity emerges near a $q=0$ QCP and soft bosonic excitations are peaked at $q=0$.
For a QCP with a finite $Q$ (e.g.
towards $(\pi,\pi)$ antiferromagnetic order), the renormalization of the current vertex is unrelated to that of the spin vertex, and in general the fully dressed superfluid stiffness scales with
$1/Z$.

The structure of the paper is as follows.
In \cref{sec:model}, we outline the
model and provide a brief review of the Eliashberg theory of superconductivity.
In \cref{sec:conductivity}, we explicitly calculate the superfluid stiffness within Eliashberg theory as the weight of the $\delta$-function in the DC conductivity, with a particular focus on the role of vertex corrections and the notion of effective Galilean invariance.
In \cref{sec:lw-stiffness},  we recontextualize Eliashberg theory in terms of the LW
functional and
employ this description to naturally obtain the superfluid stiffness, including vertex corrections.
Finally, in \cref{sec:hs-description}, we make contact with the HS
description of Eliashberg theory and present the associated phase action, including the important role played by coupling of phase and amplitude
gap fluctuations.
Some technical aspects are discussed in Appendices A-E. In particular, in Appendix B we explicitly derive Ward identities for charge, spin, and momentum for a superconductor with a frequency-dependent gap.

\section{Model}
\label{sec:model}
We consider a model of fermions
described by a Matsubara action
\begin{multline}
	S = -\sum_{k} \bar{\psi}_{k\sigma}(i \epsilon_n - \xi_{\mathbf{k}})\psi_{k\sigma}\\
	- \frac{1}{2} \int dx dx' V(x-x') \bar{\psi}_\sigma(x)\psi_\sigma(x) \bar{\psi}_{\sigma'}(x')\psi_{\sigma'}(x')
	\label{eq:model}
\end{multline}
where the effective interaction $V(x-x')$ is mediated by a soft dynamical boson. The notations are $k=(\epsilon_n, \mathbf{k})$ and $x=(\tau, \mathbf{x})$.
\begin{figure}
	\centering
	\includegraphics[width=0.6\columnwidth]{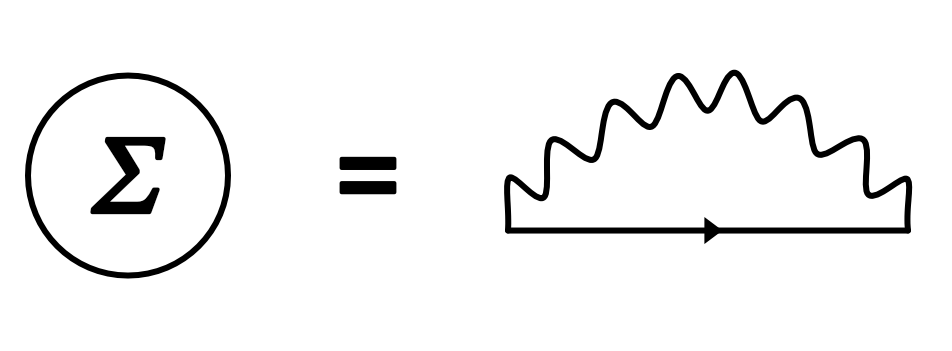}
	\caption{Self-energy diagram within the Eliashberg framework. The solid line is the full Nambu Green's function, and the wavy line is the interaction $V$.
		Vertex corrections to the self-energy are neglected as they are small for the typical frequency and momentum scales contributing to $\hat{\Sigma}$~\cite{Chubukov2004,Metlitski2015a,Sur2016,ZhangVertex2024}.
	}
	\label{fig:self-energy}
\end{figure}
We assume the fermion dispersion $\xi$ and interaction $V$  to be rotationally invariant.
We define Nambu spinors $\Psi(x) = (\psi_\uparrow(x), \bar{\psi}_\downarrow(x))$ with Green's function $\hat{\mathcal{G}}(x, x') \equiv - \braket{\Psi(x) \bar{\Psi}^T(x')}$.
Eliashberg theory approximates the matrix self-energy by the 1-loop self-consistent expression, \cref{fig:self-energy},
\begin{multline}
	\hat{\Sigma}(x, x')
	= V(x-x')\hat{\tau}_3 \hat{\mathcal{G}}(x, x') \hat{\tau}_3\\
	=
	\begin{pmatrix}
		-i\Sigma(x, x') + \chi(x,x') & \phi(x,x')                    \\
		\phi^*(x', x)                & -i\Sigma(x, x') - \chi(x, x')
	\end{pmatrix}
\end{multline}
where $\phi$ represents the pairing vertex while $\Sigma$ and $\chi$ are respectively the odd and even parts of the normal state self-energy~\footnote{We have assumed time reversal symmetry so that the normal state self-energy of the two spin-species are equal.}.
For a particle-hole symmetric system, $\chi$ can be neglected (see \cref{sec:qc-with-super}), and we do so henceforth.
The self-energy can be compactly expressed as
\begin{equation}
	\hat{\Sigma}(x, x') = -i\Sigma(x, x') \hat{\tau}_0 + \phi(x,x') \hat{\tau}_1.
\end{equation}
For a translationally invariant system, the equations for $\Sigma$ and real pairing amplitude $\phi$ can be written in momentum space
\begin{equation}
	\hat{\Sigma}(k) = T\sum_{k'}V(k-k')\hat{\tau}_3 \hat{\mathcal{G}}(k)\hat{\tau}_3.
	\label{eq:self-energy}
\end{equation}
The dependence of the self-energy on the magnitude of momentum is weak within Eliashberg theory, due to the separation of momentum scales between fermions and bosons (see below), allowing us to approximate
\begin{equation}
	\hat{\Sigma}(k) \to \hat{\Sigma}_n(\mathbf{k}_F), \quad
	V(k-k') \to V_{n-n'}(\mathbf{k}_F - \mathbf{k}'_F),
	\label{eq:scale-separation}
\end{equation}
Then
\begin{equation}
	\hat{\Sigma}_n(\mathbf{k}_F)
	= -i \Sigma_n
	\hat{\tau}_0
	+ \phi_n
	\hat{\tau}_1.
	\label{eq:sigmahatdef}
\end{equation}
Within the same approximation, the Nambu Green's function takes the form
\begin{equation}
	\mathcal{G}(k) = \frac{-i(\epsilon_n + \Sigma_n)
		\hat{\tau}_0
		- \xi_\mathbf{k}\hat{\tau}_3 - \phi_n \hat{\tau}_1}{(\epsilon_n + \Sigma_n)^2 + \xi_\mathbf{k}^2 + \phi_n^2}.
	\label{eq:nambu-green}
\end{equation}
We can now directly perform the integral over $\xi_\mathbf{k}$ and obtain the Fermi-surface projected Eliashberg equations:
\begin{equation}
	i\hat{\Sigma}_n(\mathbf{k}_F)\hat{\tau}_3 = \pi\nu T \sum_{n'} \oint_\text{FS} \frac{d \mathbf{k}'_F}{S_{d-1}} V_{n-n'}(|\mathbf{k}_F -\mathbf{k}'_F|) \hat{g}_{n'}(\mathbf{k}'_F).
	\label{eq:scqc}
\end{equation}
with $S_{n}$ the surface area of the $n$-sphere and $\nu$ the density of states per spin at the Fermi surface
(we keep dimension $d$ arbitrary, but will later apply the results to $d=2$).   Here, $\hat{g}_n(\mathbf{k}_F)$ is the
$\xi$-integrated Green's function
weighted with $\hat{\tau}_3$~\cite{Eilenberger1968,Larkin1969,Eliashberg1972}
\begin{equation}
	\hat{g}_n(\mathbf{k}_F) \equiv \frac{i}{\pi} \int d\xi \hat{\tau}_3 \hat{\mathcal{G}}_n(\xi, \mathbf{k}_F)
	\equiv g_n(\mathbf{k}_F)\hat{\tau}_3 + f_n(\mathbf{k}_F)\hat{\tau}_2.
	\label{eq:gqc}
\end{equation}

For simplicity of presentation we consider s-wave superconductivity, in which case we obtain the isotropic Eliashberg equations
\begin{equation}
	\begin{gathered}
		\tilde{\Sigma}_n = \epsilon_n + \pi \nu T \sum_{n'} V^{l=0}_{n-n'} \underbrace{\frac{\tilde{\Sigma}_{n'}}{\sqrt{\tilde{\Sigma}_{n'}^2 + \phi_{n'}^2}}}_{g_n}\\
		\phi_n = \pi \nu T \sum_{n'} V^{l=0}_{n-n'} \underbrace{\frac{\phi_{n'}}{\sqrt{\tilde{\Sigma}_{n'}^2 + \phi_{n'}^2}}}_{f_n}
	\end{gathered}
	\label{eq:SP}
\end{equation}
where we have defined $\tilde{\Sigma}_n = \epsilon_n + \Sigma_n$ and the Fermi surface average of the interaction
\begin{equation}
	V^{l=0}_m \equiv \oint_{FS} \frac{d\mathbf{k}_F}{S_{d-1}}\oint_{FS} \frac{d\mathbf{k}'_F}{S_{d-1}} V_m(|\mathbf{k}_F - \mathbf{k}'_{F}|).
	\label{eq:V0}
\end{equation}
It will also be convenient to define the related quantities~\footnote{Note that,
	strictly speaking,  $Z^{-1}_n$ is \emph{not} the
	quasiparticle residue $Z^{-1}_\text{res} \equiv  1 +
		(\partial \Sigma/\partial\epsilon_n)_{\epsilon_n \to 0}$.}
\begin{equation}
	Z_n \equiv 1 + \frac{\Sigma_n}{\epsilon_n} \equiv \frac{\tilde{\Sigma}_n}{\epsilon_n},\quad \Delta_n \equiv \frac{\phi_n}{Z_n},
	\label{eq:defZD}
\end{equation}
which obey equations
\begin{equation}
	\begin{gathered}
		\Delta_n = \pi\nu T\sum_{n'}V^{l=0}_{n-n'} \frac{\Delta_{n'} - \frac{\epsilon_{n'}}{\epsilon_n}\Delta_n}{\sqrt{\epsilon_{n'}^2 + \Delta_{n'}^2}}\\
		Z_n = 1 + \frac{\pi \nu T}{\epsilon_n} \sum_{n'} V_{n-n'}^{l=0} \frac{\epsilon_{n'}}{\sqrt{\epsilon_{n'}^2 + \Delta_{n'}^2}}.
	\end{gathered}
	\label{eq:ZD}
\end{equation}
We see that there is only one self-consistent equation for $\Delta_n$, while $Z_n$ is a functional of $\Delta_n$~\cite{Abanov2020}.
\cref{eq:SP,eq:ZD} are the central equations that define the equilibrium theory.

There are two approximations used in derivation of the Eliashberg equations.
First, \cref{eq:scale-separation} is valid when bosons are slow modes compared to fermions, i.e., for the same frequency, a typical bosonic momentum is much larger than a typical fermionic momentum. This approximation
is justified when the fermion-boson coupling is much smaller than the Fermi energy.
Second, vertex corrections are neglected. For fermions interacting by exchanging soft collective bosons, these corrections  are, in most cases, $O(1)$ parameter-wise,  but are small numerically~\cite{Chubukov2004,Metlitski2015a,Sur2016,ZhangVertex2024}.
We emphasize in this regard that a typical frequency and a typical momentum of a soft boson in the self-energy diagram are such that $v_F q \gg \omega$.
In this limit, a correction to the boson-fermion vertex in the self-energy diagram is related to the derivative of the self-energy over the momentum, which is at most logarithmic at a QCP, and is~\cite{Metlitski2015a}
weaker than the derivative over frequency, which has a power-law divergence at a QCP.
This reasoning, however, does not hold for the corrections to the external current and density vertices, as for them the incoming momentum $q=0$, while $\omega$ is finite.
In this situation, vertex corrections are generally of order of the frequency derivative of the self-energy at $k=k_F$ and are large and singular near a QCP.

\section{Superfluid weight in the conductivity}
\label{sec:conductivity}
\begin{figure}
	\centering
	\includegraphics[width=\columnwidth]{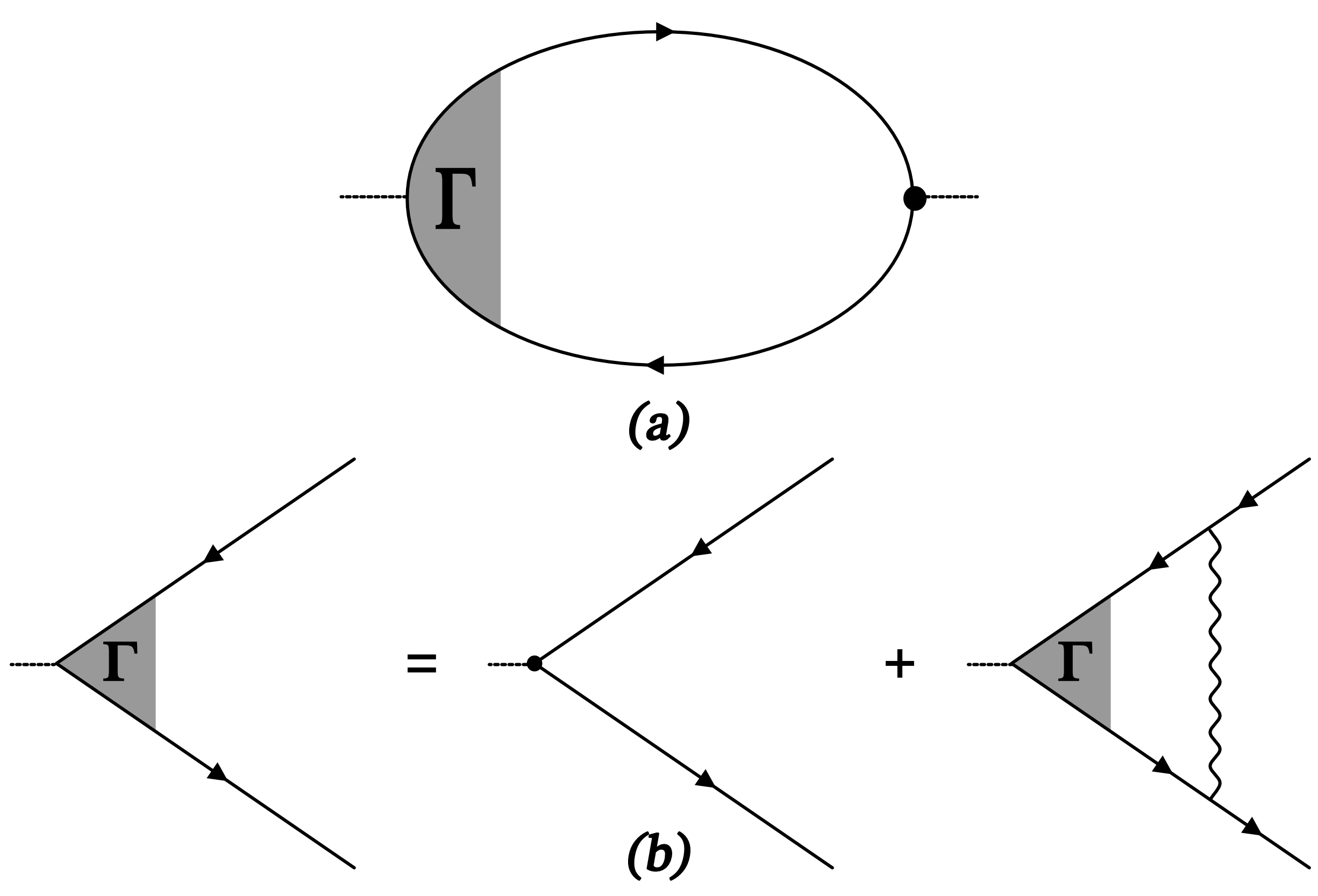}
	\caption{Diagrams contributing to the optical conductivity:
		(a) Paramagnetic velocity-velocity bubble determining the weight of the $\delta$-function in the optical conductivity.
		The dot represents the bare current vertex, while the shaded vertex represents the renormalized current vertex.
		\label{fig:current-curent}
		(b) Bethe-Salpeter equation for the renormalized current vertex.
		The dot is the bare current vertex $\mathbf{v}_F$, and the shaded vertex is the renormalized vertex $\mathbf{v}_F \hat{\Gamma}_n$.
		The thick lines are the full Green's functions of the theory, whereas the wavy line is the interaction $V(k-k')$.
		\label{fig:bethe-salpeter}
	}
\end{figure}
Given a solution to \cref{eq:SP,eq:ZD} we may calculate the conductivity in the superconducting state, from which the superfluid stiffness can be extracted.
In this section, we outline the calculation of the superfluid stiffness from the conductivity within the Eliashberg paradigm, comparing the generic result with that for an exactly Galilean invariant system.

The optical conductivity can be expressed in terms of the retarded velocity-velocity correlator $J$~\cite{Altland2010}.
In the superconducting state $\sigma'(\omega)$ has a delta function piece
\begin{equation}
	\sigma'(\omega) = e^2
	\pi\delta(\omega)
	\Re J^R(\omega, \mathbf{q}=0) + \cdots
\end{equation}
We then identify the superfluid stiffness via $D_s \equiv -
	\Re J^R(\omega\to0, \mathbf{q}=0)$.
The low energy velocity-velocity correlator is expressed in terms of the Nambu Green's functions $\hat{\mathcal{G}}$ by the diagram
in \cref{fig:current-curent}a as~\footnote{
	Below we employ the computational scheme in which we first integrate over the dispersion $\xi_k$  and then over frequency. In this scheme,
	the diamagnetic term is canceled by the high-energy contribution from the fermion bubble.
	For this reason we focus only on the low energy paramagnetic velocity-velocity correlator.}
\begin{equation}
	\hat{J}(Q)= -T\sum_k \tr\left[\boldsymbol{\gamma}_{\mathbf{k}}\hat{\mathcal{G}}_{
		K+Q}\hat{\boldsymbol{\Gamma}}_{
		K+Q,
		K}\hat{\mathcal{G}}_{
		K}\right]
	\label{eq:parabubble}
\end{equation}
where $Q = (i\Omega_m, 0)$
and $K = (i\epsilon_n, \mathbf{k})$.
Here $\boldsymbol{\gamma}_\mathbf{k}$ is the bare velocity vertex and
$\boldsymbol{\Gamma}_{K+Q,K}$ the renormalized velocity vertex within the ladder approximation, \cref{fig:bethe-salpeter}b, satisfying the Bethe-Salpeter equation
\begin{equation}
	\hat{\boldsymbol{\Gamma}}_{K+Q, K} = \hat{\boldsymbol{\gamma}}_{\mathbf{k}} + T \sum_{K'} V_{K-K'}\hat{\tau}_3\hat{\mathcal{G}}_{K'+Q}\hat{\boldsymbol{\Gamma}}_{K'+Q, K'}\hat{\mathcal{G}}_{K'}\hat{\tau}_3.
	\label{eq:BSj}
\end{equation}
The vertex correction is evaluated within the ladder approximation, consistent with the Eliashberg scheme for calculation of the self-energy~\footnote{Within our treatment we do not consider the back action of superconductivity on the bosonic action.}.
Non-ladder vertex correction diagrams, e.g., crossed diagrams, are suppressed to the same degree as vertex corrections to the self-energy.
The bare velocity vertex is the conventional
$\hat{\gamma}_\mathbf{k} =
	\nabla_{\mathbf{k}}\xi_{\mathbf{k}}\hat{\tau}_0$.
Near the Fermi surface this is simply $\hat{\boldsymbol{\gamma}}_{\mathbf{k}} =
	\mathbf{v}_F \hat{\tau}_0$.
For a rotationally symmetric interaction, in the $\mathbf{q}\to0$ limit, the renormalized current vertex must also be proportional to $\mathbf{v}_F$,
allowing us to split the renormalized vertex into a product of
$\mathbf{v}_F$ and a
rotational scalar, which only depends on frequency:
$\hat{\boldsymbol{\Gamma}}_{K+Q, K} =
	\mathbf{v}_F \hat{\Gamma}_{n+m, n}$, where $n$ and $m$ stand for $\epsilon_n$ and $\Omega_m$.
The matrix $\hat{\Gamma}_{n+m,n}$ obeys
the Bethe-Salpeter equation in the form
\begin{multline}
	\hat{\Gamma}_{n+m,n} = \hat{\tau}_0\\
	+ \nu T\sum_{n'}V^{l=1}_{n-n'}\int d\xi_k\hat{\tau}_3\hat{\mathcal{G}}_{n'+m} (\xi_k) \hat{\Gamma}_{n'+m, n'}\hat{\mathcal{G}}_{n'} (\xi_k)\hat{\tau}_3
	\label{eq:BS1}
\end{multline}
where we have defined the generalized $l=1$ harmonic of the interaction (cf.\ \cref{eq:V0})
\begin{multline}
	\left(\frac{v_F^2}{d} \delta_{ij}\right)V^{l=1}_m  \\
	\equiv \oint_{FS} \frac{d\mathbf{k}_F}{S_{d-1}}\oint_{FS} \frac{d\mathbf{k}'_F}{S_{d-1}} \mathbf{v}_{Fj} \mathbf{v}_{Fj}'V_m(|\mathbf{k}_F - \mathbf{k}'_{F}|).
	\label{eq:V1}
\end{multline}

Within the Eliashberg theory, particle-hole symmetry restricts solutions of \cref{eq:BS1} to be of the form
\begin{equation}
	\Gamma = \Gamma^{(0)} \hat{\tau}_0 + \Gamma^{(1)} \hat{\tau}_1
	\label{eq:nnnn_1}
\end{equation}
(see \cref{sec:bubbles} for details).
There is no coupling to the phase sector, and we are able to safely take the limit
$\Omega_m\to0$ (at $T=0$) without encountering any non-analyticity.
In terms of the renormalized vertex
$\hat{\Gamma}_n = \hat{\Gamma}_{n,n}$
the general expression for the superfluid stiffness is
\begin{equation}
	D_s = \frac{v_F^2}{d} \nu T\sum_n\left(
	\Pi^{00}_n \Gamma^{(0)}_n + \Pi^{01}_n \Gamma^{(1)}_n
	\right),
\end{equation}
where
\begin{equation}
	\Pi^{\mu\nu}_n \equiv \int d\xi \tr\left[\hat{\tau}^\mu \hat{\mathcal{G}}_{n}(\xi)\hat{\tau}^\nu\hat{\mathcal{G}}_n(\xi)\right].
\end{equation}
Explicitly evaluating the fermonic bubbles one finds
\begin{equation}
	\Pi^{00}_n = \frac{2\pi \Delta^2_n}{Z_n (\epsilon_n^2 + \Delta_n^2)^{3/2}},\quad
	\Pi^{01}_n = \frac{i2\pi \Delta_n \epsilon_n}{Z_n (\epsilon_n^2 + \Delta_n^2)^{3/2}}
	\label{eq:pi-n}
\end{equation}
so that
\begin{equation}
	D_s = \frac{v_F^2}{d} 2\pi \nu T\sum_n
	\frac{\Delta_n}{Z_n (\epsilon_n^2 + \Delta_n^2)^{3/2}}\left(
	\Delta_n \Gamma^{(0)}_n + i \epsilon_n \Gamma^{(1)}_n
	\right).
	\label{eq:Dsgeneral}
\end{equation}
\Cref{eq:Dsgeneral} is a general result for the superfluid stiffness within Eliashberg theory.
At $T \to 0$, $T \sum_n \to \int d\epsilon_n/(2\pi)$.

In the Galilean invariant case, there is a special Ward identity
relating the
fully renormalized current vertex to the self-energy as
\begin{equation}
	\hat{\Gamma}_{n+m,n} \equiv
	1 + i\frac{\hat{\Sigma}_{n+m} -\hat{\Sigma}_{n} }{\Omega_m},
	\label{eq:Ward_id}
\end{equation}
This relation is obtained from a combination of the Ward identity for conservation of momentum,
and the identity $\mathbf{j} = e (\mathbf{k}/m)$ allowing the renormalized current vertex to be expressed
in terms of the renormalized momentum vertex (see \cref{sec:ward-mom}).
At $\Omega_m \to 0$, this reduces to
\begin{equation}
	\hat{\Gamma}_{n+m,n} =\hat{\Gamma}_{n} \equiv
	1 + i\frac{\partial \hat{\Sigma}_{n}}{\partial \epsilon_n}
\end{equation}
In components: $\Gamma^{(0)}_n = 1 + \frac{\partial \Sigma_{n}}{\partial \epsilon_n}$ and
$\Gamma^{(1)}_n = i \frac{\partial \hat{\phi}_{n}}{\partial \epsilon_n}$
Using
these formulas, we obtain
\begin{multline}
	D_s = D^{Gal}_s =
	\frac{v_F^2}{d}2\pi \nu T\sum_n
	\frac{\Delta_n}{Z_n (\epsilon_n^2 + \Delta_n^2)^{3/2}}\\
	\times
	\left(
	\Delta_n \left[1 + \frac{\partial \Sigma_n}{\partial \epsilon_n}\right]
	- \epsilon_n \frac{\partial \phi_n}{\partial \epsilon_n}
	\right).
	\label{eq:DbarSigmaPhi}
\end{multline}
We now use \cref{eq:defZD} and rewrite
\begin{equation}
	\frac{\partial \phi_n}{\partial \epsilon_n}
	= \frac{\Delta_n}{\epsilon_n} \left(1 + \frac{\partial \Sigma_n}{\partial \epsilon_n}\right) - \frac{\Delta_n}{\epsilon_n}Z_n+ Z_n \frac{\partial \Delta_n}{\partial \epsilon_n}.
	\label{eq:deriv-phi}
\end{equation}
Inserting this into
\cref{eq:DbarSigmaPhi}, we obtain
\begin{equation}
	D^{Gal}_s = \frac{2\pi\nu v_F^2}{d} T\sum_n \frac{\Delta_n}{(\epsilon_n^2 + \Delta_n^2)^{3/2}}
	\left(
	\Delta_n
	- \epsilon_n \frac{\partial \Delta_n}{\partial \epsilon_n}
	\right)
	\label{eq:DscaseC}
\end{equation}
At $T=0$, replacing $2\pi T\sum_n$ by $\int d \epsilon_n$, we obtain
\begin{equation}
	D^{Gal}_s = \frac{
		\nu v_F^2}{d} \int d \epsilon_n \frac{\Delta_n}{(\epsilon_n^2 + \Delta_n^2)^{3/2}}
	\left(
	\Delta_n
	- \epsilon_n \frac{\partial \Delta_n}{\partial \epsilon_n}
	\right)
	\label{eq:DscaseC_1}
\end{equation}
The
integrand
is a total derivative:
\begin{equation}
	D^{Gal}_s =  \frac{v_F^2}{d}
	\nu
	\int d \epsilon_n
	\frac{d}{d \epsilon_n} \left(\frac{\epsilon_n}{\sqrt{\epsilon_n^2 + \Delta_n^2}}\right).
\end{equation}
Evaluating the integral we then obtain
\begin{equation}
	D^{Gal}_s =
	\frac{2
		\nu v_F^2}{d}  = \frac{n}{m}.
	\label{eq:Dsgal}
\end{equation}
This result implies that in an interacting Galilean-invariant system, the superfluid stiffness retains its bare value~\cite{Leggett1968a}.

To understand how and when \cref{eq:Dsgeneral} differs from the Galilean invariant result in a generic case, when there is no Ward identity relating $\hat{\Gamma}_{n+m,n}$  to the self-energy, we
recall that there are Ward identities for a generic system of interacting fermions associated with global charge conservation and global spin conservation.
The latter, for the vector of matrix spin vertex $\sigma^i_{\alpha \beta} \hat{\Gamma}^{sp}_{n+m,n}$, is of interest to us.
Specifically, the matrix $\hat{\Gamma}^{sp}_{n+m,n}$ obeys the Bethe-Salpeter equation
\begin{multline}
	\hat{\Gamma}^{(sp)}_{n+m,n} = 1
	+ \nu T\sum_{n'}V^{l=0}_{n-n'}\\
	\times\int d\xi_k\hat{\tau}_3\hat{\mathcal{G}}_{n'+m} (\xi_k) \hat{\Gamma}^{sp}_{n'+m, n'}\hat{\mathcal{G}}_{n'} (\xi_k) \hat{\tau}_3.
	\label{eq:BS-gap}
\end{multline}
whose solution is the same as for $\hat{\Gamma}_{n+m,n}$ in the Galilean-invariant case:
\begin{equation}
	\hat{\Gamma}^{sp}_{n+m,n} \equiv
	1 + i\frac{\hat{\Sigma}_{n+m} -\hat{\Sigma}_{n} }{\Omega_m}
	\label{eq:Ward_id_1}
\end{equation}
(see \cref{sec:Ward} for details). In the limit $\Omega_m \to 0$, when $\hat{\Gamma}^{sp}_{n+m,n} = \hat{\Gamma}^{sp}_{n}$, this reduces to
$\Gamma^{sp, (0)}_n = 1 + \frac{\partial \Sigma_{n}}{\partial \epsilon_n}$ and
$\Gamma^{sp,(1)}_n = i \frac{\partial \hat{\phi}_{n}}{\partial \epsilon_n}$.
We emphasize that these relations hold for both the Galilean-invariant case and non-Galilean-invariant case.
We also note that \cref{eq:Ward_id_1} holds only for the spin vertex. For the charge vertex, the equation is somewhat different (see \cref{sec:Ward}).

Comparing with \cref{eq:BS1} for
the current vertex $\hat{\Gamma}_{n+m,n}$, we see that the only difference in these equations is that the equation for $\hat{\Gamma}^{(sp)}$ involves the $l=0$ harmonic while the one for $\hat{\Gamma}$ involves the $l=1$ harmonic.
As a consequence, the superfluid stiffness in a non-Galilean system retains its free-fermion value $n/m$ when the harmonics $V^{l=0,1}_n$ are \emph{equal for all frequencies}.
We call such a system  \emph{effectively Galilean-invariant}.

There is one fundamental difference between an \emph{effectively Galilean-invariant} and a truly Galilean-invariant system. In the first, the
cancellation between fermionic $Z$ and vertex correction occurs between terms involving only quasiparticles in the vicinity of the Fermi surface.
In a generic Galilean-invariant system, the
special
Ward identity establishes the relation between properties of the system near and far away from the surface.
So while $D_s = n/m$ in an \emph{effectively Galilean-invariant} system, the
reason why interaction-driven corrections cancel out is in general quite different from that in a truly Galilean-invariant system.

We now investigate in more detail how the vertex corrections
restore
the Galilean value of $D_s$ when $V^{l=0}_n = V^{l=1}_n$.
The role of the vertex corrections in the effectively Gailean invariant case can be elucidated
by the following three cases
\begin{itemize}
	\item[(A)] A frequency independent self-energy (such as from an instantaneous interaction)
	      and $\Delta \sim \text{const}$.
	      This is the BCS case.
	\item[(B)] A
	      matrix self-energy of the form $\Sigma_n \sim \epsilon_n, \Delta \sim \text{const}$.
	      This is the case of
	      superconductivity out of a
	      Fermi liquid away from a QCP.
	\item[(C)] A matrix self-energy in which
	      both $\Delta_n$ and $Z_n$ are strongly frequency dependent.
	      This is the case of
	      superconductivity out of
	      a non-FL at a QCP.
\end{itemize}

For case (A), $Z=1$ and both vertex corrections
$\Gamma^{(0)}-1$ and $\Gamma^{(1)}$ vanish, giving
\begin{equation}
	D_s = D_s^{(0)} = \frac{v_F^2}{d}2\pi \nu T\sum_n
	\frac{\Delta^2}{(\epsilon_n^2 + \Delta^2)^{3/2}} \overset{T=0}{\longrightarrow} \frac{n}{m}
	\label{eq:Dsbcs}
\end{equation}
Indeed, one
can
easily
verify that $Z =\Gamma^{(0)}=1$ and $\Gamma^{(1)}=0$ is
the solution of \cref{eq:ZD,eq:BS1}
for any instantaneous interaction, and thus all BCS-like local interactions are effectively Galilean invariant.
For the vertex correction, this
follows from the fact that all components of
$\int d \epsilon_{n'} (\hat{G}_{n',n'})^2$ vanish, either because the integrand is odd in $\epsilon_{n'}$ or because it can be re-expressed such that
the
both fermionic poles lie in the same half plane, and the integral vanishes after
closing the integration contour in the other half-plane.

For case (B), the expression for $D_s$ is
\begin{equation}
	D_s = \frac{v_F^2}{d}2\pi \nu T\sum_n
	\frac{\Delta^2}{(\epsilon_n^2 + \Delta^2)^{3/2}} \left(\frac{1 + \partial \Sigma_n/\partial \epsilon_n}{Z}\right)
	\label{eq:Dsbcs_1}
\end{equation}
The constant factor in the last bracket
cancels out because for $\Sigma_n \propto \epsilon_n$,
$Z = 1 + \Sigma_n/\epsilon_n = 1 + \partial \Sigma_n/\partial \epsilon_n$.

For case (C),
$\Sigma_n/\epsilon_n  \neq (\partial \Sigma_n/\partial \epsilon_n)$ and thus $Z$ and $1 + \partial \Sigma_n/\partial \epsilon_n$ no longer cancel. One needs to include the frequency derivative of the pairing vertex
on \emph{equal footing} with $\partial \Sigma_n/\partial \epsilon_n$ to get the cancellation of $Z$
and reproduce $D_s = n/m$.

We also note that
for all cases (A)--(C) the relation $D_s = n/m$ holds
independent of the fermionic dispersion.
This is, of course, only approximately true as in linearizing about the Fermi surface we have neglected corrections of order $\Delta/E_F$ to $D_s$.
These corrections cancel out only in the truly Galilean-invariant case, where the relation $D_s = n/m$ is exact.

We now consider how the stiffness gets modified when
$V^{l=0}_{n-n'} \neq V^{l=1}_{n-n'}$.
We define  $V^{l=1}_{n-n'} = V^{l=0}_{n-n'} + \delta V_m$ and $\hat {\Gamma}_n = \hat{\Gamma}^{(sp)}_n + \delta \hat{\Gamma}_n$ (in the limit when external bosonic frequency $\Omega_m \to 0$).
The vertex $\delta \hat{\Gamma}_n$ obeys the modified Bethe-Salpeter equation
\begin{multline}
	\delta \hat{\Gamma}_{n} = \nu T\sum_{n'}
	\left(V^{l=0}_{n-n'}\int d\xi\hat{\tau}_3\hat{\mathcal{G}}_{k'}\delta\hat{\Gamma}_{n'}\hat{\mathcal{G}}_{k'}\hat{\tau}_3 \right.\\
	\left.+ \delta V_{n-n'}\int d\xi\hat{\tau}_3\hat{\mathcal{G}}_{k'}\hat{\Gamma}^{(sp)}_{n'}\hat{\mathcal{G}}_{k'}\hat{\tau}_3
	\right).
	\label{eq:BSdelta}
\end{multline}
Splitting $\delta \hat{\Gamma}_n$ into components
we obtain for the stiffness
\begin{multline}
	D_s = D^{Gal}_s + \frac{v_F^2}{d} 2\pi \nu T\sum_n
	\frac{\Delta_n}{Z_n (\epsilon_n^2 + \Delta_n^2)^{3/2}}\\
	\times\left(
	\Delta_n \delta \Gamma^{(0)}_n + i \epsilon_n \delta \Gamma^{(1)}_n\right)
	\label{eq:Dbardelta}
\end{multline}
If
the difference between $V^{l=1}_{n-n'}$ and $V^{l=0}_{n-n'}$ is small
for all relevant frequencies,
this will be a small correction of order $\delta V$.
This is
the case for interactions which are dominated by small angle scattering, as then
scattered particles do not
distinguish between different harmonics.
That small angle scattering leads to approximate relations between the renormalized current and spin vertices has previously been appreciated in the normal state~\cite{Castellani1994,Metzner1998}.

As an example, consider an interaction mediated by a propagating boson with mass $\omega_D$ and
dispersion $(cq)^2$:
\begin{equation}
	V_m (q) = \frac{g^2\chi_0}{\Omega_m^2 + \omega_D^2 + c^2q^2}.
	\label{eq:Vbos}
\end{equation}
For fermions on the Fermi surface $q^2 = \abs{\mathbf{k}_F - \mathbf{k}'_F}^2 = 2k_F^2(1 - \cos\theta)$,  where $\theta$ is the angle between $\mathbf{k}_F$ and $\mathbf{k}'_F$.
We can then write
\begin{equation}
	V_m
	(q) = V_m (\theta) =\frac{g^2\chi_0}{2c^2k_F^2} \frac{1}{a_m - \cos \theta}
\end{equation}
with $a_m = 1 + (\Omega_m^2 + \omega_D^2)/(2c^2k_F^2)$
and express
\begin{equation}
	V^{l=0}_m = \frac{g^2 \chi_0}{2 c^2k_F^2}
	\oint_{FS}\frac{d\mathbf{k}_F}{S_{d-1}}
	\oint_{FS}\frac{d\mathbf{k}'_F}{S_{d-1}}
	\frac{1}{a_m- \cos(\theta-\theta')}
\end{equation}
and
\begin{equation}
	V^{l=1}_m = d\frac{g^2 \chi_0}{2 c^2k_F^2}
	\oint_{FS}\frac{d\mathbf{k}_F}{S_{d-1}}
	\oint_{FS}\frac{d\mathbf{k}'_F}{S_{d-1}}
	\frac{\cos\theta \cos\theta'}{a_m- \cos(\theta-\theta')}.
\end{equation}
In $d=2$ we have
\begin{equation}
	V^{l=0}_m = \frac{g^2 \chi_0}{2 c^2k_F^2}\frac{1}{\sqrt{a_m^2 - 1}}, \quad
	V^{l=1}_m = \frac{g^2 \chi_0}{2 c^2k_F^2}\left(
	\frac{a_m}{\sqrt{a_m^2 - 1}} - 1
	\right)
	\label{ch_e1}
\end{equation}
so
\begin{multline}
	\delta V_m =
	\frac{g^2 \chi_0}{2 c^2k_F^2}\left(
	\sqrt{\frac{a_m - 1}{a_m + 1}} - 1
	\right)\\
	=
	-\frac{g^2 \chi_0}{2 c^2k_F^2}\left(1-
	\frac{1}{\sqrt{1+\frac{4 c^2 k_F^2}{\Omega^2_m + \omega_D^2}}}\right)
\end{multline}
and
\begin{equation}
	\delta V_m/V_m^{l=0} = \frac{-2}{1 + \sqrt{\frac{a_m - 1}{a_m + 1}}} = \frac{-1}{1 + \sqrt{1+\frac{4 c^2 k_F^2}{\Omega^2_m + \omega_D^2}}}
\end{equation}

The relevant frequencies $\Omega_m$ are of order $\Delta_m$. The characteristic scale for the latter is the gap function at zero frequency at $T=0$, which we label simply by $\Delta$.
We see that $\delta V_m/V_m^{l=0}$ is small when the velocity $c$ is large enough such that
$ck_F \gg (\Delta^2 + \omega_D^2)^{1/2}$.
This is
the limit of small angle scattering.
We furthermore note that in this limit, $\delta V_m$  is determined by scattering to large angles and remains non-singular at a QCP  even if we  set $\Omega_m \to 0$.
As a consequence, $\delta D_s = D_s - D^{Gal}_s$ also remains non-singular.
For a Galilean-invariant system, this non-singular $\delta D_s$ cancels exactly with contributions coming from interactions with non-critical bosons.

A near cancellation between $\Gamma$ and $Z$ factors in $D_s$ for small-angle scattering $\theta$ is similar to the near-cancellation between self-energy and Maki-Thompson contributions to optical conductivity in the normal state, in a similar situation of small momentum scattering (these are the insertions of self-energy and vertex corrections into the conductivity bubble)~\cite{Landau1957,Pitaevskii1960,Chubukov2017}.
Like there, in our case the net result for the difference between $D_s$ and $D^{Gal}_s$ contains the additional factor $1- \cos{\theta} \approx \theta^2/2$ compared to what one would get by including only $Z$ or only $\Gamma$.
Furthermore, for the truly Galilean-invariant case, the already reduced contribution to the optical conductivity
cancels out by additional, Aslamazov-Larkin-type diagrams~\cite{Holstein1964,Aslamazov1968,Chubukov2017}.
The same happens in our case - for a Galilean-invariant system the
already reduced  $D_s - D^{Gal}_s$ is canceled out by other
contributions to $\delta(\omega)$, term in the conductivity, likely also Aslamazov-Larkin-type contributions.
It is also possible that for a convex Fermi surface there is an additional reduction of $D_s - D^{Gal}_s$
when all contributions to the $\delta (\omega)$ term in the conductivity are added together~\cite{Dressel2006,Pal2012}.
We do not dwell on this issue here.

At $ck_F \sim \Delta $,  $\delta V_m/V_m^{l=0}  = O(1)$ i.e., $D_s$ differs from $D^{Gal}_s$.
A particularly extreme example where cancellation is absent is the case of $c=0$, when  $V^{l=1}=0$.
This case describes, in particular, the pairing mediated by a soft Einstein phonon.
At $T=0$, we have
\begin{equation}
	D_s = D^{Gal}_s
	\int d\epsilon
	\frac{\Delta^2(\epsilon)}{Z(\epsilon) (\epsilon^2 + \Delta(\epsilon)^2)^{3/2}} \sim D^{Gal}_s/Z(\Delta)
\end{equation}
Near a QCP, $Z(\Delta)$
is large~\cite{Zhang2022} and
$D_s$ is substantially smaller than $D^{Gal}$.
For the phonon pairing, $D^{Gal}_s \sim E_F$, while $Z(\Delta) \sim {\bar g}^2/(\omega_D \Delta)$, where ${\bar g} = (g^2 \chi_0 m)^{1/2}$.
At small $\omega_D$, $Z (\Delta) \gg 1$.
The actual  stiffness is $D_s \sim E_F \Delta \omega_D/{\bar g}^2$.
Eliashberg theory for electron-phonon interactions is valid as long as
Eliashberg parameter
$\lambda_E = {\bar g}^2/\omega_D E_F$ remains small.
Using that at small $\omega_D$,
$T_c$ and $\Delta$ are both of order $\bar{g}$~\cite{Marsiglio1991,Combescot1996},
the stiffness can be re-expressed as $D_s \sim T_c/\lambda_E$.  We see that, as long as Eliashberg theory is under control, the dressed stiffness remains larger than $T_c$. In this situation, phase fluctuations are weak and Eliashberg $T_c$ nearly coincides with the actual $T_c$.  However, at the boundary of applicability of Eliashberg theory, $D_s$ becomes comparable to $T_c$ and phase fluctuations cannot be neglected.

\section{Superfluid stiffness in the Eliashberg-Luttinger-Ward description}
\label{sec:lw-stiffness}
While the superfluid stiffness appears naturally as a transport property in the conductivity, it can also be obtained directly from the thermodynamic properties of the system.
In particular, it parametrizes the free energy cost associated with twisting the phase boundary conditions of the superconducting state~\cite{Leggett1998,Paramekanti1998,Prokofev2000}.
In this section, we
obtain the superfluid stiffness directly from the
LW variational free energy for the Green's function
in the Nambu representation.
Our particular interest here is understand how the renormalization of the current vertex appears in this approach.
We show that it emerges
naturally already within the one-loop approximation
because of the change of the self-energy due to the phase twist.
We argue that the emergence of corrections to the current vertex
is a general feature of the linear response in the LW formalism, reflecting
the conserving nature of the approach.

Luttinger and Ward showed that a many body system can be described by the variational free energy\cite{Luttinger1960}
\begin{equation}
	\beta \Omega
	[\hat{\mathcal{G}}] =  -\Tr\ln(-\hat{\mathcal{G}}^{-1}) - \Tr[ \hat{G}_0^{-1}\hat{\mathcal{G}}] + \Phi[\hat{\mathcal{G}}]
\end{equation}
where $\mathcal{G}$, the fully dressed Green's function, is to be minimized over,
and $\Phi[\hat{\mathcal{G}}]$ is the LW functional, which can be obtained
diagrammatically as the sum of all two-particle irreducible vacuum skeleton diagrams.
This description has the properties
\begin{itemize}
	\item The equilibrium Green's function $\mathcal{G}_\text{eq}$ minimizes
	      $\Omega$
	\item The self-energy is the functional derivative of the LW
	      functional $\Phi$,
	      $\hat{\Sigma} = \delta\Phi/\delta\mathcal{\hat{G}}$
	\item The minimal value of
	      $\Omega$
	      is the equilibrium free energy, $F_\text{eq} = \Omega[\mathcal{G}_\text{eq}]$
\end{itemize}
The variational free energy $\Omega$  is also known as
the Baym-Kadanoff functional~\cite{Baym1961a,Baym1962} and
is very closely related to the two-particle irreducible effective action, in that $\Gamma^{(2\text{PI})} = \beta\Omega_{LW}$ on the Matsubara axis~\cite{[{For a pedagogical treatment of the nPI action method see e.g. }] [{}] Berges2004}.

Eliashberg theory  corresponds to the 1-loop approximation for the diagrammatic series for the
LW functional $\Phi[\hat{\mathcal{G}}]$~\cite{Eliashberg1960,Benlagra2011,Zhang2023}.
Within the one-loop approximation,
\begin{equation}
	\Phi[\hat{\mathcal{G}}] = \frac{1}{2}\int dx dy
	V(x-y) \Tr[\hat{\tau}_3\hat{\mathcal{G}}(x, y)\hat{\tau}_3 \hat{\mathcal{G}}(y, x)] .
\end{equation}
Minimizing the free energy leads to the Eliashberg equations \cref{eq:self-energy} for the matrix self-energy.

The superfluid stiffness of a superconductor can be obtained by considering the energy cost associated with phase twists of the ground state.
Since the generator of the broken $U(1)$ symmetry is simply $\hat{\tau}_3$ in the Nambu basis we consider the free energy of the
superconducting state as a function of a phase twist
\begin{equation}
	\Psi(x) \to e^{i\mathbf{Q}\cdot \mathbf{r}\hat{\tau}_3}\Psi(x)
\end{equation}
imposed
on the Nambu spinors.
In terms of the LW
variational free energy we define a modified functional
\begin{equation}
	\Omega_Q[\hat{\mathcal{G}}(x-x')] \equiv  \Omega[e^{i\mathbf{Q} \cdot \mathbf{r} \hat{\tau}_3} \hat{\mathcal{G}}(x-x') e^{-i\mathbf{Q} \cdot \mathbf{r}'\hat{\tau}_3}]
\end{equation}
to be minimized over Green's functions with self energies of the form
\footnote{This is what makes the this functional correspond to twisted boundary conditions.}
\begin{equation}
	\hat{\Sigma}_n(\mathbf{k}_F)
	= -i \Sigma_n(\mathbf{k}_F)
	\hat{\tau}_0
	+ \phi_n(\mathbf{k}_F)\hat{\tau}_1.
\end{equation}
The superfluid stiffness -- twice the coefficient of the
$Q^2$ term in $F_Q \equiv \Omega_Q[\hat{\mathcal{G}}_\text{eq}(x-x')]$, can then be obtained as
\begin{equation}
	D_s \equiv \left.\frac{d^2}{dQ^2} F_Q\right|_{Q\to0}.
	\label{eq:LWDs}
\end{equation}
Functionally, the relation between $\Omega$ and $\Omega_Q$ is that we replace
\begin{equation}
	\hat{G}_0^{-1} \to \hat{G}_0^{-1} - \mathbf{v}_F \cdot \mathbf{Q} - \frac{Q^2}{2 m} \hat{\tau}_3 + \cdots
\end{equation}
in the LW
variational free energy.
Within our evaluation scheme (performing the integration over
$\xi_k$ first) the diamagnetic term $\propto Q^2\hat{\tau}_3$ can be neglected
(see~\cref{sec:quasi-classical-lw}).
The $Q^2$ in the action is then entirely due to the source term $\mathbf{v}_F \cdot \mathbf{Q}$.
Noting this, we can straightforwardly evaluate the derivatives in \cref{eq:LWDs} using the saddle-point equation
and obtain
\begin{multline}
	D_s = - T \sum_k \tr\left(\frac{d \hat{G}_0^{-1}}{dQ} \frac{d\hat{\mathcal{G}}}{dQ}\right)\\
	= -i \pi \nu T \sum_n \oint_{FS} \frac{d \mathbf{k}_F}{S_{d-1}}\mathbf{v}_F\tr\left(\hat{\tau}_3 \frac{d\hat{g}}{dQ}\right)
\end{multline}
where in the second equality we have used the
definition of the $\xi_k$-integrated Green's function, \cref{eq:gqc}.
We now define the first order variation of $\hat{g}$ due to $\mathbf{Q}$ via
\begin{equation}
	\hat{g}_n(\mathbf{k}_F) = \hat{\bar{g}}_n + i \mathbf{v}_F\cdot \mathbf{Q} \delta \hat{g}_n  + \cdots
	\label{eq:delta-g}
\end{equation}
in terms of the $Q=0$ solution $\hat{\bar{g}}_n$.
This allows us to compactly express the superfluid stiffness as
\begin{equation}
	D_s = \frac{v_F^2}{d}\pi \nu T \sum_n \tr\left(\hat{\tau}_3 \delta\hat{g}_n\right)
	= \frac{v_F^2}{d}2\pi \nu T \sum_n \delta g_n
	.
\end{equation}
What remains is to calculate $\delta\hat{g}_n$.
We start by noting that the
integration over the dispersion $\xi_k$
can be performed for arbitrary $\hat{\Sigma}_n(\mathbf{k}_F)$
and yields (see \cref{sec:qc-with-super})
\begin{equation}
	\hat{g}_n(\mathbf{k}_F) =
	\frac{\Upsilon_n(\mathbf{k}_F)\hat{\tau}_3+ \phi_n(\mathbf{k}_F)\hat{\tau}_2}{
		\sqrt{\Upsilon_n(\mathbf{k}_F)^2   + \phi_n(\mathbf{k}_F)^2}
	},
	\label{eq:qc-g-with-Q}
\end{equation}
where we have defined $\Upsilon = \varpi + \Sigma_n$ and $\varpi = \epsilon_n + i \mathbf{v}_F \cdot \mathbf{Q}$.
We now introduce, by analogy with \cref{eq:ZD},
\begin{equation}
	Z_n(\mathbf{k}_F) \equiv \frac{\Upsilon_n(\mathbf{k}_F)}{\varpi_n(\mathbf{k}_F)},\quad
	\Delta_n(\mathbf{k}_F) \equiv \frac{\phi_n(\mathbf{k}_F)}{Z_n(\mathbf{k}_F)}.
	\label{eq:ZDQ}
\end{equation}
Using these notations, we
express $\hat{g}_n$ in a form independent of $Z$ as
\begin{equation}
	\hat{g}_n(\mathbf{k}_F) =
	\frac{\varpi_n(\mathbf{k}_F)\hat{\tau}_3+ \Delta_n(\mathbf{k}_F)\hat{\tau}_2}{
		\sqrt{\varpi_n(\mathbf{k}_F)^2   + \Delta_n(\mathbf{k}_F)^2}
	}.
	\label{eq:qc-g-with-Q-delta}
\end{equation}
We see that the gap equation separates into a self-consistency condition for $\Delta$ and functional definition of $Z$ in terms of $\Delta$,  as in the isotropic case.
It is now straightforward to obtain the first order correction to the
$\xi_k$-integrated Green's function $\delta g_n$ by defining $\Delta_n(\mathbf{k}_F) = \bar{\Delta}_n + i\mathbf{v}_F \cdot \mathbf{Q} \delta \Delta_n + \cdots$, with $\bar{\Delta}_n$ the equilibrium solution.
Expanding \cref{eq:qc-g-with-Q-delta} to first order in $\mathbf{v}_F \cdot \mathbf{Q}$,
we obtain
\begin{equation}
	\delta g_n = \frac{\bar{\Delta}_n}{(\epsilon_n^2 + \bar{\Delta}_n^2)^{3/2}}
	\left(\bar{\Delta}_n - \epsilon_n \delta \Delta_n\right)
	\label{eq:delta-g-expr}
\end{equation}
and therefore
\begin{equation}
	D_s = \frac{v_F^2}{d} 2\pi\nu T \sum_n \frac{\bar{\Delta}_n}{(\epsilon_n^2 + \bar{\Delta}_n^2)^{3/2}}
	\left(\bar{\Delta}_n - \epsilon_n \delta \Delta_n\right).
	\label{eq:Ds-deltaD}
\end{equation}
Note the similarity to \cref{eq:Dsgeneral}.

We now make explicit the relation between the variation of the self-energy due to $\mathbf{Q}$
and the renormalized current vertex $\hat{\Gamma}_n$ which appears in
\cref{eq:Dsgeneral}
in the previous section.
Similar to \cref{eq:deriv-phi}, we can use \cref{eq:ZDQ} to reexpress $\delta\Delta$ in terms of $\delta\Sigma$ and $\delta\phi$ via
\begin{equation}
	\delta \phi_n
	= \frac{\bar{\Delta}_n}{\epsilon_n} \left(1 + \delta\Sigma_n\right) - \frac{\bar{\Delta}_n}{\epsilon_n}\bar{Z}_n+ \bar{Z}_n \delta \Delta_n,
	\label{eq:delta-phi}
\end{equation}
Using \cref{eq:delta-phi} we
rewrite the superfluid stiffness as
\begin{multline}
	D_s = \frac{v_F^2}{d} 2\pi\nu T \sum_n \frac{\bar{\Delta}_n}{\bar{Z}_n(\epsilon_n^2 + \bar{\Delta}_n^2)^{3/2}}\\
	\times
	\left(\bar{\Delta}_n  + \bar{\Delta}_n\delta \Sigma_n - \epsilon_n \delta \phi_n\right)
	\label{eq:Dsgenerallw}
\end{multline}
We now expand the Nambu self-energy,  \cref{eq:scqc}, to first order in $\mathbf{v}_F \cdot \mathbf{Q}$ as
\begin{equation}
	\hat{\Sigma}_n(\mathbf{k}_F) = \hat{\bar{\Sigma}}_n + i \mathbf{v}_F \cdot \mathbf{Q} \delta \hat{\Sigma}_n + \cdots
\end{equation}
Equating the first order terms using \cref{eq:delta-g,eq:qc-g-with-Q} and splitting $\delta \hat{\Sigma}_n$ in components as
\begin{equation}
	\delta \hat{\Sigma}_n = -i \delta \Sigma_n \hat{\tau}_0 +
	\delta \phi_n \hat{\tau}_1,
\end{equation}
we find
\begin{equation}
	i \delta \hat{\Sigma}_n \hat{\tau}_3 = \pi\nu T \sum_{n'}
	V^{l=1}_{n-n'}\left(
	\frac{\partial\hat{g}_{n'}}{\partial \Upsilon_{n'}} (1 + \delta\Sigma_{n'})
	+ \frac{\partial\hat{g}_{n'}}{\partial \phi_{n'}} \delta \phi_{n'}
	\right)
	\label{eq:first-order}
\end{equation}
The definition of the $\xi_k$-integrated Green's function~\cref{eq:gqc} implies the identities~\footnote{Order of limits does not matter here as this is a gapped state.},
\begin{equation}
	\begin{gathered}
		\frac{\partial\hat{\bar{g}}}{\partial\Upsilon} = \frac{1}{\pi} \int d \xi
		\hat{\tau}_3 \hat{\mathcal{G}}(\xi)\hat{\mathcal{G}}(\xi)
		, \\
		\frac{\partial\hat{\bar{g}}}{\partial\phi} = \frac{i}{\pi} \int d \xi
		\hat{\tau}_3 \hat{\mathcal{G}}_n(\xi)\hat{\tau}_1\hat{\mathcal{G}}_n(\xi).
	\end{gathered}
\end{equation}
This allows us to rewrite~\cref{eq:first-order} as
\begin{multline}
	1 + i \delta \hat{\Sigma}_n
	= 1 + \nu T \sum_{n'}
	V^{l=1}_{n-n'}\\
	\times \int d\xi\hat{\tau}_3\hat{\mathcal{G}}_{n'}(\xi)\left[
		1 + i \delta\hat{\Sigma}_{n'}\right]
	\hat{\mathcal{G}}_{n'}(\xi) \hat{\tau}_3.
	\label{eq:first-order-bs}
\end{multline}
\Cref{eq:first-order-bs} is identical
to \cref{eq:BS1} with the identification $\hat{\Gamma}_n \equiv 1 + i \delta\hat{\Sigma}_n$,
and we may rewrite \cref{eq:Dsgenerallw} as
\begin{multline}
	D_s = \frac{v_F^2}{d} 2\pi\nu T \sum_n \frac{\bar{\Delta}_n}{\bar{Z}_n(\epsilon_n^2 + \bar{\Delta}_n^2)^{3/2}}\\
	\times
	\left(\bar{\Delta}_n\Gamma^0_n + i \epsilon_n \Gamma^1_n\right)
\end{multline}
in agreement with \cref{eq:Dsgeneral}.
All the results of \cref{sec:conductivity} then follow.

We see
from the above analysis that in the LW formalism the
correction to the current vertex  is equivalent to the first order change in the self-energy due to the phase twist $\exp(i\mathbf{Q}\cdot\mathbf{r})$ (up to a constant factor).
This is a general feature of linear response in the LW
formalism, and it reflects
the conserving nature of the LW (and Baym-Kadanoff) approach.

\section{Hubbard-Stratonovich description and the phase action}
\label{sec:hs-description}
As a final perspective, we now employ the HS
formulation of Eliashberg theory\cite{Protter2021,Yuzbashyan2022} to derive the superfluid stiffness in the context of the phase action.
We start by presenting and commenting on the final result and then provide its derivation.
The action for the phase mode $\theta(i\Omega_m, \mathbf{q})$ to order $q^2$ is given by
\begin{equation}
	S_\theta = \frac{1}{2} \sum_q \theta_{-q}
	\left(
	2\nu
	\Omega_m^2 + D_s q^2
	\right)
	\theta_q
	\label{eq:Stheta}.
\end{equation}
This
form is identical to the BCS
phase action, however $D_s$ is the fully renormalized stiffness.
Note that the
prefactor of the $\Omega^2_m$ term remains the same as in BCS theory.
We argue below that this term comes from high energies, where fermions are free quasiparticles.

We now derive~\cref{eq:Stheta} starting with the HS
decoupling of~\cref{eq:model} in the Nambu basis
\begin{multline}
	S_\text{bos}[\hat{\Sigma}] =-\frac{1}{2}\int d\tau d\tau' d\mathbf{x}\\ \times \frac{1}{V(x- x')}\tr\left[\hat{\Sigma}(x, x')\hat{\tau}_3\hat{\Sigma}(x', x)\hat{\tau}_3\right]\\
	- \Tr\ln\left[-\beta(\hat{G}^{-1}-\hat{\Sigma})\right].
	\label{eq:sbosmat}
\end{multline}
One can verify that the saddle-point equation of \cref{eq:sbosmat} are the Eliashberg equations \cref{eq:self-energy}.
The phase mode $\theta$ enters as parametrization of the HS
field
\begin{equation}\label{eq:phasemode}
	\hat{\Sigma}_{\tau\tau'}(\mathbf{x}) = e^{i\theta_\tau(\mathbf{x}) \hat{\tau}_3}\hat{\Sigma}_L(x-x')e^{-i\theta_{\tau'}(\mathbf{x})\hat{\tau}_3 }
\end{equation}
where $\hat{\Sigma}_L$ contains only
longitudinal fluctuations
around saddle-point solution:
\begin{equation}
	\hat{\Sigma}_L = \hat{\Sigma}_\text{sp} - i\delta\Sigma\hat{\tau}_0 +\delta\phi \hat{\tau}_1 \equiv \hat{\Sigma}_\text{sp} + \delta\hat{\Sigma}_L.
	\label{eq:sigma-l}
\end{equation}
We
compute the phase action
by making use of the Gauge-invariance of the theory~\cite{Altland2010}.
Let us define $\hat{U} = e^{i\theta\hat{\tau}_3}$
such that
$\hat{\Sigma} = \hat{U} \hat{\Sigma}_L\hat{U}^\dagger$.
The first term of the bosonic action \cref{eq:sbosmat} is invariant under application of $\hat{U}$.
For the trace-log term, we may use the cyclic-property of the trace to rewrite
it
in terms of
${\hat \Sigma}_L$ and the gauge-transformed quantity $\hat{U}^\dagger\mathcal{G}^{-1}\hat{U}$.
One can verify that this amounts to replacing
the partial derivatives in the inverse Green's function with covariant derivatives
\begin{equation}
	\partial_\tau \to D_\tau = \partial_\tau + i\partial_\tau \theta\hat{\tau}_3,\quad \nabla \to \mathbf{D} = \nabla + i \nabla\theta\hat{\tau}_3.
\end{equation}
The action is then compactly written
as
\begin{equation}
	S_\text{bos}= S_\text{sp} + S_{HS}[\delta\hat{\Sigma}_L] -\Tr\ln\left\{-\beta \left(\hat{\mathcal{G}}^{-1}
	[D_\tau, \mathbf{D}]-\delta\hat{\Sigma}_L\right)\right\}
\end{equation}
where~\footnote{As discussed before, the diamagnetic term can safely be dropped when performing integration over the momentum first.}
\begin{equation}
	\hat{\mathcal{G}}^{-1}
	[D_\tau, \mathbf{D}]
	\approx \hat{\mathcal{G}}^{-1}_\text{sp}
	- \left(i \partial_\tau \theta \hat{\tau}_3
	+ \mathbf{v}_F \cdot \nabla \theta \right),
\end{equation}
and $\hat{\mathcal{G}}^{-1}_\text{sp}$ is the inverse of
the saddle point Green's function
(the
solution of the Eliashberg equations).
Performing a second order expansion in
derivatives  and in
longitudinal fluctuations leads to the gaussian action
\begin{equation}
	S
	_\text{bos} = S_\theta^{(b)} + S_L^{(b)} + S_c
\end{equation}
where $S^{(b)}_\theta$ is the bare phase action, $S_L^{(b)}$ the bare
action for the longitudinal mode,
and $S_c$ the coupling term.
The bare phase action is
\begin{equation}
	S^{(b)}_\theta =
	\frac{1}{2}\sum_q \left(\kappa^{(b)}\Omega^2_m  + D_s^{(b)} q^2\right)|\theta_q|^2
\end{equation}
where the constants $\kappa^{(b)}$ and $D_s^{(b)}$ are given by
\begin{equation}
	\begin{gathered}
		\kappa^{(b)} = -T \sum_k \tr\left[\hat{\mathcal{G}}_\text{sp}(k)\hat{\tau}_3\right]^2,\\
		D_s^{(b)} = \frac{v_F^2}{d} \nu
		\sum_k \tr\left[\hat{\mathcal{G}}_\text{sp}(k)\right]^2,
		\label{eq:phase-coefficients}
	\end{gathered}
\end{equation}
and we recall that
$k = (\xi_k, \epsilon_n)$ and $\sum_k = T \sum_n \int d \xi_k$.

In explicit form, we have for $\kappa^{(b)}$
\begin{multline}
	\kappa^{(b)} = -\nu\int \frac{d\epsilon_n}{2\pi} \int_{-\Lambda}^\Lambda d\xi_k \frac{\tr[(- i\tilde{\Sigma}_n - \xi_k \hat{\tau}_3 - \phi_n \hat{\tau}_1)\hat\tau_3]^2}{(\tilde{\Sigma}^2_n + \phi^2_n + \xi^2_k)^2}\\
	= 2\nu\int \frac{d\epsilon_n}{2\pi} \int d\xi_k \frac{\tilde{\Sigma}^2_n - \xi^2_k + \phi^2_n}{(\tilde{\Sigma}^2_n + \phi^2_n + \xi^2_m)^2}.
	\label{eq:kappa}
\end{multline}
It is natural to do the integration over $\xi_k$ first as this integral can be evaluated exactly.
This integration, however, should be done with care as the full integral over $\xi_k$ and $\epsilon_n$ is not uniformly convergent.
To
regulate the integral,
we introduce a UV cutoff $\Lambda$, and then take it to infinity at the end of the calculation~(see Ref.~\onlinecite{Maslov2010} and \cref{sec:Ward,sec:dynamic}).
The integration is dominated by energies $|\xi_k| \sim \Lambda$,
for which fermions are essentially free particles, and yields $\kappa^{(b)} =2\nu$.

For $D^{(b)}_s$ we have
\begin{multline}
	D^{(b)}_s =  \frac{v^2_F}{d} \nu\int \frac{d\epsilon_n}{2\pi} \int d\xi_k \frac{\tr[(- i\tilde{\Sigma}_n - \xi_k \hat{\tau}_3 - \phi_n \hat{\tau}_1)]^2}{(\tilde{\Sigma}{_n}^2 + \phi^2_n + \xi^2_k)^2}\\
	= 2 \frac{v^2_F}{d} \nu\int \frac{d\epsilon_n}{2\pi} \int d\xi_k \frac{\xi^2_k + \phi^2_n - {\tilde \Sigma}^2_n}{(\tilde{\Sigma}^2_n + \phi^2_n + \xi^2_m)^2}.
	\label{eq:ch_e2}
\end{multline}
Here we replaced $\int_{-\Lambda}^\Lambda d \xi_k$ by $\int_{-\infty}^\infty d \xi_k$ as
the integral over $|\xi_k| > \Lambda$ cancels out with the diamagnetic contribution.  Evaluating the integral over $\xi$, we obtain
\begin{equation}
	D^{(b)}_s =\frac{v^2_F}{d} \nu \int d \epsilon_n \frac{\Delta^2_n}{Z_n (\Delta^2_n + \epsilon^2_n)^{3/2}}
\end{equation}
This is the expression for the stiffness without vertex corrections.

Next, the longitudinal action is
\begin{multline}
	S_L=
	-\frac{1}{2}T^3\sum_{kk'q}V^{-1}_{k-k'}
	\tr\left[\delta\hat{\Sigma}_{k,k+q}\hat{\tau}_3\delta\hat{\Sigma}_{k'+q, k'}\hat{\tau}_3\right]\\
	+
	\frac{1}{2}T^2\sum_{kk'}
	\tr[\hat{\mathcal{G}}_\text{sp}(k)\delta\hat{\Sigma}_{kk'}\hat{\mathcal{G}}_{\text{sp}}(k')\delta\hat{\Sigma}_{k'k}]\\
	\equiv \frac{1}{2}T^3\sum_{kk'q}\delta\Sigma^\mu_{k,k+q} [T^{-1}_{kk'q}]^{\mu\nu}\delta\Sigma^\nu_{k'+q,k'},
\end{multline}
and the coupling term is
\begin{multline}
	S_c = T \sum_{kq}
	\theta_q \left(
	\Omega_m \tr\left[
		\hat{\mathcal{G}}_{\text{sp}}(k+q)
		\hat{\tau}^3
		\hat{\mathcal{G}}_{\text{sp}}(k)\delta\hat{\Sigma}_{k,k+q}
		\right]\right.\\
	\left.+
	i \mathbf{v}_F \cdot \mathbf{q}
	\tr\left[
		\hat{\mathcal{G}}_{\text{sp}}(k+q)
		\hat{\mathcal{G}}_{\text{sp}}(k)\delta\hat{\Sigma}_{k,k+q}
		\right]
	\right)\\
	\equiv T \sum_{kq}\theta_q \delta\Sigma^\mu_{k, k+q}\left(
	\Omega_m
		(C^\omega)^\mu_{k, k+q} + i\mathbf{v}_F \cdot \mathbf{q} \cdot (C^{q})^\mu_{k, k+q}
	\right),
	\label{eq:Scoupling}
\end{multline}
where
we have expanded
$\delta \hat{\Sigma}$ in Pauli matrices $\delta \hat{\Sigma} = \sum_\mu \delta \Sigma^\mu \hat{\tau}^\mu$ and introduced the couplings
\begin{equation}
	\begin{gathered}
		(C^\omega)^\mu_{k, k+q} =  \frac{1}{2}\tr\left[
			\hat{\mathcal{G}}_{\text{sp}}(k+q)
			\hat{\tau}^3
			\hat{\mathcal{G}}_{\text{sp}}(k)\hat{\tau}^\mu
			\right]\\
		(C^q)^\mu_{k, k+q}
		=
		\frac{1}{2}\tr\left[
			\hat{\mathcal{G}}_{\text{sp}}(k+q)
			\hat{\mathcal{G}}_{\text{sp}}(k)\hat{\tau}^\mu
			\right].
	\end{gathered}
	\label{eq:couplings}
\end{equation}
Upon integrating out the longitudinal modes $\delta\hat{\Sigma}_L$ the effective phase action can be written in terms of bare constants $\kappa^{(b)}, D^{(b)}_s$ and vertex corrections $\delta \kappa, \delta D_s$.
\begin{multline}
	S_\theta =
	\frac{1}{2}\sum_q \left([\kappa^{(b)} + \delta \kappa] \Omega_m^2 \right. \\
	\left.+ [D^{(b)}_s \delta^{ij} +\delta D^{ij}_s] (\mathbf{q})_i (\mathbf{q})_j\right)|\theta_q|^2
	+ O(q^4)
	\label{eq:Stheta-with-vertex}
\end{multline}
where
\begin{equation}
	\begin{gathered}
		\delta\kappa = \lim_{q\to0} T^2\sum_{kk'}(C^\omega)^\mu_{k, k+q}T^{\mu\nu}_{k,k',q} (C^\omega)^\nu_{k'+q, k},\\
		\delta D^{ij}_s= \lim_{q\to0} T^2\sum_{kk'}\mathbf{v}_F \mathbf{v}'_F(i(C^q)^\mu_{k, k+q})T^{\mu\nu}_{k,k',q} (i(C^q)^\nu_{k'+q, k}).
	\end{gathered}
	\label{eq:delta-coeff}
\end{equation}
Since
$T^{\mu \nu}_{k,k',q}$ is
non-singular at $q \to 0$, we can safely set $q=0$ in the integrands.
One can verify that $\lim_{q\to0} (C^\omega)^\mu_{k+q, k} = 0$ (see \cref{sec:bubbles}), hence
$\delta \kappa = 0$.  For $\delta D^{ij}_s$ we obtain
\begin{equation}
	\delta D_s^{ij} = -\frac{v_F^2\nu^2}{d}\delta^{ij} \sum_{nn'}
	\Pi^{0\mu}_n T^{l=1}_{\mu\nu}(\epsilon_n, \epsilon_n')\Pi^{\nu0}_{n'}
	\label{eq:T-vertex-relation}
\end{equation}
where $\Pi^{\mu\nu}_n$ is the same
as
in \cref{eq:BSdelta}, and we have defined, in analogy with $V^{l=1}$,
\begin{multline}
	\left(\frac{v_F^2}{d} \delta_{ij}\right)T^{l=1}_{\mu\nu}(\epsilon_n, \epsilon_n')  \\
	\equiv \oint_{FS} \frac{d\mathbf{k}_F}{S_{d-1}}\oint_{FS} \frac{d\mathbf{k}'_F}{S_{d-1}} \mathbf{v}_{Fj} \mathbf{v}_{Fj}'T_{\mu\nu}(\epsilon_n, \mathbf{k}_F;\epsilon_n', \mathbf{k}'_F).
	\label{eq:T1}
\end{multline}

\begin{figure}
	\centering
	\includegraphics[width=\columnwidth]{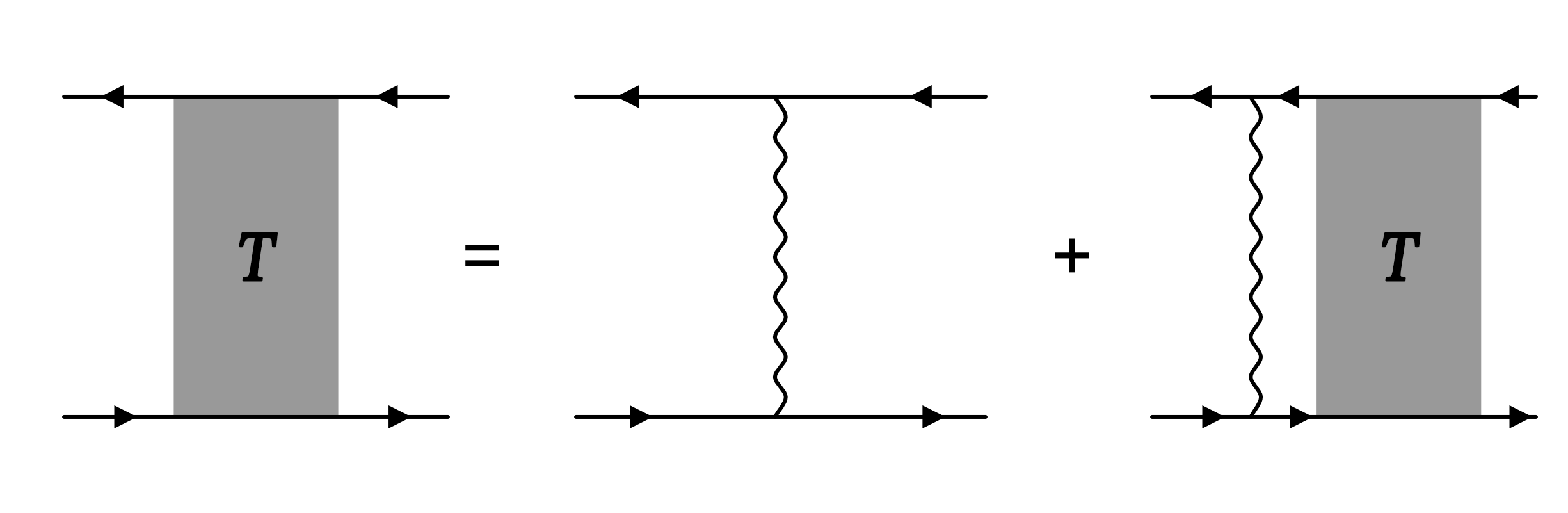}
	\caption{Bethe-Salpeter equation for the T-matrix in the ladder approximation. Solid lines are the full Nambu Green's functions.
		When restricted to the longitudinal sector $T$ is the collective mode propagator for Gaussian longitudinal fluctuations.}
	\label{fig:t-matrix-equation}
\end{figure}

\begin{figure}
	\centering
	\includegraphics[width=\columnwidth]{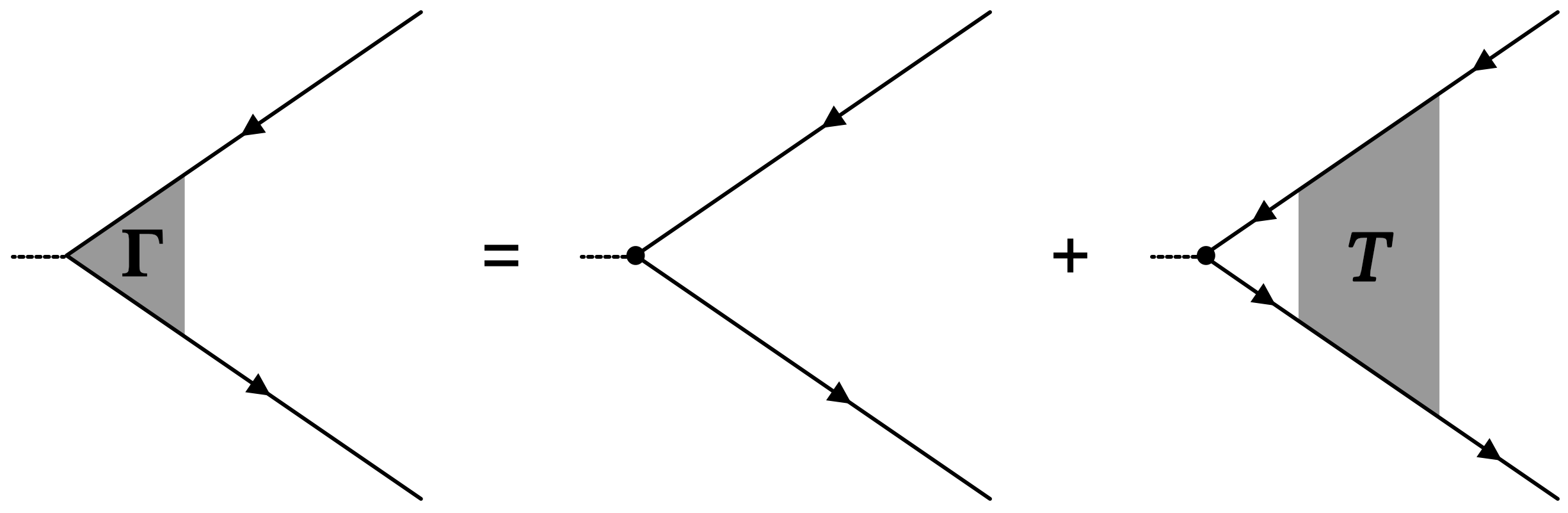}
	\caption{Relation between the Bethe-Salpeter equation for the renormalized current vertex and the T-matrix in the ladder approximation.
		Explicitly, the renormalized vertex is a quasiparticle contribution containing the bare vertex, and a vertex correction coming from the collective modes.}
	\label{fig:t-vertex-relation}
\end{figure}

The relation between the $T$ and $V$ is shown diagrammatically in \cref{fig:t-vertex-relation}: $V$ is the interaction and $T$ is the full $T-$matrix in the longitudinal channel -- the propagator of longitudinal fluctuations.
One can see by inserting the Bethe-Salpeter equation for the T-matrix, \cref{fig:t-matrix-equation}, into this relation that the Bethe-Salpeter equation for the vertex \cref{eq:BSj} is obtained.
We thus find that \cref{eq:T-vertex-relation} is the contribution to
stiffness from corrections to the current vertex,
and then combining \cref{eq:T-vertex-relation,eq:ch_e2} we reproduce \cref{eq:Dsgeneral} for the full $D_s$.
From this perspective, vertex corrections to superfluid stiffness involve
fluctuations in the $l=1$ longitudinal sector, although in the far off-shell region (i.e., far from the pole in the $T$-matrix).




To recapitulate, we have shown that within the
Eliashberg theory, the phase action is
generically of the BCS-like form \cref{eq:Stheta},
the only difference is in the value of the superfluid stiffness $D_s$.
This superfluid stiffness contains the renormalization of the effective mass (the $Z$-factor) and the renormalization from the corrections to the current vertex.  The mass renormalization factor is present already in the bare stiffness computed using HS
decoupling.  Vertex corrections arise when we include the coupling to longitudinal gap fluctuations.

\section{Conclusion}
In this work, we have calculated the superfluid stiffness for a family of
2D non-Galilean invariant models within the Eliashberg approximation.
We showed explicitly, by calculating the delta functional contribution to the conductivity, that in some cases the stiffness approaches its Galilean invariant value $D^{gal}_s = n/m$, despite the absence of Galilean invariance in the model.
In particular, when the $l=0$ and $l=1$
harmonics of the interaction on the Fermi surface are identical, the renormalization of the current vertex is fully determined by the Ward identity for the spin vertex, up to corrections of order $O(\Delta/E_{F})$.
In this situation, the frequency-dependent renormalization of the current vertex cancels out the frequency-dependent renormalization of the quasiparticle mass, and the stiffness remains the same as in the
Galilean-invariant case.
We labeled such systems as having \emph{effectively Galilean invariant} superfluid response.

As an example,  we considered a set of models with boson-mediated interaction in the density-density channel,
strongly peaked at zero momentum transfer, and isotropic but otherwise arbitrary fermionic dispersion.
For such systems, the $l=0$ and $l=1$ harmonics of the interaction are nearly identical and differ by $O(\theta^{2}_{sc})$, where $\theta_{sc}$ is a characteristic scattering angle.
We showed that these systems are effectively Galilean invariant
with  $D_{s} \approx n/m + O(\theta^{2}_{sc})$.
For a truly Galilean-invariant system, the relation $D^{gal}_s =n/m$
is
restored by going beyond the single boson exchange and including Aslamazov-Larkin type diagrams.
We also argued that for arbitrary dispersion the $O(\theta^{2}_{sc})$  term in $D_s$ vanishes only when the boson velocity is taken to infinity, corresponding to an instantaneous action.

We discussed one qualitative difference between an effectively Galilean invariant system and a truly Galilean-invariant one.
In a Galilean-invariant system, the relation $D_{s}=n/m$ is due to the existence of a special Ward identity relating the renormalized current vertex and spin-vertices \emph{exactly}.
This Ward identity results from the combination of the the Ward identity for momentum conservation and the precise relation $\mathbf{j} = e\mathbf{k}/m$ and thus depends on the behavior of particles both near and far from the Fermi surface.
In contrast,
the stiffness of an  \emph{effectively Galilean invariant} system
approaches $n/m$
by fine tuning of the low-energy interaction parameters of the model so that the relation between current and spin is approximately satisfied.
Thus, while the value of the stiffness is approximately the same, the underlying physics
is generally quite different.

We also argued that for both Galilean-invariant and non-Galilean-invariant systems with a frequency dependent gap function,  one must
include the contribution to the stiffness
from the anomalous component of the renormalized current vertex, which is given by the frequency derivative of the pairing vertex.
This contribution must be included on an equal footing with the usual renormalizations to the normal current vertex.
The presence of the anomalous contribution to $D_s$  reflects the fact that in the superconducting state, in addition to the usual diagram renormalizing the normal current vertex, one must take into account the doppler shift of the pairing vertex due to the flow of the condensate.

To further elucidate the nature of the vertex corrections we presented complementary perspectives on the stiffness by obtaining the above results from the LW
functional and the HS
decoupling of our model.

In the LW
description of Eliashberg theory, corresponding to keeping only the lowest order diagram in the LW
functional, the correct prescription for calculating linear-response
is to minimize the free energy in presence of the external fields and take derivatives of the minimal free energy
over the fields
to get the associated susceptibilities.
Then at the end of the calculation one may set the external fields to zero.
This is the sense in which the LW
formalism produces `conserving approximations' when the LW functional is truncated at any order.
Performing the calculation in this way, we showed that the required vertex corrections to the external current vertex appear naturally as the shift of the self-energy due to the phase winding $\hat{\Gamma} \sim (\partial \hat{\mathcal{G}}^{-1}/dQ)$, exactly reproducing the results of the diagrammatic calculation.

In the HS description,
we extracted the stiffness from the phase action of an Eliashberg superconductor.
Using the Gauge-invariance of the action, we showed that the Gaussian action for the phase sector includes the bare phase action as well as a coupling to the $l=1$ longitudinal modes.
Upon integrating out the longitudinal modes we showed that the phase action within Eliashberg theory takes the generic form
$S = \frac{1}{2} \int d \tau d\mathbf{r} \left(2\nu |\partial_{\tau} \theta|^{2} + D_{s}|\nabla\theta|^{2}\right)$.
Here, $D_{s}$ is the same stiffness as obtained in the previous sections, with the vertex corrections arising from the longitudinal mode propagators evaluated at $\mathbf{q}=0,i\Omega_{m}\to0$.
On the other hand, the coefficient of the $(\partial_{t}\theta)^{2}$ is unrenormalized from its bare value, reflecting its origin as coming from fermions away from the Fermi surface which are agnostic to emergence of a pairing vertex at low energy.

For clarity and simplicity of presentation this work focused on $s$-wave superconductivity in rotationally symmetric systems.
The general considerations still apply when either of these constraints are relaxed, but the calculations become more involved as one needs to evaluate products of velocities and form factors of a non-$s$-wave gap along the Fermi surface.

The key result of our work is that an effectively Galilean-invariant value of the stiffness in a non-Galilean-invariant system requires a specific relationship between the low energy interaction channels of the system, which is not guaranteed by the symmetries of the system.
Indeed interaction via Einstein phonons is an example of system which strongly violates these conditions.
Therefore, we expect that generically the superfluid stiffness of a quantum critical superconductor, where the pairing vertex is strongly frequency dependent and the $Z$-factor large, may be strongly reduced from the Galilean invariant value $D^{gal}_{s} \approx n/m$.  We discuss specific examples in a separate paper~\cite{ChubukovGamma2024},
where we analyze the stiffness for underlying quantum-critical models for the $\gamma-$model description of quantum-critical superconductivity~\cite{Zhang2022}.

\begin{acknowledgments}
	The authors would like to thank P.\ Lee and J. Schmalian for stimulating discussions.
 The research  was supported by the U.S. Department of Energy, Office of Science, Basic Energy Sciences, under Award No.\ DE-SC0014402.  This work was completed while the co-authors attended a workshop at KITP in Santa Barbara, CA. KITP is  supported in part by grant NSF PHY-1748958.

\end{acknowledgments}

\bibliography{references.gen}

\begin{thebibliography}{60}%
\makeatletter
\providecommand \@ifxundefined [1]{%
 \@ifx{#1\undefined}
}%
\providecommand \@ifnum [1]{%
 \ifnum #1\expandafter \@firstoftwo
 \else \expandafter \@secondoftwo
 \fi
}%
\providecommand \@ifx [1]{%
 \ifx #1\expandafter \@firstoftwo
 \else \expandafter \@secondoftwo
 \fi
}%
\providecommand \natexlab [1]{#1}%
\providecommand \enquote  [1]{``#1''}%
\providecommand \bibnamefont  [1]{#1}%
\providecommand \bibfnamefont [1]{#1}%
\providecommand \citenamefont [1]{#1}%
\providecommand \href@noop [0]{\@secondoftwo}%
\providecommand \href [0]{\begingroup \@sanitize@url \@href}%
\providecommand \@href[1]{\@@startlink{#1}\@@href}%
\providecommand \@@href[1]{\endgroup#1\@@endlink}%
\providecommand \@sanitize@url [0]{\catcode `\\12\catcode `\$12\catcode
  `\&12\catcode `\#12\catcode `\^12\catcode `\_12\catcode `\%12\relax}%
\providecommand \@@startlink[1]{}%
\providecommand \@@endlink[0]{}%
\providecommand \url  [0]{\begingroup\@sanitize@url \@url }%
\providecommand \@url [1]{\endgroup\@href {#1}{\urlprefix }}%
\providecommand \urlprefix  [0]{URL }%
\providecommand \Eprint [0]{\href }%
\providecommand \doibase [0]{https://doi.org/}%
\providecommand \selectlanguage [0]{\@gobble}%
\providecommand \bibinfo  [0]{\@secondoftwo}%
\providecommand \bibfield  [0]{\@secondoftwo}%
\providecommand \translation [1]{[#1]}%
\providecommand \BibitemOpen [0]{}%
\providecommand \bibitemStop [0]{}%
\providecommand \bibitemNoStop [0]{.\EOS\space}%
\providecommand \EOS [0]{\spacefactor3000\relax}%
\providecommand \BibitemShut  [1]{\csname bibitem#1\endcsname}%
\let\auto@bib@innerbib\@empty
\bibitem [{\citenamefont {Scalapino}\ \emph {et~al.}(1993)\citenamefont
  {Scalapino}, \citenamefont {White},\ and\ \citenamefont
  {Zhang}}]{Scalapino1993a}%
  \BibitemOpen
  \bibfield  {author} {\bibinfo {author} {\bibfnamefont {D.~J.}\ \bibnamefont
  {Scalapino}}, \bibinfo {author} {\bibfnamefont {S.~R.}\ \bibnamefont
  {White}},\ and\ \bibinfo {author} {\bibfnamefont {S.}~\bibnamefont {Zhang}},\
  }\bibfield  {title} {\bibinfo {title} {Insulator, metal, or superconductor:
  {{The}} criteria},\ }\href {https://doi.org/10.1103/PhysRevB.47.7995}
  {\bibfield  {journal} {\bibinfo  {journal} {Phys. Rev. B}\ }\textbf {\bibinfo
  {volume} {47}},\ \bibinfo {pages} {7995} (\bibinfo {year}
  {1993})}\BibitemShut {NoStop}%
\bibitem [{\citenamefont {Paramekanti}\ \emph {et~al.}(1998)\citenamefont
  {Paramekanti}, \citenamefont {Trivedi},\ and\ \citenamefont
  {Randeria}}]{Paramekanti1998}%
  \BibitemOpen
  \bibfield  {author} {\bibinfo {author} {\bibfnamefont {A.}~\bibnamefont
  {Paramekanti}}, \bibinfo {author} {\bibfnamefont {N.}~\bibnamefont
  {Trivedi}},\ and\ \bibinfo {author} {\bibfnamefont {M.}~\bibnamefont
  {Randeria}},\ }\bibfield  {title} {\bibinfo {title} {Upper bounds on the
  superfluid stiffness of disordered systems},\ }\href
  {https://doi.org/10.1103/PhysRevB.57.11639} {\bibfield  {journal} {\bibinfo
  {journal} {Phys. Rev. B}\ }\textbf {\bibinfo {volume} {57}},\ \bibinfo
  {pages} {11639} (\bibinfo {year} {1998})}\BibitemShut {NoStop}%
\bibitem [{\citenamefont {Benfatto}\ \emph {et~al.}(2001)\citenamefont
  {Benfatto}, \citenamefont {Toschi}, \citenamefont {Caprara},\ and\
  \citenamefont {Castellani}}]{Benfatto2001}%
  \BibitemOpen
  \bibfield  {author} {\bibinfo {author} {\bibfnamefont {L.}~\bibnamefont
  {Benfatto}}, \bibinfo {author} {\bibfnamefont {A.}~\bibnamefont {Toschi}},
  \bibinfo {author} {\bibfnamefont {S.}~\bibnamefont {Caprara}},\ and\ \bibinfo
  {author} {\bibfnamefont {C.}~\bibnamefont {Castellani}},\ }\bibfield  {title}
  {\bibinfo {title} {Phase fluctuations in superconductors: {{From Galilean}}
  invariant to quantum {{XY}} models},\ }\href
  {https://doi.org/10.1103/PhysRevB.64.140506} {\bibfield  {journal} {\bibinfo
  {journal} {Phys. Rev. B}\ }\textbf {\bibinfo {volume} {64}},\ \bibinfo
  {pages} {140506} (\bibinfo {year} {2001})}\BibitemShut {NoStop}%
\bibitem [{\citenamefont {Chubukov}\ \emph {et~al.}(2016)\citenamefont
  {Chubukov}, \citenamefont {Eremin},\ and\ \citenamefont
  {Efremov}}]{Chubukov2016}%
  \BibitemOpen
  \bibfield  {author} {\bibinfo {author} {\bibfnamefont {A.~V.}\ \bibnamefont
  {Chubukov}}, \bibinfo {author} {\bibfnamefont {I.}~\bibnamefont {Eremin}},\
  and\ \bibinfo {author} {\bibfnamefont {D.~V.}\ \bibnamefont {Efremov}},\
  }\bibfield  {title} {\bibinfo {title} {Superconductivity versus bound-state
  formation in a two-band superconductor with small {{Fermi}} energy:
  {{Applications}} to {{Fe}} pnictides/chalcogenides and doped {{SrTiO3}}},\
  }\href {https://doi.org/10.1103/PhysRevB.93.174516} {\bibfield  {journal}
  {\bibinfo  {journal} {Phys. Rev. B}\ }\textbf {\bibinfo {volume} {93}},\
  \bibinfo {pages} {174516} (\bibinfo {year} {2016})}\BibitemShut {NoStop}%
\bibitem [{\citenamefont {Ghosal}\ \emph {et~al.}(2001)\citenamefont {Ghosal},
  \citenamefont {Randeria},\ and\ \citenamefont {Trivedi}}]{Ghosal2001}%
  \BibitemOpen
  \bibfield  {author} {\bibinfo {author} {\bibfnamefont {A.}~\bibnamefont
  {Ghosal}}, \bibinfo {author} {\bibfnamefont {M.}~\bibnamefont {Randeria}},\
  and\ \bibinfo {author} {\bibfnamefont {N.}~\bibnamefont {Trivedi}},\
  }\bibfield  {title} {\bibinfo {title} {Inhomogeneous pairing in highly
  disordered {\emph{s}} -wave superconductors},\ }\href
  {https://doi.org/10.1103/PhysRevB.65.014501} {\bibfield  {journal} {\bibinfo
  {journal} {Phys. Rev. B}\ }\textbf {\bibinfo {volume} {65}},\ \bibinfo
  {pages} {014501} (\bibinfo {year} {2001})}\BibitemShut {NoStop}%
\bibitem [{\citenamefont {Seibold}\ \emph {et~al.}(2012)\citenamefont
  {Seibold}, \citenamefont {Benfatto}, \citenamefont {Castellani},\ and\
  \citenamefont {Lorenzana}}]{Seibold2012a}%
  \BibitemOpen
  \bibfield  {author} {\bibinfo {author} {\bibfnamefont {G.}~\bibnamefont
  {Seibold}}, \bibinfo {author} {\bibfnamefont {L.}~\bibnamefont {Benfatto}},
  \bibinfo {author} {\bibfnamefont {C.}~\bibnamefont {Castellani}},\ and\
  \bibinfo {author} {\bibfnamefont {J.}~\bibnamefont {Lorenzana}},\ }\bibfield
  {title} {\bibinfo {title} {Superfluid {{Density}} and {{Phase Relaxation}} in
  {{Superconductors}} with {{Strong Disorder}}},\ }\href
  {https://doi.org/10.1103/PhysRevLett.108.207004} {\bibfield  {journal}
  {\bibinfo  {journal} {Phys. Rev. Lett.}\ }\textbf {\bibinfo {volume} {108}},\
  \bibinfo {pages} {207004} (\bibinfo {year} {2012})}\BibitemShut {NoStop}%
\bibitem [{Note99()}]{Note99}%
  \BibitemOpen
  \bibinfo {note} {A more accurate expression is $T_c \sim E_F/\log {\log
  {E_0/E_F}}$}\BibitemShut {NoStop}%
\bibitem [{\citenamefont {Randeria}\ \emph {et~al.}(1989)\citenamefont
  {Randeria}, \citenamefont {Duan},\ and\ \citenamefont
  {Shieh}}]{Randeria1989}%
  \BibitemOpen
  \bibfield  {author} {\bibinfo {author} {\bibfnamefont {M.}~\bibnamefont
  {Randeria}}, \bibinfo {author} {\bibfnamefont {J.-M.}\ \bibnamefont {Duan}},\
  and\ \bibinfo {author} {\bibfnamefont {L.-Y.}\ \bibnamefont {Shieh}},\
  }\bibfield  {title} {\bibinfo {title} {Bound states, {{Cooper}} pairing, and
  {{Bose}} condensation in two dimensions},\ }\href
  {https://doi.org/10.1103/PhysRevLett.62.981} {\bibfield  {journal} {\bibinfo
  {journal} {Phys. Rev. Lett.}\ }\textbf {\bibinfo {volume} {62}},\ \bibinfo
  {pages} {981} (\bibinfo {year} {1989})}\BibitemShut {NoStop}%
\bibitem [{\citenamefont {Valentinis}\ \emph {et~al.}(2023)\citenamefont
  {Valentinis}, \citenamefont {Inkof},\ and\ \citenamefont
  {Schmalian}}]{Valentinis2023}%
  \BibitemOpen
  \bibfield  {author} {\bibinfo {author} {\bibfnamefont {D.}~\bibnamefont
  {Valentinis}}, \bibinfo {author} {\bibfnamefont {G.~A.}\ \bibnamefont
  {Inkof}},\ and\ \bibinfo {author} {\bibfnamefont {J.}~\bibnamefont
  {Schmalian}},\ }\href {https://doi.org/10.48550/arXiv.2302.13134} {\bibinfo
  {title} {Correlation between phase stiffness and condensation energy across
  the non-{{Fermi}} to {{Fermi-liquid}} crossover in the
  {{Yukawa-Sachdev-Ye-Kitaev}} model on a lattice}} (\bibinfo {year} {2023}),\
  \Eprint {https://arxiv.org/abs/2302.13134} {arxiv:2302.13134 [cond-mat]}
  \BibitemShut {NoStop}%
\bibitem [{\citenamefont {Leggett}(1965)}]{Leggett1965}%
  \BibitemOpen
  \bibfield  {author} {\bibinfo {author} {\bibfnamefont {A.~J.}\ \bibnamefont
  {Leggett}},\ }\bibfield  {title} {\bibinfo {title} {Theory of a {{Superfluid
  Fermi Liquid}}. {{I}}. {{General Formalism}} and {{Static Properties}}},\
  }\href {https://doi.org/10.1103/PhysRev.140.A1869} {\bibfield  {journal}
  {\bibinfo  {journal} {Phys. Rev.}\ }\textbf {\bibinfo {volume} {140}},\
  \bibinfo {pages} {A1869} (\bibinfo {year} {1965})}\BibitemShut {NoStop}%
\bibitem [{\citenamefont {Metlitski}\ \emph {et~al.}(2015)\citenamefont
  {Metlitski}, \citenamefont {Mross}, \citenamefont {Sachdev},\ and\
  \citenamefont {Senthil}}]{Metlitski2015a}%
  \BibitemOpen
  \bibfield  {author} {\bibinfo {author} {\bibfnamefont {M.~A.}\ \bibnamefont
  {Metlitski}}, \bibinfo {author} {\bibfnamefont {D.~F.}\ \bibnamefont
  {Mross}}, \bibinfo {author} {\bibfnamefont {S.}~\bibnamefont {Sachdev}},\
  and\ \bibinfo {author} {\bibfnamefont {T.}~\bibnamefont {Senthil}},\
  }\bibfield  {title} {\bibinfo {title} {Cooper pairing in non-{{Fermi}}
  liquids},\ }\href {https://doi.org/10.1103/physrevb.91.115111} {\bibfield
  {journal} {\bibinfo  {journal} {Phys. Rev. B}\ }\textbf {\bibinfo {volume}
  {91}},\ \bibinfo {pages} {115111} (\bibinfo {year} {2015})}\BibitemShut
  {NoStop}%
\bibitem [{\citenamefont {Abanov}\ and\ \citenamefont
  {Chubukov}(2020)}]{Abanov2020}%
  \BibitemOpen
  \bibfield  {author} {\bibinfo {author} {\bibfnamefont {A.}~\bibnamefont
  {Abanov}}\ and\ \bibinfo {author} {\bibfnamefont {A.~V.}\ \bibnamefont
  {Chubukov}},\ }\bibfield  {title} {\bibinfo {title} {Interplay between
  superconductivity and non-{{Fermi}} liquid at a quantum critical point in a
  metal. {{I}}. {{The}} {$\gamma$} model and its phase diagram at {{T}} = 0 :
  {{The}} case 0 {$<$} {$\gamma$} {$<$} 1},\ }\href
  {https://doi.org/10.1103/PhysRevB.102.024524} {\bibfield  {journal} {\bibinfo
   {journal} {Phys. Rev. B}\ }\textbf {\bibinfo {volume} {102}},\ \bibinfo
  {pages} {024524} (\bibinfo {year} {2020})}\BibitemShut {NoStop}%
\bibitem [{\citenamefont {Chubukov}\ \emph {et~al.}(2004)\citenamefont
  {Chubukov}, \citenamefont {P{\'e}pin},\ and\ \citenamefont
  {Rech}}]{Chubukov2004}%
  \BibitemOpen
  \bibfield  {author} {\bibinfo {author} {\bibfnamefont {A.~V.}\ \bibnamefont
  {Chubukov}}, \bibinfo {author} {\bibfnamefont {C.}~\bibnamefont
  {P{\'e}pin}},\ and\ \bibinfo {author} {\bibfnamefont {J.}~\bibnamefont
  {Rech}},\ }\bibfield  {title} {\bibinfo {title} {Instability of the
  {{Quantum-Critical Point}} of {{Itinerant Ferromagnets}}},\ }\href
  {https://doi.org/10.1103/physrevlett.92.147003} {\bibfield  {journal}
  {\bibinfo  {journal} {Phys. Rev. Lett.}\ }\textbf {\bibinfo {volume} {92}},\
  \bibinfo {pages} {147003} (\bibinfo {year} {2004})}\BibitemShut {NoStop}%
\bibitem [{\citenamefont {Sur}\ and\ \citenamefont {Lee}(2016)}]{Sur2016}%
  \BibitemOpen
  \bibfield  {author} {\bibinfo {author} {\bibfnamefont {S.}~\bibnamefont
  {Sur}}\ and\ \bibinfo {author} {\bibfnamefont {S.-S.}\ \bibnamefont {Lee}},\
  }\bibfield  {title} {\bibinfo {title} {Anisotropic non-{{Fermi}} liquids},\
  }\href {https://doi.org/10.1103/PhysRevB.94.195135} {\bibfield  {journal}
  {\bibinfo  {journal} {Phys. Rev. B}\ }\textbf {\bibinfo {volume} {94}},\
  \bibinfo {pages} {195135} (\bibinfo {year} {2016})}\BibitemShut {NoStop}%
\bibitem [{\citenamefont {Zhang}\ \emph {et~al.}({\natexlab{a}})\citenamefont
  {Zhang}, \citenamefont {Raines},\ and\ \citenamefont
  {Chubukov}}]{ZhangVertex2024}%
  \BibitemOpen
  \bibfield  {author} {\bibinfo {author} {\bibfnamefont {S.-S.}\ \bibnamefont
  {Zhang}}, \bibinfo {author} {\bibfnamefont {Z.~M.}\ \bibnamefont {Raines}},\
  and\ \bibinfo {author} {\bibfnamefont {A.~V.}\ \bibnamefont {Chubukov}},\
  }\bibfield  {title} {\bibinfo {title} {To {{Appear}}}}
  ({\natexlab{a}})\BibitemShut {NoStop}%
\bibitem [{Note1()}]{Note1}%
  \BibitemOpen
  \bibinfo {note} {We have assumed time reversal symmetry so that the normal
  state self-energy of the two spin-species are equal.}\BibitemShut {Stop}%
\bibitem [{\citenamefont {Eilenberger}(1968)}]{Eilenberger1968}%
  \BibitemOpen
  \bibfield  {author} {\bibinfo {author} {\bibfnamefont {G.}~\bibnamefont
  {Eilenberger}},\ }\bibfield  {title} {\bibinfo {title} {Transformation of
  {{Gorkov}}'s equation for type {{II}} superconductors into transport-like
  equations},\ }\href {https://doi.org/10.1007/bf01379803} {\bibfield
  {journal} {\bibinfo  {journal} {Z. F{\"u}r Phys. Hadrons Nucl.}\ }\textbf
  {\bibinfo {volume} {214}},\ \bibinfo {pages} {195} (\bibinfo {year}
  {1968})}\BibitemShut {NoStop}%
\bibitem [{\citenamefont {Larkin}\ and\ \citenamefont
  {Ovchinnikov}(1969)}]{Larkin1969}%
  \BibitemOpen
  \bibfield  {author} {\bibinfo {author} {\bibfnamefont {A.~I.}\ \bibnamefont
  {Larkin}}\ and\ \bibinfo {author} {\bibfnamefont {Y.~N.}\ \bibnamefont
  {Ovchinnikov}},\ }\bibfield  {title} {\bibinfo {title} {Quasiclassical method
  in the theory of superconductivity},\ }\href@noop {} {\bibfield  {journal}
  {\bibinfo  {journal} {Sov Phys JETP}\ }\textbf {\bibinfo {volume} {28}},\
  \bibinfo {pages} {1200} (\bibinfo {year} {1969})}\BibitemShut {NoStop}%
\bibitem [{\citenamefont {Eliashberg}(1972)}]{Eliashberg1972}%
  \BibitemOpen
  \bibfield  {author} {\bibinfo {author} {\bibfnamefont {{\relax
  GM}.}~\bibnamefont {Eliashberg}},\ }\bibfield  {title} {\bibinfo {title}
  {Inelastic electron collisions and nonequilibrium stationary states in
  superconductors},\ }\href@noop {} {\bibfield  {journal} {\bibinfo  {journal}
  {Sov Phys JETP}\ }\textbf {\bibinfo {volume} {34}},\ \bibinfo {pages} {668}
  (\bibinfo {year} {1972})}\BibitemShut {NoStop}%
\bibitem [{Note2()}]{Note2}%
  \BibitemOpen
  \bibinfo {note} {Note that, strictly speaking, $Z^{-1}_n$ is \protect \emph
  {not} the quasiparticle residue $Z^{-1}_\protect \text {res} \equiv 1 +
  (\partial \Sigma /\partial \epsilon _n)_{\epsilon _n \to 0}$.}\BibitemShut
  {Stop}%
\bibitem [{\citenamefont {Altland}\ and\ \citenamefont
  {Simons}(2010)}]{Altland2010}%
  \BibitemOpen
  \bibfield  {author} {\bibinfo {author} {\bibfnamefont {A.}~\bibnamefont
  {Altland}}\ and\ \bibinfo {author} {\bibfnamefont {B.~D.}\ \bibnamefont
  {Simons}},\ }\href@noop {} {\emph {\bibinfo {title} {Condensed {{Matter Field
  Theory}}}}}\ (\bibinfo  {publisher} {{Cambridge University Press}},\ \bibinfo
  {year} {2010})\BibitemShut {NoStop}%
\bibitem [{Note3()}]{Note3}%
  \BibitemOpen
  \bibinfo {note} {Below we employ the computational scheme in which we first
  integrate over the dispersion $\xi _k$ and then over frequency. In this
  scheme, the diamagnetic term is canceled by the high-energy contribution from
  the fermion bubble. For this reason we focus only on the low energy
  paramagnetic velocity-velocity correlator.}\BibitemShut {Stop}%
\bibitem [{Note4()}]{Note4}%
  \BibitemOpen
  \bibinfo {note} {Within our treatment we do not consider the back action of
  superconductivity on the bosonic action.}\BibitemShut {Stop}%
\bibitem [{\citenamefont {Leggett}(1968)}]{Leggett1968a}%
  \BibitemOpen
  \bibfield  {author} {\bibinfo {author} {\bibfnamefont {A.}~\bibnamefont
  {Leggett}},\ }\bibfield  {title} {\bibinfo {title} {Inequalities,
  instabilities, and renormalization in metals and other fermi liquids},\
  }\href {https://doi.org/10.1016/0003-4916(68)90304-7} {\bibfield  {journal}
  {\bibinfo  {journal} {Annals of Physics}\ }\textbf {\bibinfo {volume} {46}},\
  \bibinfo {pages} {76} (\bibinfo {year} {1968})}\BibitemShut {NoStop}%
\bibitem [{\citenamefont {Castellani}\ \emph {et~al.}(1994)\citenamefont
  {Castellani}, \citenamefont {Di~Castro},\ and\ \citenamefont
  {Metzner}}]{Castellani1994}%
  \BibitemOpen
  \bibfield  {author} {\bibinfo {author} {\bibfnamefont {C.}~\bibnamefont
  {Castellani}}, \bibinfo {author} {\bibfnamefont {C.}~\bibnamefont
  {Di~Castro}},\ and\ \bibinfo {author} {\bibfnamefont {W.}~\bibnamefont
  {Metzner}},\ }\bibfield  {title} {\bibinfo {title} {Dimensional crossover
  from {{Fermi}} to {{Luttinger}} liquid},\ }\href
  {https://doi.org/10.1103/PhysRevLett.72.316} {\bibfield  {journal} {\bibinfo
  {journal} {Phys. Rev. Lett.}\ }\textbf {\bibinfo {volume} {72}},\ \bibinfo
  {pages} {316} (\bibinfo {year} {1994})}\BibitemShut {NoStop}%
\bibitem [{\citenamefont {Metzner}\ \emph {et~al.}(1998)\citenamefont
  {Metzner}, \citenamefont {Castellani},\ and\ \citenamefont
  {Di~Castro}}]{Metzner1998}%
  \BibitemOpen
  \bibfield  {author} {\bibinfo {author} {\bibfnamefont {W.}~\bibnamefont
  {Metzner}}, \bibinfo {author} {\bibfnamefont {C.}~\bibnamefont
  {Castellani}},\ and\ \bibinfo {author} {\bibfnamefont {C.}~\bibnamefont
  {Di~Castro}},\ }\bibfield  {title} {\bibinfo {title} {Fermi systems with
  strong forward scattering},\ }\href {https://doi.org/10.1080/000187398243528}
  {\bibfield  {journal} {\bibinfo  {journal} {Advances in Physics}\ }\textbf
  {\bibinfo {volume} {47}},\ \bibinfo {pages} {317} (\bibinfo {year}
  {1998})}\BibitemShut {NoStop}%
\bibitem [{\citenamefont {Landau}(1957)}]{Landau1957}%
  \BibitemOpen
  \bibfield  {author} {\bibinfo {author} {\bibfnamefont {{\relax
  LD}.}~\bibnamefont {Landau}},\ }\bibfield  {title} {\bibinfo {title}
  {Oscillations in a fermi liquid},\ }\href@noop {} {\bibfield  {journal}
  {\bibinfo  {journal} {Sov. Phys. Jetp-Ussr}\ }\textbf {\bibinfo {volume}
  {5}},\ \bibinfo {pages} {101} (\bibinfo {year} {1957})}\BibitemShut {NoStop}%
\bibitem [{\citenamefont {Pitaevskii}(1960)}]{Pitaevskii1960}%
  \BibitemOpen
  \bibfield  {author} {\bibinfo {author} {\bibfnamefont {{\relax
  LP}.}~\bibnamefont {Pitaevskii}},\ }\bibfield  {title} {\bibinfo {title} {On
  the superfluidity of liquid {{He-3}}},\ }\href@noop {} {\bibfield  {journal}
  {\bibinfo  {journal} {Sov. Phys. JETP-USSR}\ }\textbf {\bibinfo {volume}
  {10}},\ \bibinfo {pages} {1267} (\bibinfo {year} {1960})}\BibitemShut
  {NoStop}%
\bibitem [{\citenamefont {Chubukov}\ and\ \citenamefont
  {Maslov}(2017)}]{Chubukov2017}%
  \BibitemOpen
  \bibfield  {author} {\bibinfo {author} {\bibfnamefont {A.~V.}\ \bibnamefont
  {Chubukov}}\ and\ \bibinfo {author} {\bibfnamefont {D.~L.}\ \bibnamefont
  {Maslov}},\ }\bibfield  {title} {\bibinfo {title} {Optical conductivity of a
  two-dimensional metal near a quantum critical point: {{The}} status of the
  extended {{Drude}} formula},\ }\href
  {https://doi.org/10.1103/PhysRevB.96.205136} {\bibfield  {journal} {\bibinfo
  {journal} {Phys. Rev. B}\ }\textbf {\bibinfo {volume} {96}},\ \bibinfo
  {pages} {205136} (\bibinfo {year} {2017})}\BibitemShut {NoStop}%
\bibitem [{\citenamefont {Holstein}(1964)}]{Holstein1964}%
  \BibitemOpen
  \bibfield  {author} {\bibinfo {author} {\bibfnamefont {T.}~\bibnamefont
  {Holstein}},\ }\bibfield  {title} {\bibinfo {title} {Theory of transport
  phenomena in an electron-phonon gas},\ }\href
  {https://doi.org/10.1016/0003-4916(64)90008-9} {\bibfield  {journal}
  {\bibinfo  {journal} {Annals of Physics}\ }\textbf {\bibinfo {volume} {29}},\
  \bibinfo {pages} {410} (\bibinfo {year} {1964})}\BibitemShut {NoStop}%
\bibitem [{\citenamefont {Aslamazov}\ and\ \citenamefont
  {Larkin}(1968)}]{Aslamazov1968}%
  \BibitemOpen
  \bibfield  {author} {\bibinfo {author} {\bibfnamefont {L.~G.}\ \bibnamefont
  {Aslamazov}}\ and\ \bibinfo {author} {\bibfnamefont {A.}~\bibnamefont
  {Larkin}},\ }\bibfield  {title} {\bibinfo {title} {Effect of fluctuations on
  the properties of a superconductor above the critical temperature},\
  }\href@noop {} {\bibfield  {journal} {\bibinfo  {journal} {Sov. Phys. Solid
  State}\ }\textbf {\bibinfo {volume} {10}},\ \bibinfo {pages} {875} (\bibinfo
  {year} {1968})}\BibitemShut {NoStop}%
\bibitem [{\citenamefont {Dressel}\ and\ \citenamefont
  {Scheffler}(2006)}]{Dressel2006}%
  \BibitemOpen
  \bibfield  {author} {\bibinfo {author} {\bibfnamefont {M.}~\bibnamefont
  {Dressel}}\ and\ \bibinfo {author} {\bibfnamefont {M.}~\bibnamefont
  {Scheffler}},\ }\bibfield  {title} {\bibinfo {title} {Verifying the {{Drude}}
  response*},\ }\href {https://doi.org/10.1002/andp.200651807-810} {\bibfield
  {journal} {\bibinfo  {journal} {Ann. Phys.}\ }\textbf {\bibinfo {volume}
  {518}},\ \bibinfo {pages} {535} (\bibinfo {year} {2006})}\BibitemShut
  {NoStop}%
\bibitem [{\citenamefont {Pal}\ \emph {et~al.}(2012)\citenamefont {Pal},
  \citenamefont {Yudson},\ and\ \citenamefont {Maslov}}]{Pal2012}%
  \BibitemOpen
  \bibfield  {author} {\bibinfo {author} {\bibfnamefont {H.~K.}\ \bibnamefont
  {Pal}}, \bibinfo {author} {\bibfnamefont {V.~I.}\ \bibnamefont {Yudson}},\
  and\ \bibinfo {author} {\bibfnamefont {D.~L.}\ \bibnamefont {Maslov}},\
  }\bibfield  {title} {\bibinfo {title} {Resistivity of
  non-{{Galilean-invariant Fermi-}} and non-{{Fermi}} liquids},\ }\href
  {https://doi.org/10.3952/physics.v52i2.2358} {\bibfield  {journal} {\bibinfo
  {journal} {Lith. J. Phys.}\ }\textbf {\bibinfo {volume} {52}},\ \bibinfo
  {pages} {142} (\bibinfo {year} {2012})}\BibitemShut {NoStop}%
\bibitem [{\citenamefont {Zhang}\ \emph {et~al.}(2022)\citenamefont {Zhang},
  \citenamefont {Wu}, \citenamefont {Abanov},\ and\ \citenamefont
  {Chubukov}}]{Zhang2022}%
  \BibitemOpen
  \bibfield  {author} {\bibinfo {author} {\bibfnamefont {S.-S.}\ \bibnamefont
  {Zhang}}, \bibinfo {author} {\bibfnamefont {Y.-M.}\ \bibnamefont {Wu}},
  \bibinfo {author} {\bibfnamefont {A.}~\bibnamefont {Abanov}},\ and\ \bibinfo
  {author} {\bibfnamefont {A.~V.}\ \bibnamefont {Chubukov}},\ }\href@noop {}
  {\bibinfo {title} {Superconductivity out of a non-{{Fermi}} liquid. {{Free}}
  energy analysis}} (\bibinfo {year} {2022}),\ \Eprint
  {https://arxiv.org/abs/2208.13888} {arxiv:2208.13888 [cond-mat]} \BibitemShut
  {NoStop}%
\bibitem [{\citenamefont {Marsiglio}\ and\ \citenamefont
  {Carbotte}(1991)}]{Marsiglio1991}%
  \BibitemOpen
  \bibfield  {author} {\bibinfo {author} {\bibfnamefont {F.}~\bibnamefont
  {Marsiglio}}\ and\ \bibinfo {author} {\bibfnamefont {J.~P.}\ \bibnamefont
  {Carbotte}},\ }\bibfield  {title} {\bibinfo {title} {Gap function and density
  of states in the strong-coupling limit for an electron-boson system},\ }\href
  {https://doi.org/10.1103/PhysRevB.43.5355} {\bibfield  {journal} {\bibinfo
  {journal} {Phys. Rev. B}\ }\textbf {\bibinfo {volume} {43}},\ \bibinfo
  {pages} {5355} (\bibinfo {year} {1991})}\BibitemShut {NoStop}%
\bibitem [{\citenamefont {Combescot}\ and\ \citenamefont
  {Varelogiannis}(1996)}]{Combescot1996}%
  \BibitemOpen
  \bibfield  {author} {\bibinfo {author} {\bibfnamefont {R.}~\bibnamefont
  {Combescot}}\ and\ \bibinfo {author} {\bibfnamefont {G.}~\bibnamefont
  {Varelogiannis}},\ }\bibfield  {title} {\bibinfo {title} {The ratio
  {{2$\Delta$}}/{{TC}} in {{Eliashberg}} theory},\ }\href
  {https://doi.org/10.1007/BF00755117} {\bibfield  {journal} {\bibinfo
  {journal} {J Low Temp Phys}\ }\textbf {\bibinfo {volume} {102}},\ \bibinfo
  {pages} {193} (\bibinfo {year} {1996})}\BibitemShut {NoStop}%
\bibitem [{\citenamefont {Leggett}(1998)}]{Leggett1998}%
  \BibitemOpen
  \bibfield  {author} {\bibinfo {author} {\bibfnamefont {A.~J.}\ \bibnamefont
  {Leggett}},\ }\bibfield  {title} {\bibinfo {title} {On the {{Superfluid
  Fraction}} of an {{Arbitrary Many-Body System}} at {{T}}=0},\ }\href
  {https://doi.org/10.1023/B:JOSS.0000033170.38619.6c} {\bibfield  {journal}
  {\bibinfo  {journal} {Journal of Statistical Physics}\ }\textbf {\bibinfo
  {volume} {93}},\ \bibinfo {pages} {927} (\bibinfo {year} {1998})}\BibitemShut
  {NoStop}%
\bibitem [{\citenamefont {Prokof'ev}\ and\ \citenamefont
  {Svistunov}(2000)}]{Prokofev2000}%
  \BibitemOpen
  \bibfield  {author} {\bibinfo {author} {\bibfnamefont {N.~V.}\ \bibnamefont
  {Prokof'ev}}\ and\ \bibinfo {author} {\bibfnamefont {B.~V.}\ \bibnamefont
  {Svistunov}},\ }\bibfield  {title} {\bibinfo {title} {Two definitions of
  superfluid density},\ }\href {https://doi.org/10.1103/PhysRevB.61.11282}
  {\bibfield  {journal} {\bibinfo  {journal} {Phys. Rev. B}\ }\textbf {\bibinfo
  {volume} {61}},\ \bibinfo {pages} {11282} (\bibinfo {year}
  {2000})}\BibitemShut {NoStop}%
\bibitem [{\citenamefont {Luttinger}\ and\ \citenamefont
  {Ward}(1960)}]{Luttinger1960}%
  \BibitemOpen
  \bibfield  {author} {\bibinfo {author} {\bibfnamefont {J.~M.}\ \bibnamefont
  {Luttinger}}\ and\ \bibinfo {author} {\bibfnamefont {J.~C.}\ \bibnamefont
  {Ward}},\ }\bibfield  {title} {\bibinfo {title} {Ground-{{State Energy}} of a
  {{Many-Fermion System}}. {{II}}},\ }\href
  {https://doi.org/10.1103/PhysRev.118.1417} {\bibfield  {journal} {\bibinfo
  {journal} {Phys. Rev.}\ }\textbf {\bibinfo {volume} {118}},\ \bibinfo {pages}
  {1417} (\bibinfo {year} {1960})}\BibitemShut {NoStop}%
\bibitem [{\citenamefont {Baym}\ and\ \citenamefont
  {Kadanoff}(1961)}]{Baym1961a}%
  \BibitemOpen
  \bibfield  {author} {\bibinfo {author} {\bibfnamefont {G.}~\bibnamefont
  {Baym}}\ and\ \bibinfo {author} {\bibfnamefont {L.~P.}\ \bibnamefont
  {Kadanoff}},\ }\bibfield  {title} {\bibinfo {title} {Conservation {{Laws}}
  and {{Correlation Functions}}},\ }\href
  {https://doi.org/10.1103/PhysRev.124.287} {\bibfield  {journal} {\bibinfo
  {journal} {Phys. Rev.}\ }\textbf {\bibinfo {volume} {124}},\ \bibinfo {pages}
  {287} (\bibinfo {year} {1961})}\BibitemShut {NoStop}%
\bibitem [{\citenamefont {Baym}(1962)}]{Baym1962}%
  \BibitemOpen
  \bibfield  {author} {\bibinfo {author} {\bibfnamefont {G.}~\bibnamefont
  {Baym}},\ }\bibfield  {title} {\bibinfo {title} {Self-{{Consistent
  Approximations}} in {{Many-Body Systems}}},\ }\href
  {https://doi.org/10.1103/PhysRev.127.1391} {\bibfield  {journal} {\bibinfo
  {journal} {Phys. Rev.}\ }\textbf {\bibinfo {volume} {127}},\ \bibinfo {pages}
  {1391} (\bibinfo {year} {1962})}\BibitemShut {NoStop}%
\bibitem [{\citenamefont {Berges}(2004)}]{Berges2004}%
  \BibitemOpen
  \bibfield  {author} {\bibinfo {author} {\bibfnamefont {J.}~\bibnamefont
  {Berges}},\ }\bibfield  {title} {\bibinfo {title} {Introduction to
  {{Nonequilibrium Quantum Field Theory}}},\ }in\ \href
  {https://doi.org/10.1063/1.1843591} {\emph {\bibinfo {booktitle} {{{AIP
  Conf}}. {{Proc}}.}}},\ Vol.\ \bibinfo {volume} {739}\ (\bibinfo  {publisher}
  {{AIP}},\ \bibinfo {address} {{Rio de Janeiro (Brazil)}},\ \bibinfo {year}
  {2004})\ pp.\ \bibinfo {pages} {3--62}\BibitemShut {NoStop}%
\bibitem [{\citenamefont {Eliashberg}(1960)}]{Eliashberg1960}%
  \BibitemOpen
  \bibfield  {author} {\bibinfo {author} {\bibfnamefont {{\relax
  GM}.}~\bibnamefont {Eliashberg}},\ }\bibfield  {title} {\bibinfo {title}
  {Interactions between electrons and lattice vibrations in a superconductor},\
  }\href@noop {} {\bibfield  {journal} {\bibinfo  {journal} {Sov. Phys. JETP}\
  }\textbf {\bibinfo {volume} {11}},\ \bibinfo {pages} {696} (\bibinfo {year}
  {1960})}\BibitemShut {NoStop}%
\bibitem [{\citenamefont {Benlagra}\ \emph {et~al.}(2011)\citenamefont
  {Benlagra}, \citenamefont {Kim},\ and\ \citenamefont
  {P{\'e}pin}}]{Benlagra2011}%
  \BibitemOpen
  \bibfield  {author} {\bibinfo {author} {\bibfnamefont {A.}~\bibnamefont
  {Benlagra}}, \bibinfo {author} {\bibfnamefont {K.-S.}\ \bibnamefont {Kim}},\
  and\ \bibinfo {author} {\bibfnamefont {C.}~\bibnamefont {P{\'e}pin}},\
  }\bibfield  {title} {\bibinfo {title} {The {{Luttinger}}{\textendash}{{Ward}}
  functional approach in the {{Eliashberg}} framework: A systematic derivation
  of scaling for thermodynamics near the quantum critical point},\ }\href
  {https://doi.org/10.1088/0953-8984/23/14/145601} {\bibfield  {journal}
  {\bibinfo  {journal} {J. Phys.: Condens. Matter}\ }\textbf {\bibinfo {volume}
  {23}},\ \bibinfo {pages} {145601} (\bibinfo {year} {2011})}\BibitemShut
  {NoStop}%
\bibitem [{\citenamefont {Zhang}\ \emph {et~al.}(2023)\citenamefont {Zhang},
  \citenamefont {Berg},\ and\ \citenamefont {Chubukov}}]{Zhang2023}%
  \BibitemOpen
  \bibfield  {author} {\bibinfo {author} {\bibfnamefont {S.-S.}\ \bibnamefont
  {Zhang}}, \bibinfo {author} {\bibfnamefont {E.}~\bibnamefont {Berg}},\ and\
  \bibinfo {author} {\bibfnamefont {A.~V.}\ \bibnamefont {Chubukov}},\
  }\bibfield  {title} {\bibinfo {title} {Free energy and specific heat near a
  quantum critical point of a metal},\ }\href
  {https://doi.org/10.1103/PhysRevB.107.144507} {\bibfield  {journal} {\bibinfo
   {journal} {Phys. Rev. B}\ }\textbf {\bibinfo {volume} {107}},\ \bibinfo
  {pages} {144507} (\bibinfo {year} {2023})}\BibitemShut {NoStop}%
\bibitem [{Note5()}]{Note5}%
  \BibitemOpen
  \bibinfo {note} {This is what makes the this functional correspond to twisted
  boundary conditions.}\BibitemShut {Stop}%
\bibitem [{Note6()}]{Note6}%
  \BibitemOpen
  \bibinfo {note} {Order of limits does not matter here as this is a gapped
  state.}\BibitemShut {Stop}%
\bibitem [{\citenamefont {Protter}\ \emph {et~al.}(2021)\citenamefont
  {Protter}, \citenamefont {Boyack},\ and\ \citenamefont
  {Marsiglio}}]{Protter2021}%
  \BibitemOpen
  \bibfield  {author} {\bibinfo {author} {\bibfnamefont {M.}~\bibnamefont
  {Protter}}, \bibinfo {author} {\bibfnamefont {R.}~\bibnamefont {Boyack}},\
  and\ \bibinfo {author} {\bibfnamefont {F.}~\bibnamefont {Marsiglio}},\
  }\bibfield  {title} {\bibinfo {title} {Functional-integral approach to
  {{Gaussian}} fluctuations in {{Eliashberg}} theory},\ }\href
  {https://doi.org/10.1103/PhysRevB.104.014513} {\bibfield  {journal} {\bibinfo
   {journal} {Phys. Rev. B}\ }\textbf {\bibinfo {volume} {104}},\ \bibinfo
  {pages} {014513} (\bibinfo {year} {2021})}\BibitemShut {NoStop}%
\bibitem [{\citenamefont {Yuzbashyan}\ and\ \citenamefont
  {Altshuler}(2022)}]{Yuzbashyan2022}%
  \BibitemOpen
  \bibfield  {author} {\bibinfo {author} {\bibfnamefont {E.~A.}\ \bibnamefont
  {Yuzbashyan}}\ and\ \bibinfo {author} {\bibfnamefont {B.~L.}\ \bibnamefont
  {Altshuler}},\ }\bibfield  {title} {\bibinfo {title} {Migdal-{{Eliashberg}}
  theory as a classical spin chain},\ }\href
  {https://doi.org/10.1103/PhysRevB.106.014512} {\bibfield  {journal} {\bibinfo
   {journal} {Phys. Rev. B}\ }\textbf {\bibinfo {volume} {106}},\ \bibinfo
  {pages} {014512} (\bibinfo {year} {2022})},\ \Eprint
  {https://arxiv.org/abs/2205.06442} {arxiv:2205.06442 [cond-mat]} \BibitemShut
  {NoStop}%
\bibitem [{Note7()}]{Note7}%
  \BibitemOpen
  \bibinfo {note} {As discussed before, the diamagnetic term can safely be
  dropped when performing integration over the momentum first.}\BibitemShut
  {Stop}%
\bibitem [{\citenamefont {Maslov}\ and\ \citenamefont
  {Chubukov}(2010)}]{Maslov2010}%
  \BibitemOpen
  \bibfield  {author} {\bibinfo {author} {\bibfnamefont {D.~L.}\ \bibnamefont
  {Maslov}}\ and\ \bibinfo {author} {\bibfnamefont {A.~V.}\ \bibnamefont
  {Chubukov}},\ }\bibfield  {title} {\bibinfo {title} {Fermi liquid near
  {{Pomeranchuk}} quantum criticality},\ }\href
  {https://doi.org/10.1103/PhysRevB.81.045110} {\bibfield  {journal} {\bibinfo
  {journal} {Phys. Rev. B}\ }\textbf {\bibinfo {volume} {81}},\ \bibinfo
  {pages} {045110} (\bibinfo {year} {2010})}\BibitemShut {NoStop}%
\bibitem [{\citenamefont {Zhang}\ \emph {et~al.}({\natexlab{b}})\citenamefont
  {Zhang}, \citenamefont {Raines},\ and\ \citenamefont
  {Chubukov}}]{ChubukovGamma2024}%
  \BibitemOpen
  \bibfield  {author} {\bibinfo {author} {\bibfnamefont {S.-S.}\ \bibnamefont
  {Zhang}}, \bibinfo {author} {\bibfnamefont {Z.~M.}\ \bibnamefont {Raines}},\
  and\ \bibinfo {author} {\bibfnamefont {A.~V.}\ \bibnamefont {Chubukov}},\
  }\bibfield  {title} {\bibinfo {title} {To {{Appear}}}}
  ({\natexlab{b}})\BibitemShut {NoStop}%
\bibitem [{\citenamefont {Lif{\v s}ic}\ \emph {et~al.}(2006)\citenamefont
  {Lif{\v s}ic}, \citenamefont {Pitaevskij}, \citenamefont {Landau},\ and\
  \citenamefont {Lifshitz}}]{Lifsic2006}%
  \BibitemOpen
  \bibfield  {author} {\bibinfo {author} {\bibfnamefont {E.~M.}\ \bibnamefont
  {Lif{\v s}ic}}, \bibinfo {author} {\bibfnamefont {L.~P.}\ \bibnamefont
  {Pitaevskij}}, \bibinfo {author} {\bibfnamefont {L.~D.}\ \bibnamefont
  {Landau}},\ and\ \bibinfo {author} {\bibfnamefont {E.~M.}\ \bibnamefont
  {Lifshitz}},\ }\href@noop {} {\emph {\bibinfo {title} {Statistical Physics.
  {{Part}} 2. {{Theory}} of the Condensed State}}},\ \bibinfo {edition}
  {reprinted}\ ed.,\ \bibinfo {series} {Course of Theoretical Physics}\ No.\
  \bibinfo {number} {by E. M. Lifshitz and L. P. Pitaevski{\u \i}; Vol.
  9[...]}\ (\bibinfo  {publisher} {{Elsevier}},\ \bibinfo {address}
  {{Oxford}},\ \bibinfo {year} {2006})\ p.\ \bibinfo {pages} {387}\BibitemShut
  {NoStop}%
\bibitem [{\citenamefont {Chubukov}\ \emph {et~al.}(2018)\citenamefont
  {Chubukov}, \citenamefont {Klein},\ and\ \citenamefont
  {Maslov}}]{Chubukov2018}%
  \BibitemOpen
  \bibfield  {author} {\bibinfo {author} {\bibfnamefont {A.~V.}\ \bibnamefont
  {Chubukov}}, \bibinfo {author} {\bibfnamefont {A.}~\bibnamefont {Klein}},\
  and\ \bibinfo {author} {\bibfnamefont {D.~L.}\ \bibnamefont {Maslov}},\
  }\bibfield  {title} {\bibinfo {title} {Fermi-{{Liquid Theory}} and
  {{Pomeranchuk Instabilities}}: {{Fundamentals}} and {{New Developments}}},\
  }\href {https://doi.org/10.1134/S1063776118110122} {\bibfield  {journal}
  {\bibinfo  {journal} {J. Exp. Theor. Phys.}\ }\textbf {\bibinfo {volume}
  {127}},\ \bibinfo {pages} {826} (\bibinfo {year} {2018})}\BibitemShut
  {NoStop}%
\bibitem [{\citenamefont {Chubukov}\ and\ \citenamefont
  {W{\"o}lfle}(2014)}]{Chubukov2014}%
  \BibitemOpen
  \bibfield  {author} {\bibinfo {author} {\bibfnamefont {A.~V.}\ \bibnamefont
  {Chubukov}}\ and\ \bibinfo {author} {\bibfnamefont {P.}~\bibnamefont
  {W{\"o}lfle}},\ }\bibfield  {title} {\bibinfo {title} {Quasiparticle
  interaction function in a two-dimensional {{Fermi}} liquid near an
  antiferromagnetic critical point},\ }\href
  {https://doi.org/10.1103/PhysRevB.89.045108} {\bibfield  {journal} {\bibinfo
  {journal} {Phys. Rev. B}\ }\textbf {\bibinfo {volume} {89}},\ \bibinfo
  {pages} {045108} (\bibinfo {year} {2014})}\BibitemShut {NoStop}%
\bibitem [{\citenamefont {Nambu}(1960)}]{Nambu1960}%
  \BibitemOpen
  \bibfield  {author} {\bibinfo {author} {\bibfnamefont {Y.}~\bibnamefont
  {Nambu}},\ }\bibfield  {title} {\bibinfo {title} {Quasi-{{Particles}} and
  {{Gauge Invariance}} in the {{Theory}} of {{Superconductivity}}},\ }\href
  {https://doi.org/10.1103/physrev.117.648} {\bibfield  {journal} {\bibinfo
  {journal} {Phys. Rev.}\ }\textbf {\bibinfo {volume} {117}},\ \bibinfo {pages}
  {648} (\bibinfo {year} {1960})}\BibitemShut {NoStop}%
\bibitem [{\citenamefont {Browne}\ and\ \citenamefont
  {Levin}(1983)}]{Browne1983}%
  \BibitemOpen
  \bibfield  {author} {\bibinfo {author} {\bibfnamefont {D.~A.}\ \bibnamefont
  {Browne}}\ and\ \bibinfo {author} {\bibfnamefont {K.}~\bibnamefont {Levin}},\
  }\bibfield  {title} {\bibinfo {title} {Collective modes in
  charge-density-wave superconductors},\ }\href
  {https://doi.org/10.1103/physrevb.28.4029} {\bibfield  {journal} {\bibinfo
  {journal} {Phys. Rev. B}\ }\textbf {\bibinfo {volume} {28}},\ \bibinfo
  {pages} {4029} (\bibinfo {year} {1983})}\BibitemShut {NoStop}%
\bibitem [{\citenamefont {Cea}\ and\ \citenamefont {Benfatto}(2014)}]{Cea2014}%
  \BibitemOpen
  \bibfield  {author} {\bibinfo {author} {\bibfnamefont {T.}~\bibnamefont
  {Cea}}\ and\ \bibinfo {author} {\bibfnamefont {L.}~\bibnamefont {Benfatto}},\
  }\bibfield  {title} {\bibinfo {title} {Nature and {{Raman}} signatures of the
  {{Higgs}} amplitude mode in the coexisting superconducting and
  charge-density-wave state},\ }\href
  {https://doi.org/10.1103/physrevb.90.224515} {\bibfield  {journal} {\bibinfo
  {journal} {Phys. Rev. B}\ }\textbf {\bibinfo {volume} {90}},\ \bibinfo
  {pages} {224515} (\bibinfo {year} {2014})}\BibitemShut {NoStop}%
\bibitem [{\citenamefont {Maiti}\ \emph {et~al.}(2017)\citenamefont {Maiti},
  \citenamefont {Chubukov},\ and\ \citenamefont {Hirschfeld}}]{Maiti2017}%
  \BibitemOpen
  \bibfield  {author} {\bibinfo {author} {\bibfnamefont {S.}~\bibnamefont
  {Maiti}}, \bibinfo {author} {\bibfnamefont {A.~V.}\ \bibnamefont
  {Chubukov}},\ and\ \bibinfo {author} {\bibfnamefont {P.~J.}\ \bibnamefont
  {Hirschfeld}},\ }\bibfield  {title} {\bibinfo {title} {Conservation laws,
  vertex corrections, and screening in {{Raman}} spectroscopy},\ }\href
  {https://doi.org/10.1103/physrevb.96.014503} {\bibfield  {journal} {\bibinfo
  {journal} {Phys. Rev. B}\ }\textbf {\bibinfo {volume} {96}},\ \bibinfo
  {pages} {014503} (\bibinfo {year} {2017})}\BibitemShut {NoStop}%
\bibitem [{\citenamefont {Fujikawa}(1979)}]{Fujikawa1979}%
  \BibitemOpen
  \bibfield  {author} {\bibinfo {author} {\bibfnamefont {K.}~\bibnamefont
  {Fujikawa}},\ }\bibfield  {title} {\bibinfo {title} {Path-{{Integral
  Measure}} for {{Gauge-Invariant Fermion Theories}}},\ }\href
  {https://doi.org/10.1103/physrevlett.42.1195} {\bibfield  {journal} {\bibinfo
   {journal} {Phys. Rev. Lett.}\ }\textbf {\bibinfo {volume} {42}},\ \bibinfo
  {pages} {1195} (\bibinfo {year} {1979})}\BibitemShut {NoStop}%
\end{thebibliography}%

\appendix
\begin{widetext}
	\section{Fermion bubbles in the limit of
	  zero
	  external momentum}
	\label{sec:bubbles}

	Consider the bare fermionic bubble in Eliashberg theory
	\begin{equation}
		\Pi^{\mu\nu}_{n+m, n} \equiv \sum_{\mathbf{k}}\tr[\hat{\tau}^\mu\hat{\mathcal{G}}_{n+m}(\mathbf{k})\hat{\tau}^\mu\hat{\mathcal{G}}_n(\mathbf{k})]
	\end{equation}
	where the Nambu Green's function is
	\begin{equation}
		\hat{\mathcal{G}}_k  = \frac{-i\tilde{\Sigma}_n -\xi\hat{\tau}_3-\phi_n\hat{\tau}_1}{D_k}, \quad D_k \equiv \tilde{\Sigma}_n^2 + \xi^2 + \phi_n^2.
	\end{equation}
	In the quasi-classical approximation this can be expressed via the integrals
	\begin{equation}
		\begin{gathered}
			I^{(1)}_{n+m,n}\equiv \nu \int d\xi \frac{1}{D_{n+m} D_{n}} = \frac{\pi\nu}{S_{n+m}S_n(S_{n+m} + S_n)}, \\
			I^{(2)}_{n+m,n}\equiv \nu \int d\xi \frac{\xi^2}{D_{n+m} D_{n}}
			= \frac{\pi\nu}{S_{n+m} + S_n}
		\end{gathered}
	\end{equation}
	where $S_n^2 = \tilde{\Sigma}_n^2 + \phi_n^2$.
	Explicitly,
	\begin{equation}
		\begin{gathered}
			\Pi^{\mu\nu}_{n+m,n} = c_{n+m, n}^{(1);\mu\nu} I^{(1)}_{n+m, n} + c_{n+m, n}^{(2);\mu\nu}I^{(2)}_{n+m, n}\\
			c^{(1);\mu\nu}_{n+m,n} \equiv \tr[\hat{\tau}^\mu(i\tilde{\Sigma}_{n+m} + \phi_{n+m}\hat{\tau}_1)\hat{\tau}^\nu(i\tilde{\Sigma}_{n} + \phi_{n}\hat{\tau}_1)]\\
			c^{(2);\mu\nu}_{n+m,n} =  \tr[\hat{\tau}^\mu\hat{\tau}_3\hat{\tau}^\nu\hat{\tau}_3]
			= 2\diag(1, -1, -1, 1)^{\mu\nu}.
		\end{gathered}
	\end{equation}
	Explicitly evaluating the traces we find for $c^{(1)}$
	\begin{gather}
		c^{(1);00}_{n+m, n}
		=
		c^{(1);11}_{n+m, n}
		= 2(\phi_{n+m}\phi_n - \tilde{\Sigma}_{n+m}\tilde{\Sigma}_n)\\
		c^{(1);01}_{n+m, n}
		=
		c^{(1);10}_{n+m, n}
		= 2i(\phi_{n+m}\tilde{\Sigma}_n + \phi_{n}\tilde{\Sigma}_{n+m})\\
		c^{(1);22}_{n+m, n}
		=
		c^{(1);33}_{n+m, n}
		= -2(\phi_{n+m}\phi_n + \tilde{\Sigma}_{n+m}\tilde{\Sigma}_n)\\
		c^{(1);23}_{n+m, n}
		=
		-c^{(1);32}_{n+m, n}
		= 2(\phi_{n+m}\tilde{\Sigma}_n - \phi_{n}\tilde{\Sigma}_{n+m})
	\end{gather}
	and the remaining elements are zero.
	Thus $\Pi$ is block diagonal
	\[
		\hat{\Pi}_{n+m, n}
		= \begin{pmatrix}
			\hat{\Pi}^L_{n+m, n} & 0                    \\
			0                    & \hat{\Pi}^T_{n+m, n}
		\end{pmatrix}
	\]
	with
	longitudinal block
	\begin{equation}
		\hat{\Pi}^L_{n+m, n} =
		\frac{2\pi\nu}{S_{n+m} S_{n} (S_{n+m} + S_n)}
		\begin{pmatrix}
			\phi_{n+m}\phi_n -\tilde{\Sigma}_{n+m}\tilde{\Sigma}_n + S_{n+m} S_n &
			i(\phi_{n+m}\tilde{\Sigma}_n + \phi_{n+m}\tilde{\Sigma}_n)             \\
			i(\phi_{n+m}\tilde{\Sigma}_n + \phi_{n+m}\tilde{\Sigma}_n)           &
			\phi_{n+m}\phi_n -\tilde{\Sigma}_{n+m}\tilde{\Sigma}_n - S_{n+m} S_n
		\end{pmatrix}
	\end{equation}
	and transverse block
	\begin{equation}
		\hat{\Pi}^T_{n+m, n} =
		\frac{2\pi\nu}{S_{n+m} S_{n} (S_{n+m} + S_n)}
		\begin{pmatrix}
			-\phi_{n+m}\phi_n -\tilde{\Sigma}_{n+m}\tilde{\Sigma}_n - S_{n+m} S_n &
			(\phi_{n+m}\tilde{\Sigma}_n - \phi_{n}\tilde{\Sigma}_{n+m})             \\
			-(\phi_{n+m}\tilde{\Sigma}_n - \phi_{n}\tilde{\Sigma}_{n+m})          &
			-\phi_{n+m}\phi_n -\tilde{\Sigma}_{n+m}\tilde{\Sigma}_n + S_{n+m} S_n
		\end{pmatrix}.
	\end{equation}
	In the $m\to0$ limit these reduce to
	\begin{equation}
		\begin{gathered}
			\hat{\Pi}^L_{n} =
			\frac{2\pi\nu}{Z_n \zeta^3_{n}}
			\begin{pmatrix}
				\Delta^2_n            &
				i\Delta_{n}\epsilon_n   \\
				i\Delta_{n}\epsilon_n &
				-\epsilon^2_n
			\end{pmatrix}\\
			\hat{\Pi}^T_{n+m, n} =
			-\frac{2\pi\nu}{Z_n\zeta_n}
			\begin{pmatrix}
				1 & 0 \\
				0 &
				0
			\end{pmatrix}
		\end{gathered}
	\end{equation}
	where $\zeta_n^2 = \epsilon_n^2 + \Delta_n^2$.

	\section{Ward identities in a superconductor with frequency dependent gap function}
	\label{sec:Ward}

	In this Appendix, we derive Ward identities
	associated with charge and spin conservation in a
	superconductor with frequency dependent gap function.

	Ward identities are special relations between vertices and self-energies, imposed by the conservation laws.
	For a system of fermions with $U(1)$ charge (gauge) symmetry and $SU(2)$ spin symmetry, they ensure that the total charge of the system (and, hence, the total number of fermions) and each component of the total spin do not change with time. In practical terms, we focus on Ward identities which
	relate two-fermion spin and charge density vertices at zero transferred  momentum and a finite transferred  frequency to the fermionic self-energy.
	The relations are particularly simple in Eliashberg-type theories, in which the self-energy $\Sigma (k, \epsilon_n)$ has much stronger dependence on frequency than on fermionic momentum, and the latter can be neglected.  On the Matsubara axis we then approximate $\Sigma (k, \epsilon_n) \approx \Sigma_n$.
	Within the same approximation,  spin and charge vertices, $\Gamma^{ch}$ and $\boldsymbol{\Gamma}^{sp}$ can also be treated as functions of frequency only.
	Each vertex depends on Matsubara frequencies and spin projections on the incoming and outgoing fermions,
	$\Gamma^{ch} = \Gamma^{ch}_{n+m \alpha,n \beta}$ and $\boldsymbol{\Gamma}^{sp} = \boldsymbol{\Gamma}^{sp}_{n+m \alpha, n \beta}$.

	For completeness, we also derive the Ward identity associated with the conservation of momentum.

	\subsection{Normal state}
	We define $\Sigma_n$ in the normal state via
	$G^{-1} (k, \epsilon_n) = i \epsilon_n - \Sigma_n - \xi_k$, where $\xi_k$ is the fermionic dispersion.
	The relations between $\Gamma^{ch}$, $\Gamma^{sp}$ and $\Sigma_n$
	are~\cite{Landau1957,Pitaevskii1960,Lifsic2006,Chubukov2018}:
	\begin{equation}
		\Gamma^{ch}_{n+m \alpha,n \beta} = \delta_{\alpha \beta} \Gamma,\quad \boldsymbol{\Gamma}^{sp}_{m+n, \alpha, m\beta} = \boldsymbol{\sigma}_{\alpha \beta}\Gamma,
		\label{eq:ch_1}
	\end{equation}
	where
	\begin{equation}
		\Gamma = 1 + i \frac{\Sigma_{n+m} -\Sigma_n}{\Omega_m}.
		\label{eq:ch_2}
	\end{equation}
	The bosonic $\Omega_m$ is the difference between outgoing and incoming fermionic frequencies.

	To set the stage for our analysis in the superconducting state, we present the diagrammatic proof of this relation.
	For this we  note that within Eliashberg theory the fermionic self-energy is obtained within the one-loop approximation, as a convolution of the fermionic propagator and the effective frequency-dependent ``local'' interaction $V^{l=0}_{n-n'}$, which is $V_{n-n'} (\mathbf{ k}_F -\mathbf{ k}'_F)$ integrated over the Fermi surface.
	Within the same computational scheme, the vertex $\Gamma^{ch}$ is obtained by summing up ladder series of vertex corrections, with the same $V^{l=0}_{n-n'}$.
	For  $\Gamma^{sp}$, the analysis is more nuanced:  ladder series hold when
	$V_{n-n'} (\mathbf{ k}_F -\mathbf{ k}'_F)$  is of density-density form, i.e.when  spin projection (up or down) is conserved along the interaction line.
	If $V_{n-n'} (\mathbf{ k}_F -\mathbf{ k}'_F)$  is a spin-spin interaction with spin $\sigma$-matrices in the vertices, one has to add additional Aslamazov-Larkin-type terms to get the proper series for $\Gamma^{sp}$~\cite{Chubukov2014}.
	For simplicity, below we assume that  the effective interaction is of the density-density type.

	\begin{figure}
		\centering
		\includegraphics[width=0.8\textwidth]{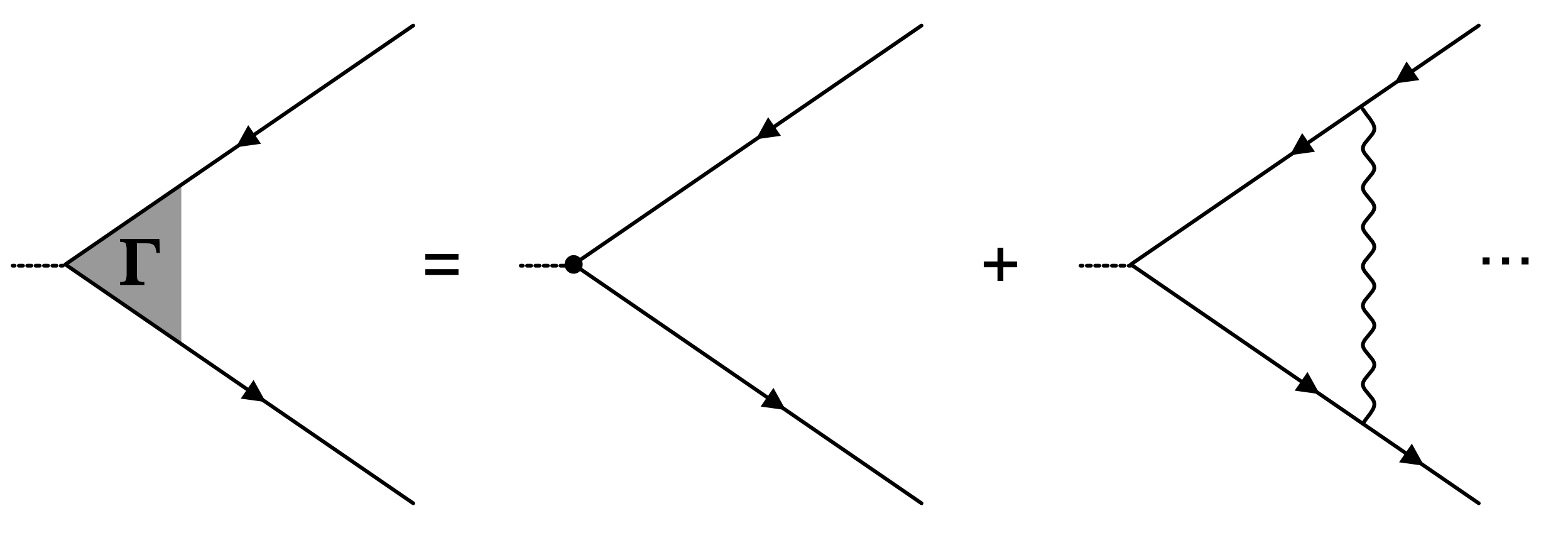}
		\caption{Ladder series for the renormalized vertex $\Gamma$, solid lines are the full Nambu Green's functions and the wavy line is the interaction.}
		\label{fig:ladder-series}
	\end{figure}
	The ladder series in
	\cref{fig:ladder-series}
	yields the following integral equation for $\Gamma_{n+m,n}$
	\begin{equation}
		\Gamma_{n+m,n} = 1 + \nu T \sum_{n'} V^{l=0}_{n-n'} \Gamma_{n'+m,n'} \int d \xi_k G_{n'+m} (\xi_k) G_{n'} (\xi_k)
		\label{eq:ch_3}
	\end{equation}
	where $\nu$ is the density of states at the Fermi level.
	The self-energy is given by
	\begin{equation}
		\Sigma_{n} = \nu T \sum_{n'} V^{l=0}_{n-n'}\int d \xi_k G_{n'} (\xi_k) = -\frac{i}{2} \nu T \sum_{n'} V^{l=0}_{n-n'} \sgn n'
		\label{eq:ch_4}
	\end{equation}
	The product of the two Green's functions in \cref{eq:ch_3} can be decoupled as
	\begin{equation}
		G_{n'+m} (\xi_k) G_{n'} (\xi_k)  = \left(G_{n'+m} (\xi_k)- G_{n'} (\xi_k)\right) \frac{i}{\Omega_m + i \left(\Sigma_{n+m}-\Sigma_{n}\right)}
		\label{eq:ch_5}
	\end{equation}
	Substituting into \cref{eq:ch_3}, we obtain
	\begin{equation}
		\Gamma_{n+m,n} = 1 + i \nu T \sum_{n'} V^{l=0}_{n-n'}\int d \xi_k \frac{\Gamma_{n'+m,n}}{\Omega_m + i \left(\Sigma_{n'+m}-\Sigma_{n'}\right)} \left(G_{n'+m} (\xi_k)- G_{n'} (\xi_k)\right)
		\label{eq:ch_6}
	\end{equation}
	One can straightforwardly verify that $\Gamma_{n+m,n}$ from
	\cref{eq:ch_2} is the solution of this equation.
	Indeed, substituting this $\Gamma_{n'+m,n'}$ into the r.h.s of \cref{eq:ch_6} we
	find  that it reduces to
	\begin{equation}
		1 + \frac{i}{\Omega_m} \nu T \sum_{n'} V^{l=0}_{n-n'}\int d \xi_k  \left(G_{n'+m} (\xi_k)- G_{n'} (\xi_k)\right).
	\end{equation}
	Using \cref{eq:ch_4}, we re-express this as
	\begin{equation}
		1 + i \frac{\Sigma_{n+m} -\Sigma_n}{\Omega_m},
	\end{equation}
	which is exactly $\Gamma_{n+m,n}$.

	Using the Ward identities, one can straightforwardly demonstrate that charge and spin correlators (the polarization bubbles) vanish at
	a zero incoming momentum and a finite incoming frequency, as should be the case for a conserved quantity $X$.
		[The choice of zero momentum and a finite frequency implies that one probes a variation of the total $X$ in the sample between different times.  For a conserved $X$, there is no such variation.]

	\begin{figure}
		\centering
		\includegraphics[height=8\baselineskip]{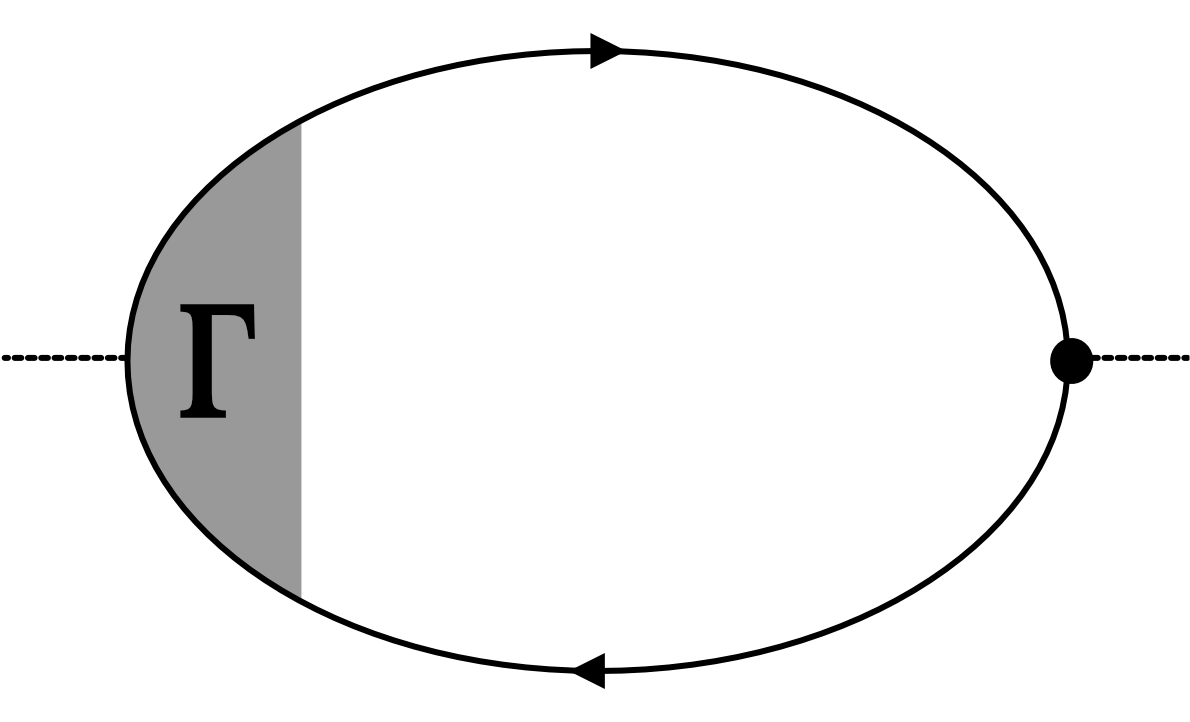}
		\caption{Dressed polarization bubble including the renomrmalized vertex $\Gamma$.}
		\label{fig:dressed-bubble}
	\end{figure}
	The fully dressed polarization bubble is shown in \cref{fig:dressed-bubble}.
	In explicit form,
	\begin{equation}
		\Pi (q=0, \Omega_m) =  \nu T \sum_{n} \int d \xi_k \Gamma_{n+m,n} G_{n+m} (\xi_k) G_{n} (\xi_k)
		\label{eq:ch_7}
	\end{equation}
	It is natural to integrate over $\xi_k$ first as this integration is straightforward.  One cannot, however, integrate over $\xi_k$  in infinite limits as at large frequencies, when $\epsilon_n > \Sigma_n$ and  $T \sum_n \to (1/2\pi) \int d \epsilon_n$, the Green's function approaches the unrenormalized form $G_{n} (\xi_k) =1/(i\epsilon_n - \xi_k)$ and the double integral $\int d \epsilon_n d \xi_k /(i\epsilon_n - \xi_k)^2$
	diverges logarithmically.
	The physically sound way to regularize the divergence  is to restrict  the $\xi_k$ integration to  $|\xi_k| < \Lambda$ and set $\Lambda \to \infty$ only at the end of the calculation.
	Carrying out the integration over $\xi_k$ this way, we obtain
	\begin{equation}
		\Pi^{ch} (q=0, \Omega_m) = \Pi^{sp} (q=0, \Omega_m) = \nu \left (1 - \frac{\Omega_m \Gamma_{n+m,n}}{\Omega_m + i \Sigma_{n+m}-\Sigma_{n}}\right) =0
		\label{eq:ch_8}
	\end{equation}
	as it should be.

	\subsection{Superconducting state}

	As in the main text, for definiteness we consider
	$s-$wave suprconductivity, in which case
	the pairing vertex $\phi_n$ and the gap function $\Delta_n$ are independent on the angle along the Fermi surface.
	We also set $T=0$ to avoid complications due to discreteness of Matsubara frequencies.
	We keep the notations $\Sigma_n$, etc, with the understanding that $\Sigma_n = \Sigma (\epsilon_n)$, where $\epsilon_n$ is a continuous variable along the Matsubara axis.

	\subsubsection{Distinction between charge and spin correlations}

	We argue below that in a superconductor charge and spin correlators have to be treated differently as the first one acquires an additional contribution from coupling to phase fluctuations.

	The distinction between spin and charge correlators can be seen already for a BCS superconductor.
	Both spin and charge polarization bubbles  have to
	vanish at zero incoming momentum and a nonzero incoming frequency $\Omega_m$, and we show below that
	this is indeed the case.  However, to prove this for the charge case, extra care is needed.

	Specifically, for a BCS superconductor, it is tempting to neglect the interaction and express spin and charge correlators as bubbles
	made of free-fermion Nambu Green's functions. For the charge bubble this gives
	\begin{equation}
		\Pi^{ch}_\text{free} (q=0, \Omega_m) =  \frac{\nu}{2\pi} \int d \epsilon_n \int^{\Lambda}_{-\Lambda} d \xi_k tr \left[\hat{\mathcal{G}}_{n+m, \alpha} (\xi_k) {\hat G}_{n,\alpha} (\xi_k)\right]
		\label{eq:ch_14}
	\end{equation}
	and for the spin case we have
	\begin{equation}
		\Pi^{sp, ii}_\text{free} (q=0, \Omega_m) =  \frac{\nu}{2\pi} \int d \epsilon_n  \int^{\Lambda}_{-\Lambda} d \xi_k tr \left[\sigma^{i}_{\alpha \beta}\hat{\mathcal{G}}_{n+m, \alpha} (\xi_k) {\hat G}_{n,\beta} \sigma^{i}_{\beta \alpha} (\xi_k)\right],
		\label{eq:ch_15}
	\end{equation}
	where $i =x,y,z$. For definiteness, we set $i=z$ below.

	Let's set continuous $\Omega_m$ to be finite but infinitesimally small. One can easily verify that for a non-zero $\Delta$, the limit $\Omega_m \to 0$ is entirely regular, and one can just set $\Omega_m =0$ in the calculations.
	Using \cref{eq:nambu-green}
	from the main text
	for the Green's function in Nambu representation and identifying $\phi_n$ in a BCS superconductor with $\Delta_n$, we obtain
	\begin{equation}
		\Pi^{ch}_\text{free} (q=0,  \Omega_m \to 0) =  \frac{\nu}{\pi} \int d \epsilon_n \int^{\Lambda}_{-\Lambda} d \xi_k
		\frac{(\xi^2_k - \Delta^2 -\epsilon^2_n)}{(\xi^2_k + \Delta^2 +\epsilon^2_n)^2}
		\label{eq:ch_14_a}
	\end{equation}
	and
	\begin{equation}
		\Pi^{sp,zz}_\text{free} (q=0, \Omega_m \to 0) =  \frac{\nu}{\pi} \int d \epsilon_n \int^{\Lambda}_{-\Lambda} d \xi_k
		\frac{(\xi^2_k + \Delta^2 -\epsilon^2_n)}{(\xi^2_k + \Delta^2 +\epsilon^2_n)^2}
		\label{eq:ch_15_a}
	\end{equation}
	In Gorkov's notations of normal and anomalous Green's functions
	$G_n (\xi_k) = -(i \epsilon_n + \xi_k)/(\xi^2_k + \epsilon^2_k + \Delta^2)$ and
	$F_n (\xi_k) = \Delta/(\xi^2_k + \epsilon^2_k + \Delta^2)$, the polarization bubbles are
	\begin{equation}
		\Pi^{ch}_\text{free} (q=0, \Omega_m \to 0)  = \frac{\nu}{\pi} \int d \epsilon_n \int^{\Lambda}_{-\Lambda} d \xi_k  ([G_n (\xi_k)]^2 - [F_n (\xi_k)]^2),
	\end{equation}
	and
	\begin{equation}
		\Pi^{sp,zz}_\text{free} (q=0, \Omega_m \to 0) = \frac{\nu}{\pi} \int d \epsilon_n \int^{\Lambda}_{-\Lambda} d \xi_k  ([G_n (\xi_k)]^2
		+
		[F_n (\xi_k)]^2),
	\end{equation}
	The two polarization bubbles differ in the sign of the $F^2$ term.

	The integration over $\epsilon_n$ and $\xi_k$ in \cref{eq:ch_14_a,eq:ch_15_a} can be carried out in any order, and the results are
	\begin{equation}
		\Pi^{sp,zz}_\text{free} (q=0, \Omega_m \to 0) =0,~~\Pi^{ch}_\text{free} (q=0, \Omega_m \to 0) = -2 \nu
		\label{eq:ch_16}
	\end{equation}
	We see that $\Pi^{sp,zz}_\text{free} (q=0 \Omega_m \to 0)$ vanishes, as expected, but $\Pi^{ch}_\text{free} (q=0 \Omega_m \to 0)$ remains finite.

	Because charge conservation must be satisfied, there must be another contribution to charge polarization, which cancels the free-fermion contribution, as was pointed out by Nambu~\cite{Nambu1960}.
	Such a contribution has been identified in other contexts as well, e.g. in the analysis of $A_{1g}$ Raman scattering in a BCS superconductor~\cite{Browne1983,Cea2014,Maiti2017}.
	The argument is that charge fluctuations are linearly coupled to phase fluctuations of the superconducting order parameter and this gives rise to the extra contribution $\Pi^{ch}_{ex} (q, \Omega_m) = S^2 (q, \Omega_m)
		\chi^{pp} (q, \Omega_m)$, where $S (q, \Omega_m)$  is the coupling and $\chi^{pp} (q, \Omega_m)$ is the  propagator of phase fluctuations.  The coupling, $S (\Omega_m)$, is generated by the triangular diagram, consisting of the original charge vertex, one normal and one anomalous Green's function, and the 4-fermion interaction $V$ (see \cref{fig:gf-diagrams}).
	This coupling vanishes at $\Omega_m =0$, but at a finite $\Omega_m$,  $S(\Omega_m) \propto V \Omega_m/\Delta$.
	Naively, this would imply that the extra contribution is irrelevant at $\Omega_m \to 0$.
	However, phase fluctuations are massless, and their propagator $\chi^{pp} (0, \Omega_m)  \propto \nu (\Delta/V \Omega_m)^2$.
	As a result, $\Pi^{ch}_{ex} (q, \Omega_m)$ is independent of $\Omega_m$ and is of order $\nu$, like the free-fermion $\Pi^{ch}_\text{free} (q=0, \Omega_m \to 0)$.

	\begin{figure}
		\centering
		\includegraphics[width=0.5\linewidth]{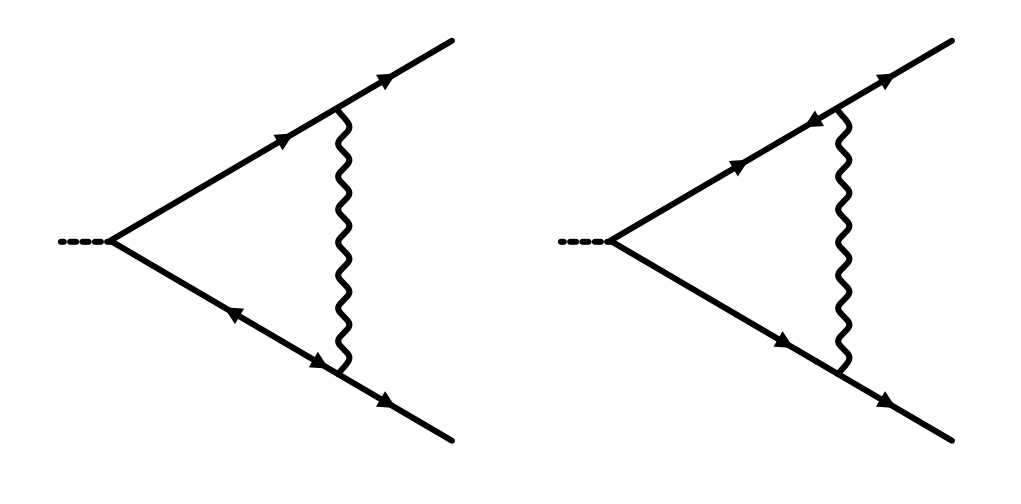}
		\caption{GF diagrams contributing to the coupling of the spin and charge vertices to the particle-particle susceptibility.
			Each contains one normal and one anomalous propagator.}
		\label{fig:gf-diagrams}
	\end{figure}
	We now compute explicitly the prefactor in $\Pi^{ch}_{ex} (q=0, \Omega_m \to 0) \sim \nu$.
	We first compute the particle-particle propagator.
	Within the ladder approximation (the same in which the BCS gap equation has been obtained)
	\begin{equation}
		\chi^{pp} (q, \Omega_m) = 2 \frac{\Pi^{pp} (q, \Omega_m)}{1 -V^{l=0} \Pi^{pp} (q, \Omega_m)},
		\label{eq:ch_17}
	\end{equation}
	where the overall factor $2$ is due to spin summation and
	\begin{equation}
		\Pi^{pp} (q, \Omega_m) = \frac{\nu}{\pi} \int d \epsilon_n \int^{\Lambda}_{-\Lambda} d \xi_k (G_n (\xi_k) G_{-n} (\xi_k) + [F_n (\xi_k)]^2),
		\label{eq:ch_18}
	\end{equation}
	where $V^{l=0}>0$ is an attractive interaction in the $s-$wave channel. Using $V^{l=0}\Pi^{pp} (0,0) =1$ and expanding in $\Omega_m$, we obtain
	\begin{equation}
		\chi^{pp} (0, \Omega_m) = 2\nu \left(\frac{2\Delta}{V^{l=0} \nu \Omega_m}\right)^2.
		\label{eq:ch_19}
	\end{equation}
	We next compute the coupling $S(\Omega_m)$. There are two topologically different triangular GF diagrams for $S(\Omega_m)$
	as shown in \cref{fig:gf-diagrams}.
	For the charge side vertex, they add up and yield
	\begin{equation}
		S(\Omega_m) = V^{l=0} \frac{\nu}{2\pi} \int d \epsilon_n \int^{\Lambda}_{-\Lambda} d \xi_k  \frac{\Delta \Omega_m}{(\xi^2_k + \epsilon^2_n + \Delta^2)^2} = \nu V^{l=0} \frac{\Omega_m}{2\Delta}
		\label{eq:ch_20}
	\end{equation}
	We then obtain
	\begin{equation}
		\Pi^{ch}_{ex} (q=0, \Omega_m \to 0) = 2\nu \left(\frac{\nu V^{l=0} \Omega_m}{2\Delta}\right)^2
		\left(\frac{2\Delta}{\nu V \Omega_m}\right)^2 = 2\nu
		\label{eq:ch_21}
	\end{equation}
	Combining with \cref{eq:ch_16}, we  find that $\Pi^{ch}_\text{free} (q=0, \Omega_m \to 0) + \Pi^{ch}_{ex} (q=0, \Omega_m \to 0) =0$, as it should be because the total charge is the conserved quantity.

	For the spin correlator, the two contributions to the coupling $S(\Omega_m)$ cancel out at order $\Omega_m$.  Then there is no additional $\Omega_m$ - independent contribution to the spin propagator, consistent with the vanishing of $\Pi^{sp,zz}_\text{free} (q=0, \Omega_m \to 0)$.
	From a physics perspective, this is a consequence of the fact that spin fluctuations do no couple linearly to phase fluctuations.

	Below we extend the analysis of a BCS superconductor to the case when the effective 4-fermion interaction is a dynamical $V_{n-n'}$. A dynamical interaction gives rise to fermionic self-energy $\Sigma_n$, and also the
	pairing vertex $\phi_n$ and the gap function $\Delta_n$ become functions of frequency.  The proof of the Ward identities in this situation is more involved, and for the charge correlator it is further involved by the necessity to include the coupling to phase fluctuations.
	For this reason, we consider Ward identities associated with spin and charge conservation separately.

	\subsubsection{Ward identity for \texorpdfstring{$\Gamma^{sp}_{n+m,n}$}{Gamma-sp n+m, n}}.

	\begin{figure}
		\centering
		\includegraphics[width=\linewidth]{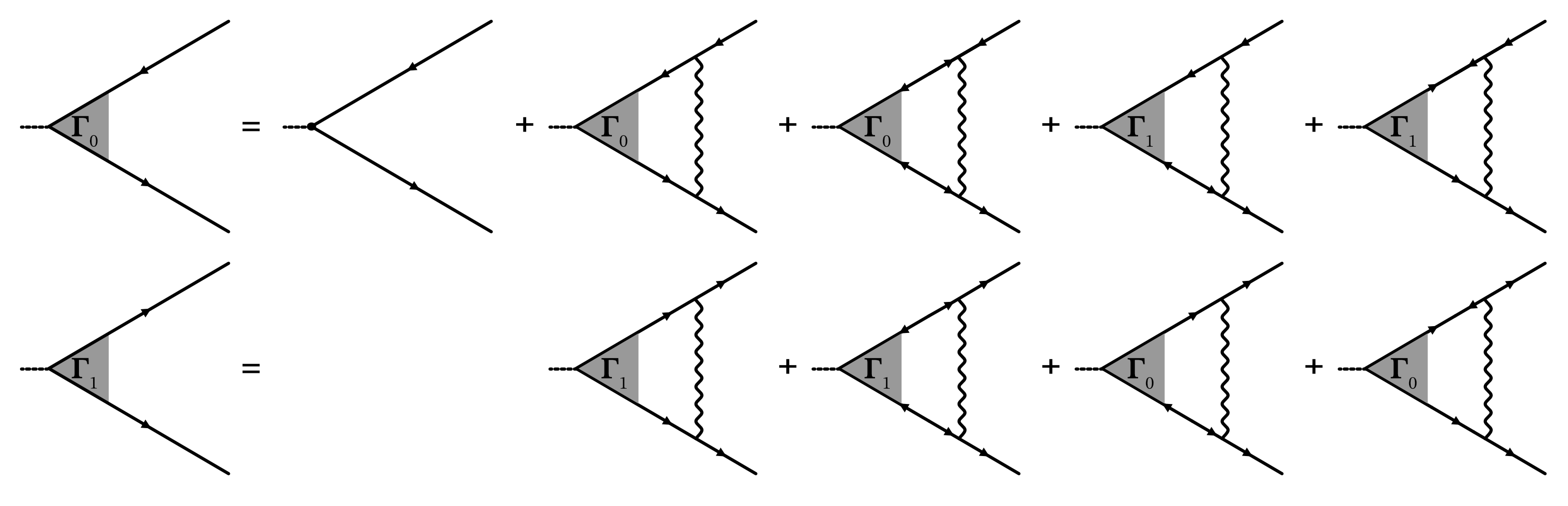}
		\caption{Bethe-Salpeter equation in the Ladder approximation for the renormalized normal $\Gamma^0$ and anomalous $\Gamma^1$ vertices.
			The spin and charge diagrams differ in their spin structure and thus the symmetry of vertices under reversal of the direction of the legs.
			In particular the relative sign of between two anomalous diagrams in each RHS differs for the spin and charge channels.
			\label{fig:spin-charge-vertices}
		}
	\end{figure}

	As before, we use matrix Nambu notations and write the self-consistent one-loop equation for the
	matrix $\hat{\Sigma}_n	= -i \Sigma_n
		\hat{\tau}_0  + \phi_n \hat{\tau}_1$
	and the ladder equation for the matrix $\hat{\Gamma}^{sp,ii}_{n+m \alpha,n \beta}	= \sigma^i_{\alpha \beta} {\hat \Gamma}_{n+m,n}$,
	where
	${\hat \Gamma}_{n+m,n} = \Gamma^{(0)}_{n+m,n}\hat{\tau}_0  + \Gamma^{(1)}_{n+m,n}\hat{\tau}_1$,
	(see
	\cref{eq:sigmahatdef,eq:nnnn_1} in the main text).
	The equations for ${\hat \Sigma}_n$ and $\hat{\Gamma}_{n+m,n}$ are formally the same as \cref{eq:ch_3,eq:ch_4},
	but now have matrix form
	\begin{equation}
		\hat{\Sigma}_n = \nu T\sum_{n'}V^{l=0}_{n-n'} \int d\xi \hat{\tau}_3 \hat{\mathcal{G}}_{k'}\hat{\tau}_3
		\label{eq:ch_9}
	\end{equation}
	and
	\begin{equation}
		\hat{\Gamma}_{n+m,n} = 1 + \nu T\sum_{n'}V^{l=0}_{n-n'}\int d\xi_k \hat{\tau}_3\hat{\mathcal{G}}_{n'+m} (\xi_k)\hat{\Gamma}_{n'+m, n'}\hat{\mathcal{G}}_{n'} (\xi_k)\hat{\tau}_3.
		\label{eq:Bs_sp}
	\end{equation}
	Splitting ${\hat \Gamma}$ into components and taking the limit of $\Omega_m \to 0$,  we obtain the set of two coupled equations
	\begin{equation}
		\begin{gathered}
			\begin{multlined}
				\Gamma^{(0)}_{n+m,n} = 1 + \frac{\nu}{2\pi} \int d \epsilon_{n'} \int^{\Lambda}_{-\Lambda} d \xi_k  \Gamma^{(0)}_{n'+m,n'}
				\frac{\xi^2_k + \phi^2_{n'} - \tilde{\Sigma}^2_{n'}}{(\xi^2_k + \phi^2_{n'} + \tilde{\Sigma}^2_{n'})^2} V^{l=0}_{n-n'}\\
				+ 2 i \frac{\nu}{2\pi} \int d \epsilon_{n'} \int^{\Lambda}_{-\Lambda} d \xi_k \Gamma^{(1)}_{n'+m,n'} \frac{\tilde{\Sigma}_{n'} \phi_{n'}}{(\xi^2_k + \phi^2_{n'} + \tilde{\Sigma}^2_{n'})^2} V^{l=0}_{n-n'}
			\end{multlined}\\
			\begin{multlined}
				\Gamma^{(1)}_{n+m,n} = \frac{\nu}{2\pi}  \int d \epsilon_{n'} \int^{\Lambda}_{-\Lambda} d \xi_k  \Gamma^{(1)}_{n'+m,n'} \frac{\xi^2_k - \phi^2_{n'} + \tilde{\Sigma}^2_{n'}}{(\xi^2_k + \phi^2_{n'} + \tilde{\Sigma}^2_{n'})^2} V^{l=0}_{n-n'}\\
				-2 i \frac{\nu}{2\pi} \int d \epsilon_{n'} \int^{\Lambda}_{-\Lambda} d \xi_k \Gamma^{(
					0)}_{n'+m,n'} \frac{\tilde{\Sigma}_{n'} \phi_{n'}}{(\xi^2_k + \phi^2_{n'} + \tilde{\Sigma}^2_{n'})^2} V^{l=0}_{n-n'}
			\end{multlined}
		\end{gathered}
		\label{eq:ch_22}
	\end{equation}
	We assume that the  dynamical interaction vanishes in the limit of large frequency transfer.  The
	double integral over $\xi_k$ and $\epsilon_{n'}$ is then ultraviolet convergent, and the integration over $\xi_k$ can be extended to infinite limits.  Integrating over $\xi_k$ in each term in \cref{eq:ch_22}, we reduce it
	to
	\begin{equation}
		\begin{gathered}
			\Gamma^{(0)}_{n+m,n} = 1 + \frac{\nu}{2} \int d \epsilon_{n'} \Gamma^{(0)}_{n'+m,n'} \frac{\Delta^2_{n'} }{Z_{n'} (\Delta^2_{n'} + \epsilon^2_{n'})^{3/2}} V^{l=0}_{n-n'}  + i \frac{\nu}{2} \int d \epsilon_{n'}  \Gamma^{(1)}_{n'+m,n'} \frac{\epsilon_{n'} \Delta_{n'}}{Z_{n'} (\epsilon^{2}_{n'}+ \Delta^2_{n'})^{3/2}} V^{l=0}_{n-n'}\\
			\Gamma^{(1)}_{n+m,n} = -\frac{\nu}{2} \int d \epsilon_{n'}  \Gamma^{(
				0)}_{n'+m,n'} \frac{\epsilon^2_{n'} }{Z_{n'}
			(\epsilon^{2}_{n'}+ \Delta^2_{n'})^{3/2}} V^{l=0}_{n-n'}  - i \frac{\nu}{2} \int d \epsilon_{n'} \Gamma^{(1)}_{n'+m,n'} \frac{\epsilon_{n'} \Delta_{n'}}{Z_{n'} (\epsilon^{2}_{n'}+ \Delta^2_{n'})^{3/2}} V^{l=0}_{n-n'}
		\end{gathered}
		\label{eq:ch_23}
	\end{equation}
	We assume and verify that the solution of these equations is the matrix extension of \cref{eq:ch_2}
	\begin{equation}
		\hat{\Gamma}_{n+m,n} = \hat{\tau}_0 + i\frac{\hat{\Sigma}_{n+m} -\hat{\Sigma}_{n}}{\Omega_m} \overset{\Omega_m \to 0}{\longrightarrow} \hat{\tau}_0 + i \frac{d\hat{\Sigma}_{n}}{d\epsilon_n}.
		\label{eq:ch_11}
	\end{equation}
	Using $\hat{\Sigma}_{n} =-i \Sigma_n \tau_0 + i \phi_n \tau_1$,
	we re-write \cref{eq:ch_11} in components
	\begin{equation}
		\Gamma^{(0)}_{n+m,n} = 1 + \frac{d{\Sigma}_{n}}{d\epsilon_n},\quad \Gamma^{(1)}_{n+m,n} = i \frac{d\phi_{n}}{d\epsilon_n}
		\label{eq:ch_12}
	\end{equation}
	Substituting these forms into the r.h.s. \cref{eq:ch_23} and using \cref{eq:deriv-phi} from the main text:
	\begin{equation}
		\frac{d \phi_n}{d \epsilon_n}
		= \frac{\Delta_n}{\epsilon_n} \left(1 + \frac{d \Sigma_n}{d \epsilon_n}\right) - \frac{\Delta_n}{\epsilon_n}Z_n+ Z_n \frac{d \Delta_n}{d \epsilon_n},
		\label{eq:ch_24_a}
	\end{equation}
	we obtain after simple algebra
	\begin{equation}
		\begin{gathered}
			\Gamma^{(0)}_{n+m,n} = 1 + \frac{\nu}{2} \int d \epsilon_{n'} \frac{\Delta_{n'} \left(\Delta_{n'} - \epsilon_{n'} \frac{d \Delta_{n'}}{d\epsilon_{n'}}\right)}{(\Delta^2_{n'} + \epsilon^2_{n'})^{3/2}} V^{l=0}_{n-n'}\\
			\Gamma^{(1)}_{n+m,n} = -i\frac{\nu}{2} \int d \epsilon_{n'} \frac{\epsilon_{n'} \left(\Delta_{n'} - \epsilon_{n'} \frac{d \Delta_{n'}}{d\epsilon_{n'}}\right)}{(\Delta^2_{n'} + \epsilon^2_{n'})^{3/2}} V^{l=0}_{n-n'}
		\end{gathered}
		\label{eq:ch_24}
	\end{equation}
	Note that the quasiparticle residue $Z_n$ cancels out between \cref{eq:ch_23,eq:ch_24}.
	Using
	\begin{equation}
		\begin{gathered}
			\frac{\Delta_{n'} \left(\Delta_{n'} - \epsilon_{n'} \frac{d \Delta_{n'}}{d\epsilon_{n'}}\right)}{(\Delta^2_{n'} + \epsilon^2_{n'})^{3/2}} = \frac{d}{d \epsilon_{n'}} \left(\frac{\epsilon_{n'}}{(\Delta^2_{n'} + \epsilon^2_{n'})^{1/2}}\right)\\
			\frac{\epsilon_{n'} \left(\Delta_{n'} - \epsilon_{n'} \frac{d \Delta_{n'}}{d\epsilon_{n'}}\right)}{(\Delta^2_{n'} + \epsilon^2_{n'})^{3/2}} = -\frac{d}{d \epsilon_{n'}} \left(\frac{\Delta_{n'}}{(\Delta^2_{n'} + \epsilon^2_{n'})^{1/2}}\right),
		\end{gathered}
		\label{eq:ch_25}
	\end{equation}
	integrating by parts, and replacing $d V_{n-n'}/d \epsilon_{n'}$ by  $- d V_{n-n'}/d \epsilon_{n}$, we obtain
	\begin{equation}
		\begin{gathered}
			\Gamma^{(0)}_{n+m,n} = 1 +  \frac{d}{d \epsilon_n} \left[\frac{\nu}{2}\int d \epsilon_{n'} \frac{\epsilon_{n'}}{(\Delta^2_{n'} + \epsilon^2_{n'})^{1/2}}\right]\\
			\Gamma^{(1)}_{n+m,n} =  \frac{d}{d \epsilon_n} \left[\frac{\nu}{2}\int d \epsilon_{n'}\frac{\Delta_{n'}}{(\Delta^2_{n'} + \epsilon^2_{n'})^{1/2}}\right]
		\end{gathered}
		\label{eq:ch_26}
	\end{equation}
	The self-energy and the pairing vertex are given by \cref{eq:SP,eq:defZD} from the main text:
	\begin{equation}
		\begin{gathered}
			\Sigma_n = \frac{\nu}{2}\int d \epsilon_{n'} \frac{\epsilon_{n'}}{(\Delta^2_{n'} + \epsilon^2_{n'})^{1/2}}\\
			\phi_n = \frac{\nu}{2}\int d \epsilon_{n'} \frac{\Delta_{n'}}{(\Delta^2_{n'} + \epsilon^2_{n'})^{1/2}}
		\end{gathered}
		\label{eq:ch_27}
	\end{equation}

	Comparing \cref{eq:ch_26,eq:ch_27}, we see that the relations \cref{eq:ch_12} are satisfied. These relations are spin Ward identities for a superconductor with frequency-dependent gap function.

	\begin{figure}
		\centering
		\includegraphics[width=0.9\linewidth]{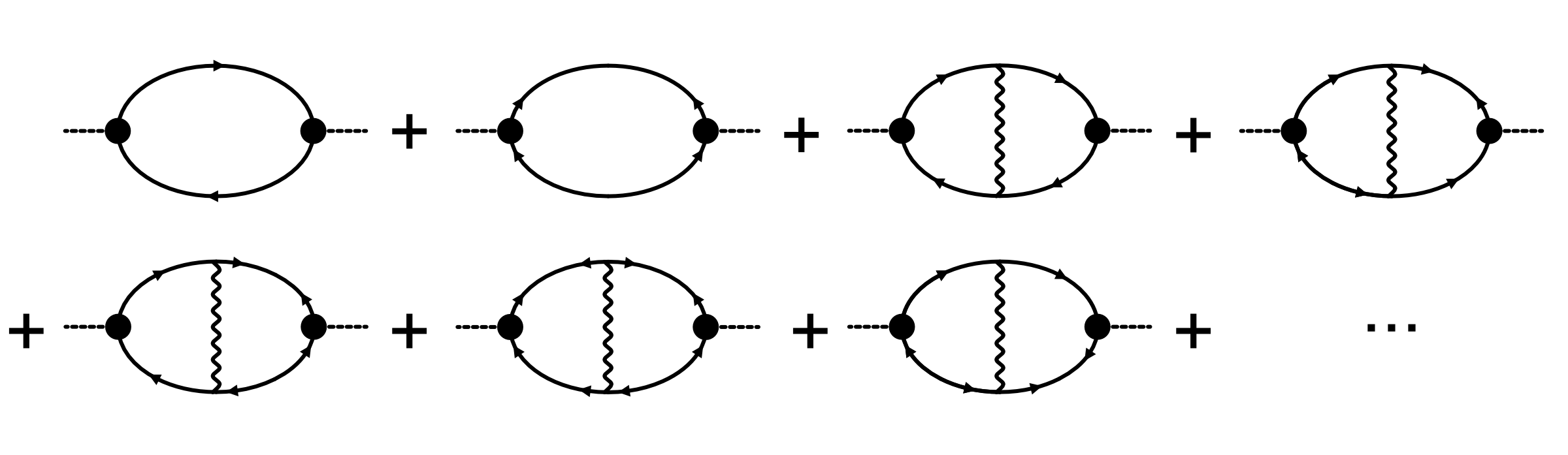}
		\caption{Ladder series contributing to the spin and charge correlators to first order in the $\mathbf{q}=0, i\Omega_{m}\to0$ limit in terms of Gorkov's normal and anomalous functions.
			For the spin correlator the side vertex is $\sigma^{z}$ while for the charge correlator is $\sigma^{0}$.
			This cause terms containing anomalous propagators at only one side vertex to differ in sign between the spin and charge series
			since $[\hat{F}, \hat{\sigma}_{0}]=0$ while $\{\hat{F},\hat{\sigma}^{z}\}$.
		}
		\label{fig:spin-ladder}
	\end{figure}

	\begin{figure}
		\centering
		\includegraphics[width=0.6\linewidth]{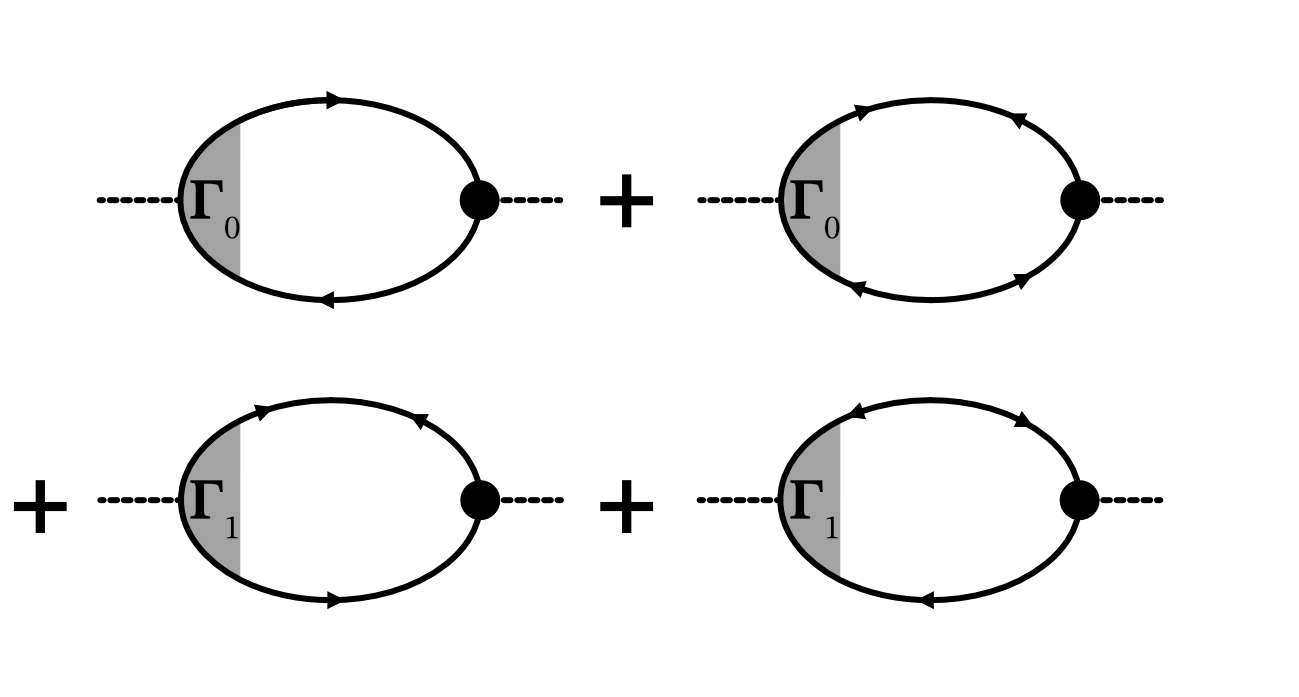}
		\caption{Bubble diagram for the renormalized spin or charge correlator at $\mathbf{q}=0$.
			The normal $\Gamma^{0}$ and anomalous $\Gamma^{1}$ vertices are solutions of the Bethe-Salpeter equations
			\cref{eq:Bs_sp,eq:Bs_ch}
			for spin and charge respectively.
			The diagrammatic formulation of the vertex renormalization is shown in \cref{fig:spin-charge-vertices}.
			\label{fig:spin-charge-bubble}
		}
	\end{figure}

	We next use these Ward identities to prove that $\Pi^{sp,ii} (q=0, \Omega_m \to 0)$ vanishes, as is required by global spin conservation.
	To first order in the interaction $V^{l=0}_{n-n'}$, the ladder diagrams for $\Pi^{sp,ii} (q, \Omega_m)$ in a superconductor are shown
	in \cref{fig:spin-ladder}.
	The full spin polarization bubble, expressed
	in terms of renormalized vertices, is shown in \cref{fig:spin-charge-bubble}.
	In analytical form,
	\begin{equation}
		\Pi^{sp,ii} (q=0, \Omega_m \to 0) = \frac{\nu}{\pi} \int d \epsilon_n \int_{-\Lambda}^{\Lambda} d \xi_k
		\frac{\xi^2_k - Z^2_n \left[ \left(\epsilon^2_n - \Delta^2_n\right) \left(1 + d \Sigma_n/d\epsilon_n\right) +2 \epsilon_n \Delta_n d \phi_n/d\epsilon_n \right]}{(\xi^2_k + Z^2_n(\Delta^2_{n'} + \epsilon^2_{n'}))^2}
		\label{eq:ch_28}
	\end{equation}
	Integrating over $\xi_k$ in the same was as we did in the normal state and then taking the limit $\Lambda \to \infty$, we obtain
	\begin{equation}
		\Pi^{sp,ii} (q=0, \Omega_m \to 0) = \nu \left[2 - \int d \epsilon_n \frac{\Delta^2_n \left(1 + d \Sigma_n/d\epsilon_n\right) - \Delta_n \epsilon_n d \phi_n/d\epsilon_n}{Z_n (\Delta^2_n + \epsilon^2_n)^{3/2}}\right]
		\label{eq:ch_29}
	\end{equation}
	Expressing $d \phi_n/d\epsilon_n$ via \cref{eq:ch_24_a}, we re-write \cref{eq:ch_29} as
	\begin{equation}
		\Pi^{sp,ii} (q=0, \Omega_m \to 0) = \nu \left[2 - \int d \epsilon_n  \frac{ \Delta_n \left(\Delta_n - \epsilon_n \frac{d \Delta_n}{d\epsilon_n}\right)}{(\Delta^2_n + \epsilon^2_n)^{3/2}}\right]
		\label{eq:ch_30}
	\end{equation}
	Using \cref{eq:ch_25} we re-express the integrand in the r.h.s. of \cref{eq:ch_30} as the full derivative:
	\begin{equation}
		\Pi^{sp,ii} (q=0, \Omega_m \to 0) = 2\nu \left[1 - \int_0^\infty d \epsilon_n \frac{d}{d \epsilon_n}
			\frac{\epsilon_n}{(\Delta^2_n + \epsilon^2_n)^{1/2}}\right]
		\label{eq:ch_31}
	\end{equation}
	We see that $\Pi^{sp,ii} (q=0, \Omega_m \to 0)$ indeed vanishes.

	\subsubsection{Ward identity for \texorpdfstring{$\Gamma^{ch}_{n+m,n}$}{Gamma-ch, n}}.

	We now
	repeat the calculations of the previous section in the charge channel,
	for the vertex
	$\hat{\Gamma}^{ch}_{n+m \alpha,n \beta}    = \delta_{\alpha \beta} {\hat \Gamma}_{n+m,n}$, ${\hat \Gamma}_{n+m,n} = \hat{\tau}_3\left(\Gamma^{(0)}_{n+m,n}\hat{\tau}_0  + \Gamma^{(1)}_{n+m,n}\hat{\tau}_1\right)$

	The
	Bethe-Salpeter equation for
	$\hat{\Gamma}_{n+m,n}$
	is
	\begin{equation}
		\hat{\Gamma}_{n+m,n} = 1 + \nu T\sum_{n'}V^{l=0}_{n-n'}\int d\xi_k \hat{\mathcal{G}}_{n'+m }(\xi_k)\hat{\tau}_3\hat{\Gamma}_{n'+m, n'}\hat{\mathcal{G}}_{n'} (\xi_k)\hat{\tau}_3.
		\label{eq:Bs_ch}.
	\end{equation}
	Again, working in the small $\Omega_{m}$ limit we can express this in components
	\begin{equation}
		\begin{gathered}
			\begin{multlined}
				\Gamma^{(0)}_{n+m,n} = 1
				+ \frac{\nu}{2\pi} \int d \epsilon_{n'} \int^{\Lambda}_{-\Lambda} d \xi_k  \Gamma^{(0)}_{n'+m,n'}
				\frac{\xi_{\mathbf{k}}^{2}-\tilde{\Sigma}_{n}^{2}-\phi_{n}^{2}}{(\tilde{\Sigma}^{2}_{n} + \phi^{2}_{n} + \xi^{2}_{\mathbf{k}})^{2}}
				V^{l=0}_{n-n'}\\
				+ i\Omega_{m}\frac{\nu}{2\pi} \int d \epsilon_{n'} \int^{\Lambda}_{-\Lambda} d \xi_k \Gamma^{(1)}_{n'+m,n'}
				\frac{
				\phi_{n} \left[1 + \frac{d\Sigma_{n}}{d\epsilon_{n}}\right]
				-
				\tilde{\Sigma}_{n} \frac{d\phi_{n}}{d\epsilon_{n}}
				}{(\tilde{\Sigma}^{2}_{n} + \phi^{2}_{n} + \xi^{2}_{\mathbf{k}})^{2}}
				V^{l=0}_{n-n'}
			\end{multlined}\\
			\begin{multlined}
				\Gamma^{(1)}_{n+m,n} =
				\frac{\nu}{2\pi} \int d \epsilon_{n'} \int^{\Lambda}_{-\Lambda} d \xi_k \Gamma^{(1)}_{n'+m,n'}
				\frac{
				1
				}{\tilde{\Sigma}^{2}_{n} + \phi^{2}_{n} + \xi^{2}_{\mathbf{k}}}
				V^{l=0}_{n-n'}\\
				-i\Omega_{m}\frac{\nu}{2\pi} \int d \epsilon_{n'} \int^{\Lambda}_{-\Lambda} d \xi_k \Gamma^{(0)}_{n'+m,n'}
				\frac{
				\phi_{n} \left[1 + \frac{d\Sigma_{n}}{d\epsilon_{n}}\right]
				-
				\tilde{\Sigma}_{n} \frac{d\phi_{n}}{d\epsilon_{n}}
				}{(\tilde{\Sigma}^{2}_{n} + \phi^{2}_{n} + \xi^{2}_{\mathbf{k}})^{2}}
				V^{l=0}_{n-n'}.
			\end{multlined}
		\end{gathered}
	\end{equation}
	Integrating  over $\xi_{\mathbf{k}}$, we
	find
	\begin{equation}
		\begin{gathered}
			\begin{multlined}
				\Gamma^{(0)}_{n+m,n} = 1
				+ i\Omega_{m}\frac{\nu}{4} \int d \epsilon_{n'} \Gamma^{(1)}_{n'+m,n'}
				\frac{
					\Delta_{n'} \left[1 + \frac{d\Sigma_{n'}}{d\epsilon_{n'}}\right]
					-
					\epsilon_{n'} \frac{d\phi_{n'}}{d\epsilon_{n'}}
				}{Z^{2}_{n}(\epsilon^{2}_{n'} + \Delta^{2}_{n'} )^{3/2}}
				V^{l=0}_{n-n'}
			\end{multlined}\\
			\begin{multlined}
				\Gamma^{(1)}_{n+m,n} =
				\frac{\nu}{2} \int d \epsilon_{n'}
				\Gamma^{(1)}_{n'+m,n'}
				\frac{
					1
				}{Z_{n'}\sqrt{\epsilon^{2}_{n'} + \Delta^{2}_{n'} }}
				V^{l=0}_{n-n'}
				\\
				-i\Omega_{m}\frac{\nu}{4} \int d \epsilon_{n'}
				\Gamma^{(0)}_{n'+m,n'}
				\frac{
					\Delta_{n'} \left[1 + \frac{d\Sigma_{n'}}{d\epsilon_{n'}}\right]
					-
					\epsilon_{n'} \frac{d\phi_{n'}}{d\epsilon_{n'}}
				}{Z^{2}_{n}(\epsilon^{2}_{n'} + \Delta^{2}_{n'} )^{3/2}}
				V^{l=0}_{n-n'}.
			\end{multlined}
		\end{gathered}
		\label{eq:ll}
	\end{equation}
	Note that there is no term with $\Gamma^{(0)}$ in the r.h.s. of the first equation.
	We again use \cref{eq:deriv-phi} to rewrite
	\begin{equation}
		\Delta_{n'} \left[1 + \frac{d\Sigma_{n'}}{d\epsilon_{n'}}\right]
		-
		\epsilon_{n'} \frac{d\phi_{n'}}{d\epsilon_{n'}}
		= Z_{n}\left(\Delta_{n}-\frac{d\Delta_{n}}{d\epsilon_{n}}\right).
	\end{equation}
	Substituting into \cref{eq:ll}, we
	obtain
	\begin{equation}
		\begin{gathered}
			\Gamma^{(0)}_{n+m,n} = 1
			+ i\Omega_{m}\frac{\nu}{4} \int d \epsilon_{n'} \Gamma^{(1)}_{n'+m,n'}
			\frac{\Delta_{n'}-\frac{d\Delta_{n'}}{d\epsilon_{n'}}}{Z_{n'}(\epsilon^{2}_{n'} + \Delta^{2}_{n'} )^{3/2}}
			V^{l=0}_{n-n'}\\
			\Gamma^{(1)}_{n+m,n} =
			\frac{\nu}{2} \int d \epsilon_{n'}
			\Gamma^{(1)}_{n'+m,n'}
			\frac{
				1
			}{Z_{n'}\sqrt{\epsilon^{2}_{n'} + \Delta^{2}_{n'} }}
			V^{l=0}_{n-n'}
			-i\Omega_{m}\frac{\nu}{4} \int d \epsilon_{n'}
			\Gamma^{(0)}_{n'+m,n'}
			\frac{\Delta_{n'}-\frac{d\Delta_{n'}}{d\epsilon_{n'}}}{Z_{n'}(\epsilon^{2}_{n'} + \Delta^{2}_{n'} )^{3/2}}
			V^{l=0}_{n-n'}.
		\end{gathered}
		\label{eq:ll_2}
	\end{equation}
	One can straightforwardly verify that at $\Omega_m \to 0$, the solution of these equations is
	\begin{equation}
		\begin{gathered}
			\lim_{\Omega_m \to 0}	\Gamma^{(0)}_{n+m,n} = 1
			+ \frac{\nu}{2} \int d \epsilon_{n'}
			\Delta_{n'}\frac{\Delta_{n'}-\frac{d\Delta_{n'}}{d\epsilon_{n'}}}{(\epsilon^{2}_{n'} + \Delta^{2}_{n'} )^{3/2}}
			V^{l=0}_{n-n'}
			= 1 + \frac{d\Sigma_{n}}{d\epsilon_n},
			\\
			\lim_{\Omega_m \to 0} \Gamma^{(1)}_{n+m,n} =
			\frac{\nu}{i\Omega_{m}} \int d \epsilon_{n'}
			\frac{
				\Delta_{n'}
			}{\sqrt{\epsilon^{2}_{n'} + \Delta^{2}_{n'} }}
			V^{l=0}_{n-n'}
			= \frac{2 \phi_n}{i \Omega_m}
		\end{gathered}
		\label{eq:ll_1}
	\end{equation}
	Note that for this solution
	the last term for $\Gamma^{1}$ can be dropped in \cref{eq:ll_1}.

	We can now verify that $\Pi^{ch}(\mathbf{q}=0,i\Omega_{m}\to0)$ vanishes as required by global charge conservation.
	The ladder diagrams contributing to the bubble are still given by \cref{fig:spin-ladder,fig:spin-charge-bubble}, but the side vertices are now spin $\delta$-functions.
	Analytically, the full $\Pi^{ch}(\mathbf{q}=0,i\Omega_{m}\to0)$ is expressed as
	\begin{equation}
		\Pi^{ch}(\mathbf{q}=0,i\Omega_{m}\to0)
		= \frac{\nu}{2\pi}\int d\epsilon_{n'} \int_{-\Lambda}^{\Lambda} d\xi_{k}
		2\left(
		\left(1 + \frac{d\Sigma_{n'}}{d\epsilon_{n'}}\right)
		\frac{(\xi_{k}^{2}-\tilde{\Sigma}^{2}_{n'} -\phi^{2}_{n'})}
		{(\tilde{\Sigma}^{2}_{n'} + \phi^{2}_{n'} + \xi^{2}_{\mathbf{k}})^{2}}
		+
		\frac{2\phi_{n'}}{i\Omega_{m}}\frac{i\Omega_{m}}{2}\frac{Z_{n'}(\Delta_{n'} - \epsilon_{n'}\frac{d\Delta_{n'}}{d\epsilon_{n'}})}{(\tilde{\Sigma}^{2}_{n'} + \phi^{2}_{n'} + \xi^{2}_{\mathbf{k}})^{2}}
		\right).
	\end{equation}
	Substituting $\Gamma^{(0)}_{n+m,n}$ and $\Gamma^{(1)}_{n+m,n}$ at $\Omega_m \to 0$ from \cref{eq:ll_1} and
	doing the integrals as in the normal state, we find
	\begin{equation}
		\Pi^{ch}(\mathbf{q}=0,i\Omega_{m}\to0)
		= \nu
		\left(-2
		+
		\int d\epsilon_{n'}
		\Delta_{n'}\frac{\Delta_{n'} - \epsilon_{n'}\frac{d\Delta_{n'}}{d\epsilon_{n'}}}{(\epsilon^{2}_{n'} + \Delta^{2}_{n'})^{3/2}}
		\right) = 0
	\end{equation}
	In the last line we used that the integrand is a total derivative.
	We thus verify that the particle density is conserved if the relation between the vertex function and the self-energy is given by \cref{eq:ll_1}.

	With some extra effort, one can extend the analysis to finite $\Omega_m$ and show that the solution of
	\cref{eq:ll_2} is
	\begin{equation}
		\hat{\Gamma}_{n+m, n} = \hat{\tau}_0 + i \frac{\hat{\tau}_3\hat{\Sigma}_{n+m}\hat{\tau}_3 - \hat{\Sigma}_n}{\Omega_m}
	\end{equation}
	or in components
	\begin{equation}
		{\Gamma}^{(0)}_{n+m, n}  = 1 + \frac{\Sigma_{n+m} - \Sigma_{n}}{\Omega_m},\quad
		{\Gamma}^{(1)}_{n+m, n} = -i\frac{\phi_{n+m} + \phi_n}{\Omega_m}.
		\label{ll_3}
	\end{equation}
	This result has been obtained by Nambu~\cite{Nambu1960} by different means.
	As we said, at
	$\Omega_m\to 0$  these relations reduce to
	\begin{equation}
		{\Gamma}^{(0)}_{n+m, n}  \overset{\Omega_m\to0}{\longrightarrow} 1 + \frac{d\Sigma_{n}}{d\epsilon_n},\quad
		{\Gamma}^{(1)}_{n+m, n} \overset{\Omega_m \to 0}{\longrightarrow} -i\frac{2\phi_{n}}{\Omega_m}.
		\label{eq:WI-cgamma}
	\end{equation}

	Comparing Ward identities for spin and charge vertices, \cref{eq:ch_12,eq:WI-cgamma}, we see that the
	ones for $\Gamma^{(0)}$ are identical, while the ones for $\Gamma^{(1)}$ are different.  In particular, for the spin vertex $\Gamma^{(1)}$ vanishes for frequency-independent gap, while for the charge vertex it remains finite and moreover is singular at $\Omega_m \to 0$. As we said, the origin for the difference is in the fact that charge fluctuations couple linearly to massless phase fluctuations and spin fluctuations do not couple to phase fluctuations.

	\subsubsection{Ward identity for momentum conservation}
	\label{sec:ward-mom}

	For completeness, we also consider the Ward identity associated with translational invariance, i.e., with conservation of the total momentum.
	Let us consider a model with action
	\begin{multline}
		S = \sum_{n,\mathbf{p}} \bar{\Psi}(\mathbf{p}, \tau)\left[ -\partial_\tau + \hat{H}(\mathbf{p})\right]\Psi(\mathbf{p}, \tau)\\
		+ \frac{1}{2}\sum_{\mathbf{k},\mathbf{k}', \mathbf{q}}\int d\tau d\tau'
		V(t-t', \mathbf{q})
		\bar{\Psi}(\mathbf{k} + \frac{\mathbf{q}}{2}, \tau)
		\hat{\tau}_3
		\Psi(\mathbf{k} - \frac{\mathbf{q}}{2}, \tau)
		\bar{\Psi}(\mathbf{k}' - \frac{\mathbf{q}}{2}, \tau')
		\hat{\tau}_3
		\Psi(\mathbf{k} + \frac{\mathbf{q}}{2}, \tau')
	\end{multline}
	where $\Psi$ are Nambu spinors and we allow for the possibility of a time-dependent interaction due to exchange of bosons.
	The
	local symmetry transformation
	associated with translational invariance and appearing in Noether's theorem is $\Psi(\mathbf{p}, \tau) = e^{i\boldsymbol{\alpha}(\tau)\cdot \mathbf{p}}\Psi'(\mathbf{p}, \tau)$ for the Nambu spinors.
	Under such a change of variables the action changes as
	\begin{equation}
		S[\bar{\psi},\psi]
		=
		S[\bar{\psi}',\psi']
		+ \delta S[\bar{\psi}', \psi] + O(\alpha^2)
	\end{equation}
	where
	\begin{multline}
		\delta S
		= \int d\tau  \boldsymbol{\alpha}(\tau) \cdot \partial_t \sum_{\mathbf{p}} \mathbf{p}\bar{\psi}(\mathbf{p}, \tau)\psi(\mathbf{p},\tau)\\
		-i \frac{1}{2}\sum_{\mathbf{k},\mathbf{k}', \mathbf{q}}d\tau d\tau'
		\mathbf{q} \cdot \left(\boldsymbol{\alpha}(\tau) - \boldsymbol{\alpha}(\tau')\right)
		V(t-t', \mathbf{q})
		\bar{\psi}(\mathbf{k} + \frac{\mathbf{q}}{2}, \tau)
		\hat{\tau}_3
		\psi(\mathbf{k} - \frac{\mathbf{q}}{2}, \tau)
		\bar{\psi}(\mathbf{k}' - \frac{\mathbf{q}}{2}, \tau')
		\hat{\tau}_3
		\psi(\mathbf{k} + \frac{\mathbf{q}}{2}, \tau').
	\end{multline}
	This defines, through Noether's theorem, the total momentum of the system $\Braket{\mathbf{P}(\tau)}$ via
	\begin{equation}
		\delta S \equiv i \int d\tau \boldsymbol{\alpha(\tau)}\cdot \partial_t\mathbf{P}(\tau).
	\end{equation}
	Note that \emph{only for an instantaneous interaction is the total fermionic momentum separately conserved}.
	This simply reflects the fact that the bosons mediating the interaction may carry momentum too.

	In the usual fashion~\cite{Fujikawa1979} one may obtain a Ward identity by considering such a symmetry transformation within the functional integral.
	Specifically consider the following expectation value, where we perform a change of coordinates in the functional integral,
	\begin{multline}
		\Braket{\Psi(\mathbf{k}, \tau) \bar{\Psi}(\mathbf{k}', \tau)}
		=
		\frac{1}{Z}\oint \mathcal{D}[\bar{\Psi}, \Psi]
		\Psi(\mathbf{k}, \tau) \bar{\Psi}(\mathbf{k}', \tau)   e^{-S[\bar{\Psi}, \Psi]}\\
		=
		\frac{1}{Z}\oint \mathcal{D}[\bar{\Psi}, \Psi]
		(1 + i \boldsymbol{\alpha}(\tau) \cdot \mathbf{k})\Psi'(\mathbf{k}, \tau) \bar{\Psi}(\mathbf{k}', \tau)
		(1 - i \boldsymbol{\alpha}(t') \cdot \mathbf{k}')
		e^{-S[\bar{\Psi}', \Psi']}(1 - \delta S[\bar{\Psi}', \Psi'])
		+ O(\alpha^2).
	\end{multline}
	Using the fact that the measure is invariant under the change of variables $\Psi\to\Psi'$ we then obtain
	\begin{equation}
		(i \boldsymbol{\alpha}(\tau) \cdot \mathbf{k}
		- i \boldsymbol{\alpha}(t') \cdot \mathbf{k}'
		)\Braket{\Psi(\mathbf{k}, \tau) \bar{\Psi}(\mathbf{k}', \tau)}
		+ O(\alpha^2)=
		\Braket{\Psi(\mathbf{k}, \tau) \bar{\Psi}(\mathbf{k}', \tau)\delta S[\bar{\Psi},\Psi]}.
		\label{eq:WI-tau}
	\end{equation}
	The expectation value on the LHS is simply the Green's function, while the RHS is related to the vertex function via the usual rule, expressed here in terms of Matsubara frequencies,
	\begin{equation}
		\Braket{\Psi_n(\mathbf{k}) \bar{\Psi}_{n'}(\mathbf{k}')
			\sum_{n'',\mathbf{p}}\bar{\Psi}_{n''}(\mathbf{p})\hat{\gamma}_{n''}(\mathbf{p})\Psi_{n''}(\mathbf{p})
		}
		\equiv \hat{\mathcal{G}}_{n}(\mathbf{k})\hat{\Gamma}_{n, n'}(\mathbf{k})\hat{\mathcal{G}}_{n'}(\mathbf{k}') \delta(\mathbf{k}-\mathbf{k}')
	\end{equation}
	where $\hat{\Gamma}$ is the fully renormalized vertex corresponding to $\hat{\gamma}$.
	Using these expectation values and writing the correlators in terms of Matsubara frequencies we then can rewrite \cref{eq:WI-tau} as
	\begin{multline}
		T \sum_n \left[
			i \boldsymbol{\alpha}_m \cdot \mathbf{k}
			\left(
			-\hat{\mathcal{G}}_{n-m}(\mathbf{k})
			\beta \delta_{n-n'-m}
			\delta(\mathbf{k}-\mathbf{k'})
			\right)
			\right]
		-T \sum_n \left[
			i \boldsymbol{\alpha}_m \cdot \mathbf{k}'
			\left(
			-\hat{\mathcal{G}}_{n}(\mathbf{k}')
			\beta \delta_{n-n'-m}
			\delta(\mathbf{k}-\mathbf{k'})
			\right)
			\right]\\
		= i \sum_m (i\Omega_m) \boldsymbol{\alpha}_m
		\hat{\mathcal{G}}_{n}(\mathbf{k})\hat{\Gamma}^\text{mom}_{n, n-m}(\mathbf{k})\hat{\mathcal{G}}_{n-m}(\mathbf{k}') \delta(\mathbf{k}-\mathbf{k}').
	\end{multline}
	We now use the fact that $\alpha_m$ is arbitrary and additionally act on both sides with the inverse of the Green's functions to arrive at the Ward identity for the momentum vertex
	\begin{equation}
		i\Omega_m\hat{\Gamma}_{n+m, n}^\text{mom}(\mathbf{k}) = \mathbf{k} \left(
		\hat{\mathcal{G}}^{-1}_{n+m}
		- \hat{\mathcal{G}}^{-1}_{n}
		\right)
	\end{equation}
	or
	\begin{equation}
		\hat{\Gamma}_{n+m, n}^\text{mom}(\mathbf{k}) = \mathbf{k} \left(
		1 + i\frac{\hat{\Sigma}^{-1}_{n+m}
			- \hat{\Sigma}^{-1}_{n}}{\Omega_m}
		\right).
		\label{eq:WI-mom}
	\end{equation}
	This relation holds for
	both Galilean-invariant and non-Galilean-invariant systems.
	However,
	in general this does not determine the form of the current vertex.
	Only \emph{in a Galilean invariant system}, is the current given by
	\begin{equation}
		\mathbf{J}(\tau) = \frac{e}{m}\mathbf{P}(\tau)
		= \sum_{\mathbf{p}}e\frac{\mathbf{p}}{m}\bar{\Psi}(\mathbf{p},\tau)\Psi(\mathbf{p},\tau)
	\end{equation}
	and thus the renormalized current vertex is determined directly from \cref{eq:WI-mom}.

	\section{\textxi-integrated Green's function with supercurrent}
	\label{sec:qc-with-super}
	For a translationally invariant state, we can express the inverse Green's function to second order in $\mathbf{Q}$ as
	\begin{equation}
		\hat{\mathcal{G}}_k^{-1} = i\tilde{\Sigma}_n - \frac{\mathbf{k}}{m_1} \cdot \mathbf{Q} - \left(\xi_\mathbf{k} + \frac{Q^2}{2m_2} + \chi_k\right) \hat{\tau}_3  -\phi_k\hat{\tau}_1
	\end{equation}
	The $\xi$-integrated Green's function is then, according to \cref{eq:gqc},
	\begin{equation}
		\hat{g}_n(\mathbf{k}_F) = \frac{1}{\pi} \int d\xi
		\frac{(\tilde{\Sigma}_n (\mathbf{k}_F) + i\mathbf{v}_F \cdot \mathbf{Q})\hat{\tau}_3 - i(\xi + \frac{Q^2}{2m_2} + \chi_n(\mathbf{k}_F))+ \phi_n(\mathbf{k}_F)\hat{\tau}_2}{
			(\tilde{\Sigma}_n (\mathbf{k}_F) + i\mathbf{v}_F \cdot \mathbf{Q})^2  +(\xi + \frac{Q^2}{2m_2} + \chi_n(\mathbf{k}_F) )^2 + |\phi_n(\mathbf{k}_F)|^2
		}.
	\end{equation}
	As long as the quasi-classical approximation holds, we can shift $\xi$ to eliminate $(Q^2/2m_2) + \chi$ leaving simply
	\begin{equation}
		\hat{g}_n(\mathbf{k}_F) =
		\frac{(\tilde{\Sigma}_n (\mathbf{k}_F) + i\mathbf{v}_F \cdot \mathbf{Q})\hat{\tau}_3+ \phi_n(\mathbf{k}_F)\hat{\tau}_2}{
			\sqrt{(\tilde{\Sigma}_n (\mathbf{k}_F) + i\mathbf{v}_F \cdot \mathbf{Q})^2   + |\phi_n(\mathbf{k}_F)|^2}
		}.
	\end{equation}

	\section{Quasi-classical free energy functional for Eliashberg theory}
	\label{sec:quasi-classical-lw}
	Starting with the inverse Green's function
	\begin{equation}
		\hat{\mathcal{G}}_k^{-1} = i\tilde{\Sigma}_n - \frac{\mathbf{k}}{m_1} \cdot \mathbf{Q} - (\xi_\mathbf{k} + \frac{Q^2}{2m_2} + \chi_k) \hat{\tau}_3  -\phi_k\hat{\tau}_1
	\end{equation}
	we can evaluate the quasi-classical free energy as a sum of a kinetic term
	\begin{equation}
		F_\text{kin} = - T\ln(-\det(-\beta \hat{\mathcal{G}}^{-1}_k))
	\end{equation}
	and a potential term
	\begin{equation}
		F_\text{pot} = -\frac{1}{2}T^2\sum_{k, k'}V_{k-k'}\tr[\hat{\tau}_3 \hat{\mathcal{G}}_k\hat{\tau}_3\hat{\mathcal{G}}_{k'}].
	\end{equation}
	Note the presence of an additional minus sign inside the log of the kinetic term, coming from the Nambu spinor measure
	\begin{equation}
		d\bar{\psi}_{k\uparrow} d\psi_{k\uparrow}
		d\bar{\psi}_{-k\downarrow} d\psi_{-k\downarrow}
		=
		-d\bar{\psi}_{k\uparrow} d\psi_{k\uparrow}
		d\psi_{-k\downarrow} d\bar{\psi}_{-k\downarrow}
		= d\bar{\Psi}_k d\Psi_k.
	\end{equation}

	\subsection{Kinetic term}
	For the kinetic term we can start by writing
	\begin{equation}
		F_\text{kin} = - T\sum_k \ln(-\beta^2 \det \hat{\mathcal{G}}^{-1}_k).
	\end{equation}
	The determinant is
	\begin{equation}
		D_k \equiv - \det \hat{\mathcal{G}}^{-1}_k
		=  \left(\tilde{\Sigma}_n + i\frac{\mathbf{k}}{m_1} \cdot \mathbf{Q}\right)^2 + \left(\xi_\mathbf{k} + \frac{Q^2}{2m_2} + \chi_k\right)^2 +\phi_k^2.
	\end{equation}
	To regulate the sum, we will first integrate over momentum within finite limits and then take the limits to infinity at the end.
	Defining $S_n(\mathbf{k}_F)^2 = \left(\tilde{\Sigma}_n + i\frac{\mathbf{k}}{m_1} \cdot \mathbf{Q}\right)^2  +\phi_k^2$
	\begin{equation}
		F_\text{kin} = - \nu T\sum_n \oint_{FS} \frac{d \mathbf{k}_F}{S_{d-1}}\int_{-\Lambda}^{\Lambda} d\xi
		\ln \frac{S_n(\mathbf{k})^2 +\left(\xi_\mathbf{k} + \frac{Q^2}{2m_2} + \chi_k\right)^2}{T^2}
		\equiv - \nu T\sum_n \oint_{FS} \frac{d \mathbf{k}_F}{S_{d-1}} I_{n}(\mathbf{k}_F)
	\end{equation}
	Let us define dimensionless variables
	\begin{equation}
		r \equiv \frac{\chi + \frac{Q^2}{2m_1} }{\Lambda},\quad
		s \equiv \frac{S}{\Lambda},\quad
		z \equiv \frac{\xi}{\Lambda}.
	\end{equation}
	which lets us write
	\begin{equation}
		I_n(\mathbf{k}_F) = \Lambda \int_{-1}^{1} dz
		\ln\left[\frac{\Lambda^2}{T^2}
			\left((z + r)^2 + s^2 \right)\right]
		= 4 \Lambda \ln\frac{\Lambda}{T} +
		\Lambda \int_{-1 + r}^{1+r} dz
		\ln \left(z^2 + s^2 \right).
	\end{equation}
	We can expand Taylor in the limits of the integral
	\begin{equation}
		\int_{-1 + r}^{1+r} dx f(x) =
		\int_{-1}^{1} dx f(x)  + r (f(1) - f(-1)) + O(r^2)
	\end{equation}
	and we find
	\begin{equation}
		I_n(\mathbf{k}_F)
		= 4 \Lambda \ln\frac{\Lambda}{T} +
		\Lambda \int_{-1}^{1} dz
		\ln \left(z^2 + s^2 \right) + O(r^2).
	\end{equation}
	We can thus safely neglect $\chi$ and $Q^2/(2m_2)$ since they are, by assumption, much smaller than $\Lambda$.
	We evaluate the remaining integral using integration by parts
	\begin{multline}
		\frac{I_n(\mathbf{k}_F) }{\Lambda}
		= 4 \ln\frac{\Lambda}{T} +
		+ \cancel{\left.z \ln(z^2 + s^2)\right|_{-1}^1}
		-\int_{-1}^{1}dz\frac{2z^2}{z^2 + s^2}\\
		= 4 \ln\frac{\Lambda}{T} +
		-2\int_{-1}^{1}dz\frac{z^2+s^2 -s^2}{z^2 + s^2}\\
		= 4 \ln\frac{\Lambda}{T} +
		-4
		+2\int_{-1/s}^{1/s}dy\frac{1}{1+y^2}\\
		= 4\ln\frac{\Lambda}{eT}
		+2s\left.\tan^{-1}y\right|_{-1/s}^{1/s}
		= 4\ln\frac{\Lambda}{eT}
		+2\pi s + O(s^2)
	\end{multline}
	In the limit of $\Lambda \to \infty$, the integral $I_n(\mathbf{k}_F)$ consists of an (infinite) constant term, which is irrelevant for the response of the system plus a a term of order $\Lambda^0$
	\begin{equation}
		\lim_{\Lambda \to \infty} I_{n}(\mathbf{k}_F)
		= C_\Lambda + 2\pi S_n(\mathbf{k}_F).
	\end{equation}
	We thus arrive at the expression for the kinetic part of the quasi-classical free energy
	\begin{equation}
		F_\text{kin} = - 2\pi\nu T \sum_n \oint_{FS}\frac{d\mathbf{k}_F}{S_{d-1}}
		\sqrt{(\tilde{\Sigma}_n(\mathbf{k}_F) + i\mathbf{v}_F \cdot \mathbf{Q})^2 + \phi_n(\mathbf{k}_F)^2}.
	\end{equation}

	\subsection{Potential term}
	The potential term is straightforwardly simplified using the definition of the $\xi$-integrated Green's function~\cref{eq:gqc}
	\begin{multline}
		F_\text{pot} = -\frac{1}{2}T^2\sum_{n, n'}
		\oint_{FS} \frac{d\mathbf{k}_F}{S_{d-1}}
		\oint_{FS} \frac{d\mathbf{k}'_F}{S_{d-1}}
		V_{n-n'}(\abs{\mathbf{k}_F - \mathbf{k}_F'})\\
		\times \nu^2 \int d\xi \int d\xi' \tr[\hat{\tau}_3 \hat{\mathcal{G}}_n(\xi, \mathbf{k}_F)\hat{\tau}_3\hat{\mathcal{G}}_{n'}(\xi', \mathbf{k}'_F)]\\
		= \nu^2\pi^2 T^2\sum_{n, n'}
		\oint_{FS} \frac{d\mathbf{k}_F}{S_{d-1}}
		\oint_{FS} \frac{d\mathbf{k}'_F}{S_{d-1}}
		V_{n-n'}(\abs{\mathbf{k}_F - \mathbf{k}_F'})\\
		\times
		\left(g_n(\mathbf{k}_F)g_{n'}(\mathbf{k}'_F) + f_n(\mathbf{k}_F) f_{n'}(\mathbf{k}'_F)\right)
	\end{multline}

	\subsection{Total quasi-classical expression}
	In combining the two terms we note that we can rewrite the kinetic part,
	\begin{multline}
		F_\text{kin}
		= - 2\pi\nu T \sum_n \oint_{FS}\frac{d\mathbf{k}_F}{S_{d-1}}
		\frac{\Upsilon_n(\mathbf{k}_F)^2 + \phi_n(\mathbf{k}_F)^2}{S_n(\mathbf{k}_F)}\\
		= - 2\pi\nu T \sum_n \oint_{FS}\frac{d\mathbf{k}_F}{S_{d-1}}
		\frac{\varpi_n(\mathbf{k}_F)\Upsilon_n(\mathbf{k}_F) + \Delta_n(\mathbf{k}_F)\phi_n(\mathbf{k}_F)}{S_n(\mathbf{k}_F)}\\
		= - 2\pi\nu T \sum_n \oint_{FS}\frac{d\mathbf{k}_F}{S_{d-1}}
		\left(
		\Upsilon_n(\mathbf{k}_F) g_n(\mathbf{k}_F)
		+ \phi_n(\mathbf{k}_F) f_n(\mathbf{k}_F)
		\right)\\
		= - 2\pi\nu T \sum_n \oint_{FS}\frac{d\mathbf{k}_F}{S_{d-1}}
		\varpi_n(\mathbf{k}_F) g_n(\mathbf{k}_F)\\
		-2\nu^2\pi^2 T^2\sum_{n, n'}
		\oint_{FS} \frac{d\mathbf{k}_F}{S_{d-1}}
		\oint_{FS} \frac{d\mathbf{k}'_F}{S_{d-1}}
		V_{n-n'}(\abs{\mathbf{k}_F - \mathbf{k}_F'})\\
		\left(g_n(\mathbf{k}_F)g_{n'}(\mathbf{k}'_F) + f_n(\mathbf{k}_F) f_{n'}(\mathbf{k}'_F)\right)
	\end{multline}
	where in the last equality we used the gap equation.
	We see that the second term is just $-2F_\text{pot}$ and thus we have
	\begin{multline}
		F =
		- 2\pi\nu T \sum_n \oint_{FS}\frac{d\mathbf{k}_F}{S_{d-1}}
		\varpi_n(\mathbf{k}_F) g_n(\mathbf{k}_F)\\
		-\nu^2\pi^2 T^2\sum_{n, n'}
		\oint_{FS} \frac{d\mathbf{k}_F}{S_{d-1}}
		\oint_{FS} \frac{d\mathbf{k}'_F}{S_{d-1}}
		V_{n-n'}(\abs{\mathbf{k}_F - \mathbf{k}_F'})\\
		\left(g_n(\mathbf{k}_F)g_{n'}(\mathbf{k}'_F) + f_n(\mathbf{k}_F) f_{n'}(\mathbf{k}'_F)\right).
	\end{multline}

	\section{Evaluation of dynamic coefficient}
	\label{sec:dynamic}
	The integral of $\xi$ may be performed immediately in equation \cref{eq:kappa}.
	By rescaling the integration variable, we may then express the frequency integral as
	\begin{equation}
		\kappa
		= 2\nu \lim_{\Lambda\to\infty} \int \frac{dz}{2\pi} \frac{2}{1+z^2Z^2(\Lambda z)}
		\label{eq:kxifirst}
	\end{equation}
	In general $Z(\epsilon)$ has the following properties
	\begin{itemize}
		\item $Z$ is an even function of frequency
		\item at large frequencies $Z$ goes to $1$, i.e.\ $\exists 0<\Omega_{FL}\ll \Lambda$ s.t. $\forall |\epsilon| > \Omega_{FL}, Z_\epsilon  - 1 \leq (\Omega_{FL}/\Lambda)$.
	\end{itemize}
	With this in mind, we split the integration into a low energy and a high energy part
	\begin{equation}
		\kappa
		\approx 2\nu \int \frac{dz}{2\pi} \frac{2}{1+z^2Z^2(\Lambda z)}
		= 4\nu \left(\int_{0}^{\Omega_{FL}/\Lambda} + \int_{\Omega_{FL}/\Lambda}^\infty\right)\frac{dz}{2\pi} \frac{2}{1+z^2Z^2(\Lambda z)}
	\end{equation}
	The first term can then be bounded by
	\begin{equation}
		\int_0^{\Omega_\text{FL}/\Lambda} dz  \frac{1}{1+z^2Z^2(\Lambda z)} \leq \frac{\Omega_\text{FL}}{\Lambda}.
	\end{equation}
	while for the second term
	\begin{multline}
		\int_{\Omega_\text{FL}/\Lambda}^\infty dz  \frac{1}{1+z^2Z^2(\Lambda z)}
		=
		\int_{\Omega_\text{FL}/\Lambda}^\infty dz  \frac{1}{1+z^2}
		\frac{1}{1+\frac{z^2}{1+z^2}(Z^2(\Lambda z)-1)}\\
		\approx
		\int_{\Omega_\text{FL}/\Lambda}^\infty dz  \frac{1}{1+z^2}
		\left(1 - \frac{z^2}{1+z^2}(Z^2(\Lambda z)-1) + \cdots\right)
		\approx
		\int_0^\infty dz  \frac{1}{1+z^2}
		+ O(\frac{\Omega_{FL}}{\Lambda}).
	\end{multline}
	We thus arrive at
	\begin{equation}
		\kappa = 2\frac{\nu}{\pi} \int_{-\infty}^\infty \frac{1}{1+z^2}
		= 2\nu
	\end{equation}
	where all other terms vanish in the $\Lambda \to \infty$ limit.
\end{widetext}
\end{document}